\documentclass[12pt]{article}
\usepackage{graphicx}
\usepackage{epsfig}
\usepackage{epstopdf}
\DeclareGraphicsExtensions{.pdf,.eps,.png,.jpg,.mps}
\usepackage{amssymb,graphicx}

\usepackage{graphicx}
\usepackage{wrapfig}
\usepackage{lscape}
\usepackage{rotating}

\usepackage{cite}

\def\lsim{\mathrel{\rlap{\lower3pt\hbox{\hskip0pt$\sim$}}
     \raise1pt\hbox{$<$}}}         
\def\gsim{\mathrel{\rlap{\lower4pt\hbox{\hskip1pt$\sim$}}
     \raise1pt\hbox{$>$}}}         

\usepackage{amsmath}
\usepackage{amsfonts}

\begin{document}
\begin{titlepage}

\centerline{\Large \bf Factor Models for Cancer Signatures}
\medskip

\centerline{Zura Kakushadze$^\S$$^\dag$\footnote{\, Zura Kakushadze, Ph.D., is the President of Quantigic$^\circledR$ Solutions LLC,
and a Full Professor at Free University of Tbilisi. Email: zura@quantigic.com} and Willie Yu$^\sharp$\footnote{\, Willie Yu, Ph.D., is a Research Fellow at Duke-NUS Medical School. Email: willie.yu@duke-nus.edu.sg}}
\bigskip

\centerline{\em $^\S$ Quantigic$^\circledR$ Solutions LLC}
\centerline{\em 1127 High Ridge Road \#135, Stamford, CT 06905\,\,\footnote{\, DISCLAIMER: This address is used by the corresponding author for no
purpose other than to indicate his professional affiliation as is customary in
publications. In particular, the contents of this paper
are not intended as an investment, legal, tax or any other such advice,
and in no way represent views of Quantigic$^\circledR$ Solutions LLC,
the website \underline{www.quantigic.com} or any of their other affiliates.
}}
\centerline{\em $^\dag$ Free University of Tbilisi, Business School \& School of Physics}
\centerline{\em 240, David Agmashenebeli Alley, Tbilisi, 0159, Georgia}
\centerline{\em $^\sharp$ Centre for Computational Biology, Duke-NUS Medical School}
\centerline{\em 8 College Road, Singapore 169857}
\medskip
\centerline{(April 28, 2016)}

\bigskip
\medskip

\begin{abstract}
{}We present a novel method for extracting cancer signatures by applying statistical risk models (http://ssrn.com/abstract=2732453) from quantitative finance to cancer genome data. Using 1389 whole genome sequenced samples from 14 cancers, we identify an ``overall" mode of somatic mutational noise. We give a prescription for factoring out this noise and source code for fixing the number of signatures. We apply nonnegative matrix factorization (NMF) to genome data aggregated by cancer subtype and filtered using our method. The resultant signatures have substantially lower variability than those from unfiltered data. Also, the computational cost of signature extraction is cut by about a factor of 10. We find 3 novel cancer signatures, including a liver cancer dominant signature (96\% contribution) and a renal cell carcinoma signature (70\% contribution). Our method accelerates finding new cancer signatures and improves their overall stability. Reciprocally, the methods for extracting cancer signatures could have interesting applications in quantitative finance.
\end{abstract}
\medskip
\end{titlepage}

\newpage
\section{Introduction and Summary}

{}One in eight human deaths is caused by cancer. Cancer stands out among diseases for it stems from somatic alterations in the genome. One common type of somatic alterations found in cancer is due to single nucleotide variations (SNVs) or alterations to single bases in the genome. These SNVs are accumulated throughout the lifetime of the cancer via exposures to different mutational processes. These processes can be endogenous to the cell such as imperfect DNA replication during cell division or spontaneous cytosine deamination \cite{Goodman}, \cite{Lindahl}. They can also be exogenous due to exposures to chemical insults or ultraviolet radiation \cite{Loeb}, \cite{Ananthaswamy}. All of these mutational processes, whether extrinsic or intrinsic, will leave evidence of their activity in the cancer genome characterized by distinctive alteration patterns or mutational signatures. From a knowledge standpoint, if one can identify all signatures and thus all mutational processes contributing to cancer, then one can begin to understand the origins of cancer and its development. From a therapeutic point of view, if there are no discernible patterns of mutations between different cancer types, then different cancers will mostly likely require their own type-specific or even patient-specific therapeutics. However, if there is a much smaller number of mutational signatures describing all or most cancer types, then a therapeutic for one cancer type with certain mutational signatures present may very well be applicable across other cancer types with the same or similar mutational signatures.\footnote{\, Another practical motivation for identifying cancer signatures is prevention, by pairing the signatures observed in cancer samples with those caused by exposure to various carcinogens.}

{}At present, the identification of mutational signatures involves analyzing SNV patterns present in a cohort of DNA sequenced whole cancer genomes. SNVs found in each cancer genome can be classified into 96 distinct mutation categories.\footnote{\, In brief, DNA is a double helix of two strands, and each strand is a string of letters A, C, G, T corresponding to adenine, cytosine, guanine and thymine, respectively. In the double helix, A in one strand always binds with T in the other, and G always binds with C. This is known as base complementarity. Thus, there are six possible base mutations C $>$ A, C $>$ G, C $>$ G, T $>$ A, T $>$ C, T $>$ G, whereas the other six base mutations are equivalent to these by base complementarity. Each of these 6 possible base mutations is flanked by 4 possible bases on each side thereby producing $4 \times 6 \times 4 = 96$ distinct mutation categories.} The data is organized into a matrix $G_{is}$, where the rows correspond to the $N=96$ mutation categories, the columns correspond to $d$ samples, and each element is a nonnegative occurrence count of a given mutation in a given sample. The commonly accepted method for extracting cancer signatures from $G_{is}$ \cite{Alexandrov.NMF} is via nonnegative matrix factorization (NMF) \cite{Paatero}, \cite{LeeSeung}. Under NMF the matrix $G$ is approximated via $G \approx W~H$, where $W_{iA}$ is an $N\times K$ matrix, $H_{As}$ is a $K\times d$ matrix, and both $W$ and $H$ are nonnegative. The appeal of NMF is its biologic interpretation whereby the $K$ columns of the matrix $W$ are interpreted as the weights with which the $K$ cancer signatures contribute into the $N=96$ mutation categories, and the columns of the matrix $H$ are interpreted as the exposures to the $K$ signatures in each sample. The price to pay for this is that NMF, which is an iterative procedure, is computationally costly and depending on the number of samples $d$ it can take days or even weeks to run it. Furthermore, it does not automatically fix the number of signatures $K$, which must either be guessed or obtained via trial and error, thereby adding to the computational cost. Additional considerations include: i) out-of-sample instability, i.e., the signatures obtained from non-overlapping sets of samples can be dramatically different; ii) in-sample instability, i.e., the signatures can have a strong dependence on the initial iteration choice; and iii) samples with low counts or sparsely populated samples (i.e., those with many zeros -- such samples are ubiquitous, e.g., in exome data) are usually deemed not too useful as they contribute to the in-sample instability.

{}Happily, a conceptually similar problem is well-studied in quantitative finance and we can simply borrow from the arsenal of tools developed there once we establish a {\em dictionary} between the biologic and finance quantities. Thus, in the quant finance context one deals with a portfolio of $N$ stocks, which are analogous to the $N =96$ mutation categories. The data consists of a time series of $d$ (e.g., daily) stock returns for each stock, so we have an $N\times d$ matrix $R_{is}$. The $d$ observations in the time series of stock returns are analogous to the $d$ samples in the cancer data. The returns $R_{is}$ are analogous to the counts $G_{is}$ except that the returns $R_{is}$ need not be positive. However, this does not affect what we wish to borrow from quantitative finance.

{}The sample {\em correlation}\footnote{\, The sample covariance matrix $C_{ij}=\sigma_i\sigma_j\Psi_{ij}$, where $\sigma_i^2$ are the sample variances and $\Psi_{ii}\equiv 1$.} matrix $\Psi_{ij}$ computed based on the time series of stock returns contains important information about the correlation structure of the returns. Its spectral decomposition via principal components provides a tool for identifying common {\em risk factors} underlying the returns, i.e., up to an error term, we have $R \approx \Omega~F$, where the columns of the $N\times K$ (so-called factor loadings) matrix $\Omega_{iA}$ are related to the first $K$ principal components of $\Psi_{ij}$, and the columns of the $K\times d$ matrix $F_{As}$ are the time series of the factor returns. In our dictionary, the matrix $\Omega$ is analogous to the matrix $W$, and the matrix $F$ is analogous to the matrix $H$.

{}So, why is this useful, especially considering that the matrices $\Omega$ and $F$ are not nonnegative in the finance context? There are two pieces of useful information we can extract from this analogy. First, algorithms for fixing the number of factors $K$ are readily available \cite{KYb}. So, if we compute the sample correlation matrix $\Psi_{ij}$ based on our occurrence count matrix $G_{is}$ and apply the methods employed in {\em statistical risk models}, we can fix the number of cancer signatures (or at least a useful expected range for it) based on purely statistical methods.\footnote{\, We review these methods in detail below. The gist of the idea is to identify $K$  which optimizes the contributions from the factors (signatures) and the error terms into the diagonal of $\Psi_{ij}$.} E.g., one such method proposed in \cite{KYb} is based on eRank (effective rank) \cite{RV} of $\Psi_{ij}$ and appears to work well for cancer signatures.

{}Second, intuitively it is clear that there is a lot of noise in the occurrence count data $G_{is}$. There are mutations that also occur in healthy humans, e.g., via imperfections in DNA repair. Furthermore, one can expect that in the presence of cancer such or similar mutations not directly associated with cancer signatures may become more ubiquitous due to disruption in the normal operation of various processes (including repair) in DNA. This ``background noise" obscures the signatures and must be identified and factored out of the data prior to attempting any signature extraction. In the context of finance this is well-known as the ``market" mode, which corresponds to the overall movement of the broad market affecting all stocks (to varying degrees) -- cash inflow (outflow) into (from) the market tends to push stock prices higher (lower). This is the market risk factor. To mitigate this risk factor, one can, e.g., hold a dollar-neutral portfolio of stocks (the same dollar holdings for long and short positions).\footnote{\, The ``market" mode is the (quasi uniform) 1st principal component of $\Psi_{ij}$: $V^{(1)}_i \approx 1/\sqrt{N}$.} And we can use this analogy for cancer signatures.

{}Based on our empirical analysis, we indeed find what we term the ``overall" mode -- the analog of the ``market" mode in finance -- in the occurrence count data. It is unequivocally present. Here is a simple way to understand this ``overall" mode. The average pair-wise correlation $\Psi_{ij}$ between different mutations ($i\not= j$) is nonzero and is in fact high for most cancer types we study. This is noise that must be factored out. If we aggregate samples by cancer type and compute the sample correlation matrix $\Psi_{ij}$ for the so-aggregated data (across the $n = 14$ cancer types we study), the average correlation is about 75\% if we base it on $G_{is}$, and over whopping 96\% if we use the log-based matrix instead (see below). Another way of thinking about this is that the occurrence counts in different samples are not normalized uniformly across all samples. Therefore, running NMF on a vanilla matrix $G_{is}$ could amount to mixing apples with oranges thereby obscuring the true underlying signatures.

\begin{figure}[ht]
\centering
\includegraphics[scale=1.0]{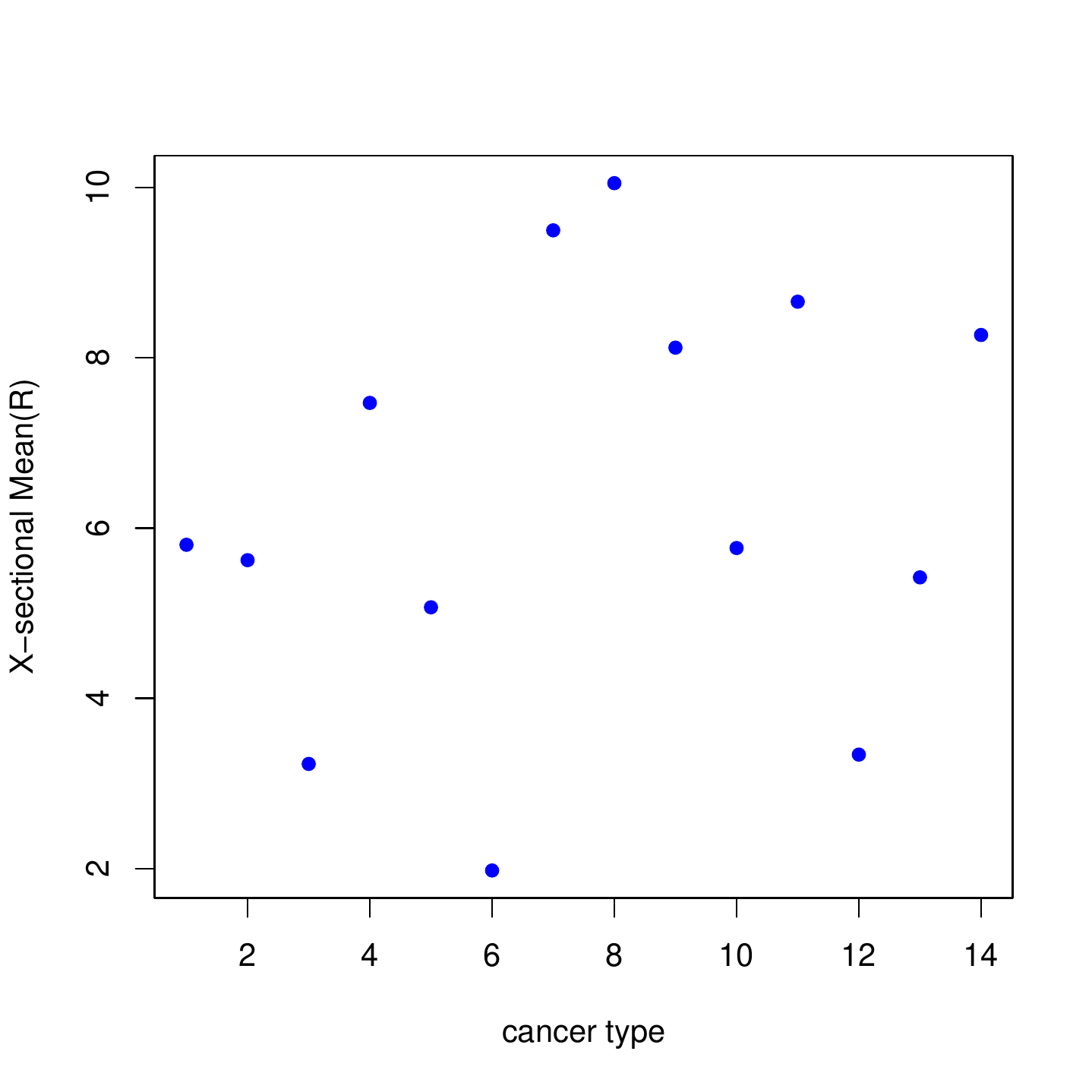}
\caption{Variability in cross-sectional means. $x$-axis: the cancer types labeled 1-14 in the order in Table \ref{table.genome.summary}. $y$-axis: the column means of the matrix $R_{is} = \ln(1 + G_{is})$.}
\label{noise.1}
\end{figure}

\begin{figure}[ht]
\centering
\includegraphics[scale=1.0]{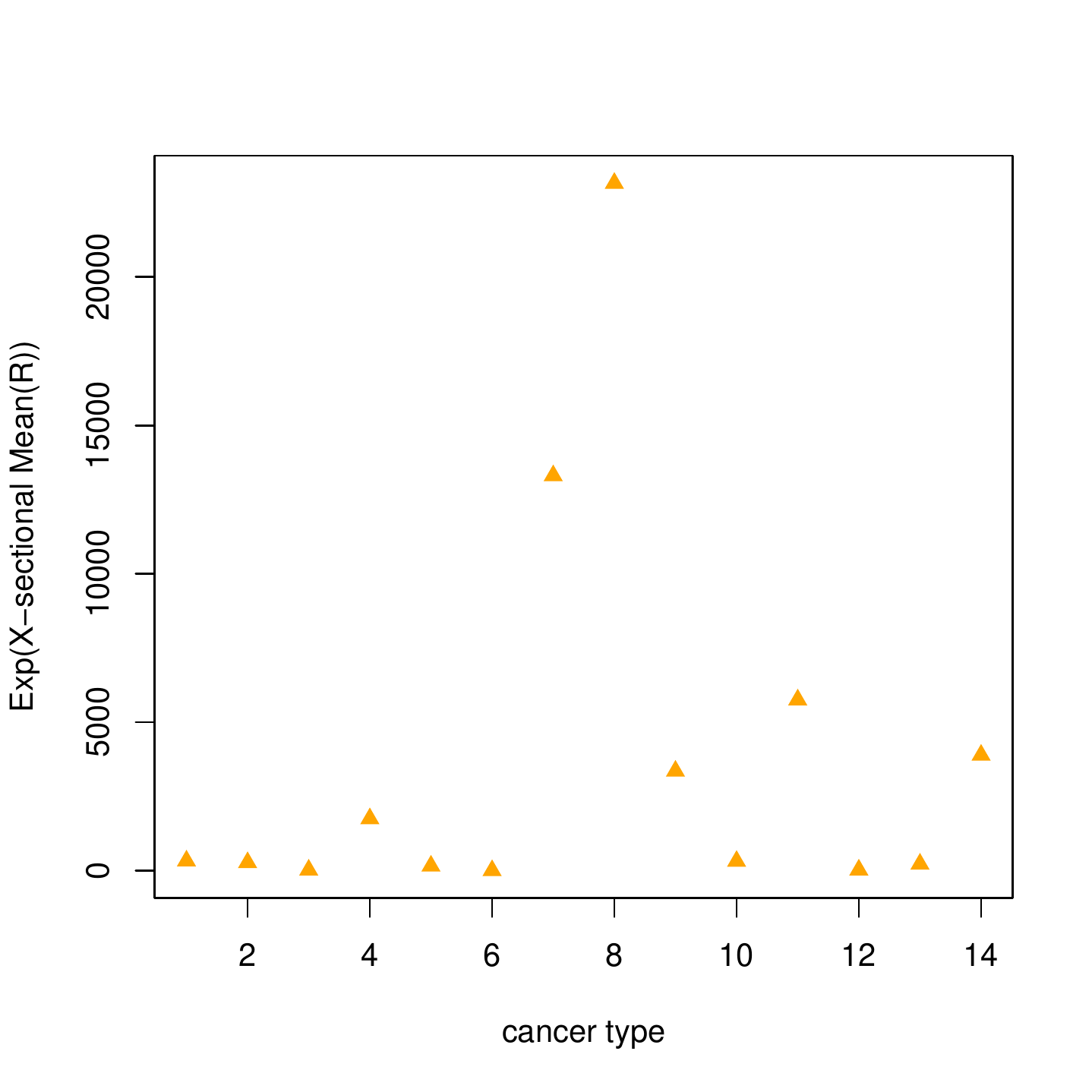}
\caption{Variability in cross-sectional means. $x$-axis: the cancer types labeled 1-14 in the order in Table \ref{table.genome.summary}. $y$-axis: exponents of the column means of the matrix $R_{is} = \ln(1 + G_{is})$.}
\label{noise.2}
\end{figure}

{}Factoring out the ``overall" mode (or ``de-noising" the matrix $G_{is}$) therefore most simply would amount to cross-sectional (i.e., across the 96 mutation categories) demeaning. Simply put, we could demean the columns of $G_{is}$. One evident issue with this is that, while the so-demeaned $G_{is}$ can be used in the context of applying statistical factor model methods to it (recall that the returns $R_{is}$ need not be positive) to fix the number of signatures, we would not be able to run NMF on such a matrix as it is no longer nonnegative. Another, more subtle issue is that distributions of counts in $G_{is}$ -- the counts being nonnegative numbers -- are not (quasi) normal but skewed, with long tails at the higher end. In fact, they are quasi log-normal, which is common for nonnegative quantities. Therefore, instead of demeaning the columns of $G$, it makes much more sense to demean the columns of $\ln(G)$ (and re-exponentiate for the purpose of running NMF). A minor hiccup is that some elements of $G_{is}$ can be 0. A simple way to deal with this is to set $R_{is} = \ln(1 + G_{is})$ and construct the correlation matrix $\Psi_{ij}$ based on $R_{is}$ (as opposed to $G_{is}$) or $R^\prime_{is}$, which is $R_{is}$ with columns demeaned -- this amounts to factoring out the ``overall" mode. We run our analysis using $R_{is}$, $R^\prime_{is}$ as well as $G_{is}$ and $G^\prime_{is}$ (which is $G_{is}$ with columns demeaned) and unequivocally find that using ``de-noised" log-based matrix $R^\prime_{is}$ works best.\footnote{\, Table \ref{table.skewness} summarizes the Mean/Median ratio and skewness for the matrices $G_{is}$ and $R_{is}$ across cancer types and mutation categories and makes the skewed nature of the counts evident. This skewness is exacerbated when we consider it across samples for many individual cancer types. Also, Figures \ref{noise.1} and \ref{noise.2} help visualize why factoring out the ``overall" mode reduces noise.}

{}So, here is a simple prescription for fixing the number of signatures using our statistical factor model based methodology. Compute $R^\prime_{is}$ as above for the occurrence counts aggregated by cancer type. I.e., $G_{is}$ is an $N\times n$ matrix, where the number of cancer types $n=14$ in our case. Compute the sample correlation matrix $\Psi_{ij}$ based on $R^\prime_{is}$, i.e., the correlations are computed across the 14 cancer types. Compute $\mbox{eRank}(\Psi_{ij})$ and round it to the nearest integer. This is the expected number of cancer signatures $K$ (excluding the ``overall" mode, which is noise). This simple procedure appears to work well for this purpose and we explain in detail why this is the case based on the statistical factor model methodology, along with another method for fixing $K$, which gives similar results. A complementary way of fixing $K$ is to compute the sample correlation matrices $[\Psi(\alpha)]_{ij}$ for each cancer type labeled by $\alpha = 1,\dots,n$ ($[\Psi(\alpha)]_{ij}$ is computed based on the samples for the $\alpha$-th cancer type), take the first principal component $[V(\alpha)]^{(1)}_i$ for each correlation matrix $[\Psi(\alpha)]_{ij}$, compute the $n\times n$ matrix of inner products $E^{(1)}_{\alpha\beta} = \sum_{i=1}^N [V(\alpha)]^{(1)}_i~[V(\beta)]^{(1)}_i$, compute $E_1 = \mbox{eRank}(E^{(1)}_{\alpha\beta})$, and identify $K$ with rounded $E_1$. This method produces essentially the same prediction for $K$ as the aforesaid method using $\mbox{eRank}(\Psi_{ij})$.

{}Once we fix the expected number of signatures $K$, we are ready to use NMF to extract cancer signatures. However, as mentioned above, running NMF on $G_{is}$ is suboptimal as it contains noise due to the ``overall" mode. A simple way to eliminate the ``overall" mode is to run NMF on the re-exponentiated matrix ${\widetilde G}_{is} = \exp(R^\prime_{is})$. Note that the elements in ${\widetilde G}_{is}$ are no longer interpreted as ``counts" -- they are fractional and low. We can include an overall normalization\footnote{\, E.g., we can take ${\widetilde G}_{is} = \exp(\mbox{Mean}(R_{is}) + R^\prime_{is})$, or ${\widetilde G}_{is} = \exp(\mbox{Median}(R_{is}) + R^\prime_{is})$, etc.\label{foot.norm}} to make it look more like the original matrix $G_{is}$, however, this does not affect the signatures extracted via NMF.\footnote{\, Technically speaking, after re-exponentiating we should subtract the extra 1 we added in the definition $R_{is} = \ln(1 + G_{is})$. This can be done using the definitions in footnote \ref{foot.norm} and the (relatively scarce) negative elements resulting from subtracting 1 should be zeroed out. However, this does not seem to affect the results much, so not to overcomplicate things we work with ${\widetilde G}_{is} = \exp(R^\prime_{is})$.} Now we are in good shape: we have the expected number of signatures $K$ and the ``de-noised" matrix ${\widetilde G}_{is}$ from which we can extract signatures via NMF.

{}Remarkably, what we find is 4 previously known signatures\footnote{\, To wit, mutational signatures 1 (spontaneous cytosine deamination), 2+13 (APOBEC mediated cytosine deamination), 4 (tobacco carcinogen related exposure) and 17 (appearing in oesophagus cancer, breast cancer, liver cancer, lung adenocarcinoma, B-cell lymphoma, stomach cancer and melanoma; mutational process unknown) of \cite{Nik-Zainal}, \cite{Alexandrov}.} plus 3 new signatures. One of the new signatures dominates liver cancer (with over 96\% contribution), with almost no peak variability. Another new signature to a lesser degree dominates renal cell carcinoma (with over 70\% contribution). The third new signature appears mostly in bone cancer, brain lower grade glioma and medulloblastoma (and also 5 other cancers to lesser degrees). We find the same signatures (plus the ``overall" mode) if we use $G_{is}$ instead of ${\widetilde G}_{is}$, but the signatures are unequivocally more stable when using ${\widetilde G}_{is}$. Simply put, removing the ``overall" mode (the noise) pays off high dividends. Now, we emphasize that our results are based on using the occurrence counts aggregated by cancer type. The advantages of this method include: i) the data is much less noisy than for samples by individual cancer type; and ii) it allows us to use {\em all} genomic data, including that with low counts. In this regard, our approach here can be readily applied to exome data, which we will report elsewhere along with extending our analysis to individual cancer types.

{}The remainder of this paper is organized as follows. In Sections \ref{sec.2}-\ref{sub.fix.K} we review the quantitative finance machinery we borrow from. Section \ref{sec.5} applies this machinery to cancer signatures. Section \ref{sec.6} discusses empirical results based on the published data for the 14 cancer types. Section \ref{sec.7} discusses our NMF results. We briefly conclude in Section \ref{sec.8}. Appendix \ref{app.IDs} lists the genome data sample IDs we use. Appendix \ref{app.code} contains our R source code for factor models. Appendix \ref{app.disc} contains some legalese.

\section{Sample Covariance Matrix}\label{sec.2}

\subsection{Sample Data}

{}In many practical applications we have $N$ objects characterized by an observable quantity, which is measured over $d$ observations for each object. The resulting data is an $N \times d$ matrix -- call it $R_{is}$ -- where the rows correspond to the objects labeled by $i=1,\dots,N$, and the columns correspond to the observations labeled by $s=1,\dots,d$. In general there can be some missing observations, i.e., NAs in $R_{is}$. However, for our purposes here it will suffice to assume that there are no NAs.

{}Here are some examples of such data. In finance we have $N$ stocks, $d$ trading days,\footnote{\, A trading day refers to a day on which the stock market is open.} and we measure daily stock returns\footnote{\, E.g., the so-called close-to-close return, i.e., the return from yesterday's closing price to today's closing price. This return can be defined as $R_{is} = P_{is} / P_{i,(s+1)} - 1$ or $R_{is} = \ln(P_{is} / P_{i,(s+1)})$ (for daily returns, usually $|R_{is}|\ll 1$, so the difference between the two definitions is mostly small). A further detail is that the closing prices $P_{is}, P_{i,(s+1)}$ are fully adjusted for any splits and dividends.} $R_{is}$. Or, e.g., $i$ labels large cities in the US (or, alternatively, zip codes), $s$ labels years, and $R_{is}$ is violent crime rate per capita. In the context of this paper, we have $N=96$ mutation types\footnote{\, We use ``mutation type" and ``mutation category" interchangeably.} occurring in various types of cancers, $d$ is the number of collected samples, and $R_{is}$ is (related to -- see below) the occurrence count for the mutation type $i$ in the sample $s$.

\subsection{Serial Covariances and Correlations}

{}We can think of the matrix $R_{is}$ as $N$ series of $d = M+1$ observations.\footnote{\, In our finance example above, $s$ labels dates in the $N$ time series. More generally, the rows of $R_{is}$ are not necessarily time series. E.g., in the context of cancer mutations, we are dealing with series of samples (without any reference to time or chronology). In what follows we will use the adjective ``serial" in the general context (be it dates, samples, etc.), not necessarily for time series.} The sample covariance matrix (SCM) is defined as an $N\times N$ matrix of pair-wise serial covariances:\footnote{\, The overall normalization of $C_{ij}$, i.e., $M$ (unbiased estimate) vs. $M+1$ (maximum likelihood estimate) in the denominator in (\ref{sample.cov.mat}), is immaterial for our purposes here. In many cases $M\gg 1$.}
\begin{equation}\label{sample.cov.mat}
 C_{ij} = {1\over M}\sum_{s=1}^{M+1} X_{is}~X_{js}
\end{equation}
where $X_{is} = R_{is} - {\overline R}_i$ are serially demeaned quantities: ${\overline R}_i = {1\over {M+1}}\sum_{s=1}^{M+1} R_{is}$. SCM contains important information about the quantity characterizing our $N$ objects, to wit: i) serial variances $C_{ii} = \sigma_i^2$, which measure serial variability; and ii) serial pair-wise correlations $\Psi_{ij}$ between different series ($i\neq j$). Here
\begin{equation}
 \Psi_{ij} = {1\over \sigma_i\sigma_j}~C_{ij}  = {1\over M}\sum_{s=1}^{M+1} Y_{is}~Y_{js}
\end{equation}
is the sample correlation matrix, and $Y_{is} = X_{is} / \sigma_i$. Note that $\Psi_{ii}\equiv 1$.

{}When $M < N$, $C_{ij}$ is singular: we have $\sum_{s=1}^{M+1} X_{is} = 0$, so only $M$ columns of the matrix $X_{is}$ are linearly independent. Let us eliminate the last column: $X_{i,M+1}=-\sum_{s=1}^M X_{is}$. Then we can express $C_{ij}$ via the first $M$ columns:
\begin{equation}\label{SCM}
 C_{ij} = \sum_{s,s^\prime=1}^{M} X_{is}~\phi_{ss^\prime}~X_{js^\prime}
\end{equation}
Here $\phi_{ss^\prime} = \left(\delta_{ss^\prime} + u_s u_{s^\prime}\right)/M$ is a nonsingular $M\times M$ matrix ($s,s^\prime = 1,\dots,M$); $u_s \equiv 1$ is a unit $M$-vector. Note that $\phi_{ss^\prime}$ is a 1-factor model (see below). Similarly,
\begin{equation}
 \Psi_{ij} = \sum_{s,s^\prime=1}^{M} Y_{is}~\phi_{ss^\prime}~Y_{js^\prime}
\end{equation}

\subsection{Out-of-sample (In)stability}

{}Suppose we compute SCM based on a set -- call it set A -- of $d$ observations. Suppose now we compute SCM based on a different set -- call it set B -- of $d$ observations such that set A and set B are non-overlapping. Typically, unless $M \gg N$, the off-diagonal elements of SCM in the two computations can be vastly different. This is known as out-of-sample instability of sample correlations.\footnote{\, This statement is often regarded as stemming from empirical evidence. However, it is well-understood theoretically. We can always rotate our serially demeaned returns $X_{is}$ to an orthogonal basis and rescale them to have unit serial variances. Then the true covariance matrix is the $N\times N$ identity matrix. Pursuant to the Bai-Yin theorem \cite{BY}, the smallest and largest eigenvalues of SCM have the limits $\lambda_{min} = (1 - \sqrt{y})^2$ and $\lambda_{max} = (1 + \sqrt{y})^2$, where $y = N/M$ is fixed and $N,M \rightarrow\infty$. So for $M,N\gg 1$ we must have $M\gg N$ for all eigenvalues to be close to 1.\label{fn.BY}} On the other hand, sample variances tend to be much more stable out-of-sample and in many cases can be reliably computed even if $M \ll N$. One way to think about this is to note that $C_{ii} = {1\over M}\sum_{s=1}^{M+1} X_{is}^2$, so if the serially demeaned quantities $X_{is}$ are (quasi-)normally distributed within each series and $M\gg 1$, then $C_{ii}$ should be relatively stable.

\subsection{Correlations, Not Covariances}

{}In many applications involving SCM, it must be invertible and, furthermore, out-of-sample stable.\footnote{\, This is required, e.g., in finance, in the context of stock portfolio optimization including mean-variance optimization \cite{Markowitz}, Sharpe ratio maximization \cite{Sharpe94}, etc.} As mentioned above, in many cases SCM does not satisfy these requirements and one replaces it with a {\em constructed} matrix such that it is positive definite and more stable. However, in practice it is convenient to model the sample correlation matrix $\Psi_{ij}$ instead of $C_{ij}$, for two reasons. First, since sample variances $C_{ii}$ are relatively stable and can be readily computed, there is no need to model them; it is the pair-wise correlations $\Psi_{ij}$ ($i\neq j$) that require modeling. Second, in many cases, the sample variances $C_{ii}$ have a skewed cross-sectional\footnote{\, Throughout herein ``cross-sectional" refers to ``across the index $i$".} (e.g., (quasi) log-normal) distribution, as is often the case with positive-valued quantities. It is therefore convenient to factor $\sigma_i$ out of SCM, i.e., to work with the sample correlation matrix $\Psi_{ij} = C_{ij}/\sigma_i\sigma_j$. Its diagonal elements are nicely uniform ($\Psi_{ii} \equiv 1$), and the off-diagonal elements $\Psi_{ij}$ ($i\neq j$) take values in $(-1, 1)$ with a tight distribution. We can think of $\Psi_{ij}$ as a sample covariance matrix for {\em normalized} quantities ${\widetilde R}_{is} = R_{is} / \sigma_i$, i.e., $\Psi_{ij} = \mbox{Cov}({\widetilde R}_i, {\widetilde R}_j) = \mbox{Cor}(R_i, R_j)$. So, in what follows we will always work with $\Psi_{ij}$ and ${\widetilde R}_{is}$, and SCM will refer to the sample correlation matrix $\Psi_{ij}$.

\section{Factor Models}\label{sec.3}
{}Factor models are a popular method for constructing a nonsingular replacement $\Gamma_{ij}$ for $\Psi_{ij}$:
\begin{equation}\label{fac.mod}
 \Gamma_{ij} = \xi_i^2~\delta_{ij} + \sum_{A,B=1}^K \Omega_{iA}~\Phi_{AB}~\Omega_{jB}
\end{equation}
Here: $\xi_i^2$ is the specific (a.k.a. idiosyncratic) variance; $\Omega_{iA}$ is an $N\times K$ factor loadings matrix; and $\Phi_{AB}$ is a $K\times K$ factor covariance matrix (FCM), $A,B=1,\dots,K$. The number of factors $K\ll N$ to have FCM more stable than SCM. I.e., the off-diagonal elements of SCM (i.e., the pair-wise correlations $\Psi_{ij}$, $i\neq j$) are modeled via contributions from $K$ factors $f_{As}$, while the diagonal elements of SCM (i.e., $\Psi_{ii}\equiv 1$) receive contributions from the factors and purely diagonal specific variances $\xi_i^2$. This corresponds to modeling ${\widetilde R}_{is}$ via a matrix $\Upsilon_{is}$ such that:
\begin{eqnarray}\label{Upsilon}
 &&\Upsilon_{is} = \chi_{is} + \sum_{A=1}^K \Omega_{iA}~f_{As}\\
 &&\mbox{Cov}(\chi_i, \chi_j) = \xi_i^2 ~\delta_{ij}\\
 &&\mbox{Cov}(\chi_i, f_A) = 0\\
 &&\mbox{Cov}(f_A, f_B) = \Phi_{AB}\\
 &&\mbox{Cov}(\Upsilon_i, \Upsilon_j) = \Gamma_{ij}
\end{eqnarray}
As above, $\mbox{Cov}(\cdot, \cdot)$ are serial covariances.\footnote{\, For notational convenience we omit the index $s$ in serial covariances $\mbox{Cov}(\cdot,\cdot)$.} A nice feature of $\Gamma_{ij}$ is that it is positive-definite (and thus invertible) if FCM is positive-definite assuming that all $\xi_i^2 > 0$.

{}We can think about (\ref{Upsilon}) as an {\em approximation} for the decomposition
\begin{equation}\label{lin.reg}
 {\widetilde R}_{is} = \varepsilon_{is} + \sum_{A = 1}^K \Omega_{iA}~f_{As}
\end{equation}
whereby ${\widetilde R}_{is}$, an $N\times d$ matrix, is assumed to essentially be described by a linear $K$-factor model, $f_{As}$ being the factors. Then $\varepsilon_{is}$ corresponds to the ``error" term, i.e., deviation from a linear factor model. Note that generally $\varepsilon_{is}\neq \chi_{is}$. Thus, generally the matrix $\mbox{Cov}(\varepsilon_i, \varepsilon_j)$ is not diagonal, nor do the covariances $\mbox{Cov}(\varepsilon_i, f_A)$ vanish. However, approximating ${\widetilde R}_{is}$ via $\Upsilon_{is}$ is useful because constructing the factor model (\ref{fac.mod}) for SCM involves defining $\Omega_{iA}$, which we then can use to further compute the factors $f_{As}$, e.g., via the least-squares method, i.e., by minimizing the ``quadratic error" $\sum_{i=1}^N \varepsilon_{is}^2 \rightarrow\mbox{min}$. This, by definition, is equivalent to a cross-sectional linear regression of ${\widetilde R}_{is}$ over $\Omega_{iA}$ (without the intercept), where $f_{is}$ are the regression coefficients, whereas $\varepsilon_{is}$ are the regression residuals.\footnote{\, For general $\Omega_{iA}$ there are subtleties that require nontrivial modifications of the regression -- see \cite{KYa} for details. For our purposes here such subtleties do not arise.} So, we need to construct $\Gamma_{ij}$.

\subsection{``Binary" and ``Analog" Factors}

{}To construct a factor model, we need to define the factor loadings $\Omega_{iA}$. In the context of the decomposition (\ref{lin.reg}), the columns of $\Omega_{iA}$ are nothing but $K$ explanatory variables. The question is how to pick them. And this is not a rhetorical question.

{}Thus, imagine that we can classify our $N$ objects via a binary taxonomy, i.e., each object belongs to one and only one ``cluster". This might be possible if the objects can be grouped into ``clusters" based on some similarity criteria. E.g., in the case of stocks, they can be grouped into sectors, industries, sub-industries, etc. If such a grouping is possible, then we can take our explanatory variables as $\Omega_{iA} = \delta_{S(i), A}$, where $S$ maps our $N$ objects into $K$ ``clusters": $S: \{1,\dots,N\} \rightarrow \{1,\dots,K\}$. I.e., $\Omega_{iA}=1$ if the object labeled by $i$ belongs to the ``cluster" labeled by $A$; otherwise, $\Omega_{iA} = 0$. These ``binary" factors are based on the objects' membership in ``clusters".

{}If a binary classification is not attainable, then we can try to use some measured or estimated properties of our objects to populate the columns of our factor loadings. We can refer to them as ``analog" factors as they typically lack any ``binary" or ``clustering" structure and characterize the entire cross-section of the $N$ objects. E.g., in the case of stocks such ``analog" factors can be based on the companies' size (market capitalization), earnings, book value, etc.\footnote{\, In the case of stocks such ``analog" factors are called ``style factors".}  Even if some ``analog" factors can be defined, they may not always be good explanatory variables \cite{KYa}, so care is in order when attempting to use them as columns in $\Omega_{iA}$.

\subsection{Statistical Factor Models}\label{sub.statfm}

{}In many applications ``binary" and ``analog" factors mentioned above are unattainable or unreliable. In this case, we can resort to statistical factor models \cite{KYb}. The idea is simple. We have our data ${\widetilde R}_{is}$. What if we construct $\Omega_{iA}$ based on this data and no other input? I.e., we have to take an $N\times d$ matrix and somehow distill it down to a smaller $N \times K$ matrix. The question is how and what should $K$ be? And this is precisely where the factor model approximation via $\Gamma_{ij}$ defined in (\ref{fac.mod}) for the sample correlation matrix $\Psi_{ij}$ becomes a useful tool.

{}The idea behind statistical factor models is simple.  Let $V_i^{(a)}$, $a=1,\dots,N$, be the principal components of $\Psi_{ij}$ forming an orthonormal basis
\begin{eqnarray}\label{Psi.eigen}
 &&\sum_{j=1}^N \Psi_{ij}~V_j^{(a)} = \lambda^{(a)}~V_i^{(a)}\\
 &&\sum_{i=1}^N V_i^{(a)}~V_i^{(b)} = \delta_{ab}
\end{eqnarray}
such that the eigenvalues $\lambda^{(a)}$ are ordered decreasingly: $\lambda^{(1)} > \lambda^{(2)} >\dots$. More precisely, some eigenvalues may be degenerate. For generic (large enough) datasets -- and this is not critical in what follows -- the positive eigenvalues are non-degenerate. However, we can have multiple null eigenvalues. Typically, the number of nonvanishing eigenvalues\footnote{\, This number can be smaller if some series ${\widetilde R}_{is}$ are 100\% pair-wise (anti-)correlated. Again, for generic datasets -- and this not critical here -- this is not the case.} is $M$, where, as above, $d = M+1$ is the number of observations in each series. So, we have (assuming $M < N$; otherwise $M$ is replaced by $N$ below):
\begin{equation}\label{CM.PC}
 \Psi_{ij} = \sum_{a = 1}^M V_i^{(a)}~\lambda^{(a)}~V_j^{(a)}
\end{equation}
This resembles a factor model (\ref{fac.mod}) with a diagonal factor covariance matrix. However, the specific variance is missing. This can be rectified by noting that higher principal components contribute in (\ref{CM.PC}) with smaller weights, that is, eigenvalues. So, we can simply keep only the first $K$ principal components in the sum in (\ref{CM.PC}), where $K < M$, and replace the diagonal contribution of the dropped $M-K$ higher principal components via the specific variance:
\begin{eqnarray}\label{PC}
 && \Gamma_{ij} = {\xi}_i^2~\delta_{ij} + \sum_{A=1}^K \lambda^{(A)}~V_i^{(A)}~V_j^{(A)}\\
 && {\xi}_i^2 = 1 - \sum_{A=1}^K \lambda^{(A)}\left(V_i^{(A)}\right)^2\label{PC.xi}
\end{eqnarray}
This corresponds to taking the factor loadings matrix and factor covariance matrix of the form
\begin{eqnarray}\label{FLM.PC}
 &&{\Omega}_{iA} = \sqrt{\lambda^{(A)}}~V_i^{(A)},~~~A=1,\dots,K\\
 &&\Phi_{AB} = \delta_{AB}\label{CorMat.PC}
\end{eqnarray}
This construction is nicely simple. However, what should $K$ be? Two simple methods for fixing $K$ are discussed in \cite{KYb}, where R source code for constructing statistical factor models is also given. We briefly review them here.

\section{Fixing Factor Number}\label{sub.fix.K}

{}When $K = M$ we have $\Gamma_{ij} = \Psi_{ij}$ (which is singular if $M < N$). Therefore, we must have $K \leq K_{max} < M$. So, what is $K_{max}$? And what is $K_{min}$ (other than the evident $K_{min} = 1$)? It might be tempting to do complicated and convoluted things. We will not do this here. Instead, we will follow a pragmatic approach. One simple (``minimization" based) algorithm was set forth in \cite{Het}. We review it below and then give yet another simple algorithm based on eRank (effective rank).

\subsection{``Minimization" Algorithm}\label{sub.min}

{}The idea is simple \cite{Het}. It is based on the observation that, as $K$ approaches $M$, $\mbox{min}({\xi}^2_i)$ goes to 0 (i.e., less and less of the total variance $\Gamma_{ii}\equiv 1$ is attributed to the specific variance, and more and more of it is attributed to the factors), while as $K$ approaches 0, $\mbox{max}({\xi}^2_i)$ goes to 1 (i.e., less and less of the total variance is attributed to the factors, and more and more of it is attributed to the specific variance). So, we can define $K$ as follows:
\begin{eqnarray}\label{K}
 &&|g(K) - 1| \rightarrow \mbox{min}\\
 \label{g}
 &&g(K) = \sqrt{\mbox{min}({\xi}^2_i)} + \sqrt{\mbox{max}({\xi}^2_i)}
\end{eqnarray}
This simple algorithm works well in practical finance applications, see \cite{Het}, \cite{KYb}. Open source R code for computing statistical factor models (\ref{PC}) utilizing this ``minimization" based algorithm for fixing the number of factors $K$ is given in Appendix A of \cite{KYb}.

\subsection{Effective Rank}\label{sub.erank}

{}Another simple method is to set \cite{KYb}\footnote{\, Here $\mbox{Round}(\cdot)$ can be replaced by $\mbox{floor}(\cdot) = \lfloor\cdot\rfloor$.}
\begin{equation}\label{eq.eRank}
 K = \mbox{Round}(\mbox{eRank}(\Psi))
\end{equation}
Here $\mbox{eRank}(Z)$ is the effective rank \cite{RV} of a symmetric semi-positive-definite (which suffices for our purposes here) matrix $Z$. It is defined as
\begin{eqnarray}
 &&\mbox{eRank}(Z) = \exp(H)\\
 &&H = -\sum_{a=1}^L p_a~\ln(p_a)\\
 &&p_a = {\lambda^{(a)} \over \sum_{b=1}^L \lambda^{(b)}}
\end{eqnarray}
where $\lambda^{(a)}$ are the $L$ {\em positive} eigenvalues of $Z$, and $H$ has the meaning of the (Shannon a.k.a. spectral) entropy \cite{Campbell60}, \cite{YGH}.

{}The meaning of $\mbox{eRank}(Z)$ is that it is a measure of the effective dimensionality of the matrix $Z$, which is not necessarily the same as the number $L$ of its positive eigenvalues, but often is lower. This is due to the fact that many series $R_{is}$ can be highly correlated (which manifests itself by a large gap in the eigenvalues -- see below) thereby further reducing the effective dimensionality of the correlation matrix.

\subsection{A Variation}\label{sub.k.prime}

{}When the average correlation\footnote{\, Instead we can define ${\overline \Psi} = {1\over N(N-1)}\sum_{i,j=1;~i\neq j}^N$. Since $N\gg 1$, the difference is immaterial. \label{psi.bar}} ${\overline \Psi} = {1\over N^2}\sum_{i,j=1}^N\Psi_{ij}$ is high, then both the ``minimization" and eRank based algorithms can produce low values of $K$ (including 1). This is because in this case $\lambda^{(1)}\gg 1$ and there is a large gap between the first and higher eigenvalues. To circumvent this, we can define $K = K^\prime + 1$, where $K^\prime$ is defined as above via the ``minimization" or eRank based algorithms for the matrix
\begin{equation}
 \Psi_{ij}^\prime = \sum_{a = 2}^M V_i^{(a)}~\lambda^{(a)}~V_j^{(a)}
\end{equation}
I.e., we simply drop the first eigenpair, determine the corresponding value of $K^\prime$, and add 1 to it. Open source R code for computing statistical factor models (\ref{PC}) for both the ``minimization" and eRank based algorithms with and without utilizing the $K^\prime$ based definition is given in Appendix A of \cite{KYb}.

\section{Application to Cancer Signatures}\label{sec.5}

{}Now we are ready to apply the above machinery to cancer signatures. Our basic data consists of a matrix -- call it $G_{is}$ -- whose elements are occurrence counts of mutation types labeled by $i = 1,\dots, N = 96$ in samples labeled by $s = 1,\dots, d$. More precisely, we can work with one matrix $G_{is}$ which combines data from different cancer types; or, alternatively, we may choose to work with individual matrices $[G(\alpha)]_{is}$, where: $\alpha = 1,\dots,n$ labels $n$ different cancer types; as before, $i=1,\dots,N=96$; and $s = 1,\dots, d(\alpha)$. Here $d(\alpha)$ is the number of samples for the cancer type labeled by $\alpha$. The combined matrix $G_{is}$ is obtained simply by appending the matrices $[G(\alpha)]_{is}$ together column-wise. We will discuss a refinement of this data structure below.

{}The simplest thing we can do is to identify the matrix $R_{is}$ in our discussion above with $G_{is}$ (or $[G(\alpha)]_{is}$).\footnote{\, The discussion below focuses on $G_{is}$ and, unless stated otherwise, also applies to $[G(\alpha)]_{is}$.} However, this may not be the most optimal choice. The issue is this. The elements of the matrix $G_{is}$ are populated by nonnegative occurrence counts. Nonnegative quantities with large numbers of samples tend to have skewed distributions with long tails at higher values. I.e., such distributions are not normal but (in many cases) roughly log-normal. One simple way to deal with this is to identify $R_{is}$ with a (natural) logarithm of $G_{is}$ (instead of $G_{is}$ itself). A minor hiccup here is that some elements of $G_{is}$ can be 0. We can do a lot of complicated and even convoluted things to deal with this issue. Here we will follow a pragmatic approach and do something simple instead -- there is so much noise in the data that doing otherwise simply does not pay off. So, we will simply take
\begin{equation}\label{log.def}
 R_{is} = \ln\left(1 + G_{is}\right)
\end{equation}
This takes care of the $G_{is} = 0$ cases; for $G_{is}\gg 1$ we have $R_{is}\approx\ln(G_{is})$, as desired.

{}Now we can construct statistical factor models for cancer signatures using the ``minimization" and eRank based methods (with or without the $K^\prime$ based variation) for fixing the number of cancer signatures $K$. In fact, for the sake of completeness and comparative purposes, below we will construct such factor models assuming both (\ref{log.def}) and $R_{is} = G_{is}$. Happily, qualitatively the results turn out to be similar.

\section{Empirical Results}\label{sec.6}
\subsection{Data Summary}\label{data.summary}

{}In our empirical analysis below we use genome data from published samples only. This data is summarized in Table \ref{table.genome.summary}, where we give total counts, number of samples and the data sources, which are as follows: A1 = \cite{Alexandrov}, A2 = \cite{Love}, B1 = \cite{Tirode}, C1 = \cite{Zhang}, D1 = \cite{Nik-Zainal}, E1 = \cite{Puente2011}, E2 = \cite{Puente2015}, F1 = \cite{Cheng}, G1 = \cite{Wang}, H1 = \cite{Sung}, H2 = \cite{Fujimoto}, I1 = \cite{Imielinski}, J1 = \cite{Jones}, K1 = \cite{Patch}, L1 = \cite{Waddell}, M1 = \cite{Gundem}, N1 = \cite{Scelo}. Sample IDs with the corresponding publication sources are given in Appendix \ref{app.IDs}.

\subsection{Genome Data Results}

{}In our genome dataset we have 14 cancer types. Using the definition (\ref{log.def}), we apply the ``minimization" and eRank based methods (with and without the $K^\prime$ based variation) for fixing the number of cancer signatures $K$. We use the R functions {\tt{\small bio.erank.pc()}} and {\tt{\small bio.cov.pc()}} in Appendix \ref{app.code} hereof, which are adapted from Appendix A of \cite{KYb}. The results are summarized in Table \ref{table.genome.fac.counts}. Unless we use the $K^\prime$ based variation, the value of $K$ tends to be low. If we combine the samples from all 14 cancer types into a single ``big" matrix (in our case, of dimension $96\times 1389$), then we get $K=2$ for the eRank based method and $K=1$ for the``minimization" based method (without the $K^\prime$ based variation). Both of these methods produce $K=1$ if we aggregate all samples within each cancer type and run them on the resulting $96\times 14$ matrix. The question is, how come?

{}The answer is quite prosaic. Table \ref{table.eig.cor} provides the average pair-wise correlation ${\overline\Psi}$ (as defined in footnote \ref{psi.bar}) and the first 5 eigenvalues of the sample correlation matrix $\Psi_{ij}$. Except for Brain Lower Grade Glioma, Esophageal Cancer and Pancreatic Cancer (for which cancer types the matrix $G_{is}$ is sparsely populated with many 0s), these average correlations are rather high and there is a large gap between the first and higher eigenvalues. Therefore, the first eigenvector dominates in the spectral decomposition (\ref{CM.PC}). Excluding it via the $K^\prime$ based variation then produces higher values of $K$. However, on general grounds we expect the higher principal components to be out-of-sample unstable. That is, if we compute them based on two or more sets of non-overlapping samples, there is no guarantee that they will be stable from set to set. Therefore, we must address the issue out-of-sample stability first.

\subsubsection{Out-of-sample (In)stability}

{}A convenient way of addressing this issue is by checking whether the first and higher principal components computed for each cancer type are stable from one cancer type to another. As above, let $[G(\alpha)]_{is}$ be the occurrence count matrix for the cancer type labeled by $\alpha$ (for our genome data $\alpha$ takes 14 values). We then compute the corresponding matrix $[R(\alpha)]_{is}$ via (\ref{log.def}) and the correlation matrix $[\Psi(\alpha)]_{ij}$. Let $[V(\alpha)]^{(a)}_i$ be the $a$-th principal component of $[\Psi(\alpha)]_{ij}$. We then define a very informative matrix of inner products
\begin{equation}
 E^{(a)}_{\alpha\beta} = \sum_{i=1}^N [V(\alpha)]^{(a)}_i~[V(\beta)]^{(a)}_i
\end{equation}
By definition, $E^{(a)}_{\alpha\alpha} \equiv 1$, and $|E^{(a)}_{\alpha\beta}| < 1$ for $\alpha\neq \beta$. This matrix can be thought of as a measure of how ``correlated" the $a$-th principal components are across different cancer types. Table \ref{table.summary.corr} gives summaries of $|E^{(a)}_{\alpha\beta}|$ for $a=1,2,3$ and $\alpha\neq \beta$ (there are $14 \times 13~/~2 = 91$ independent values for each $a$). For illustrative purposes, in the fourth row we also give a summary of a similar matrix of inner products based on a union of the second and third principal components. From Table \ref{table.summary.corr} it is evident that the first principal component is extremely stable from one cancer type to another. However, higher principal components appear to be rather unstable. In this regard it is informative to compute the eRank of the matrix $E^{(a)}_{\alpha\beta}$ (using the {\tt{\small calc.erank()}} subfunction in the {\tt{\small bio.erank.pc()}} function in Appendix \ref{app.code} hereof). For each $a$ this is a measure of how independent of each other the principal components $[V(\alpha)]^{(a)}_i$ are across the 14 cancer types: the lower the eRank, the less independent they are, and the more stable they are from one cancer type to another. So, for $E_a = \mbox{eRank}(E^{(a)}_{\alpha\beta})$ we get $E_1 = 1.31$, $E_2 = 9.49$, $E_3 = 10.59$, and $E_{2+3} = 15.54$, where $E_{2+3}$ is based on the union of the second and third principal components as above. Based on the foregoing, it appears that higher principal components are highly out-of-sample unstable. Put differently, higher (than the first) principal components computed for one cancer type apparently have little predictive power for other cancer types.\footnote{\, To be clear, this does not mean that higher principal components are out-of-sample stable within each cancer type. We will return to this issue below.}

\subsubsection{The ``Overall" Mode}

{}The first principal component is highly stable from one cancer type to another. The values of $E^{(1)}_{\alpha\beta}$ in Table \ref{table.summary.corr} are mostly above 90\%. This implies that we have a dominant ``overall" mode. In finance the analog of this is the so-called ``market" mode\footnote{\, See, e.g., \cite{CFM} and references therein.} corresponding to the overall movement of the broad market, which affects all stocks (to varying degrees) -- cash inflow (outflow) into (from) the market tends to push stock prices higher (lower). This is the market risk factor. To mitigate this risk factor, one can, e.g., hold a dollar-neutral portfolio of stocks (i.e., the same dollar holdings for long and short positions). And we can draw from this analogy.

{}We can think of the ``overall" mode as follows. We can always write the sample correlation matrix as
\begin{equation}
 \Psi_{ij} = \left(1 - \rho\right)\delta_{ij} + \rho ~u_i~ u_j + \Delta_{ij} = \Psi_{ij}^\prime +\Delta_{ij}
\end{equation}
Here $\rho = {1\over N(N-1)}\sum_{i,j=1;~i\neq j}^N\Psi_{ij}$ is the average pair-wise correlation, $u_i\equiv 1$ is the unit $N$-vector, and $\sum_{i,j=1}^N \Delta_{ij} = 0$. In the zeroth approximation we can drop $\Delta_{ij}$, i.e., $\Psi_{ij}\approx \Psi^\prime_{ij}$. Note that $\Psi^\prime_{ij}$ is a 1-factor model. Its first principal component $U^{(1)}_i = u_i/\sqrt{N}$. It describes the ``overall" mode, i.e., the average correlation of all mutation types.\footnote{\, Note that the eigenvalue of $\Psi^\prime_{ij}$ corresponding to $U^{(1)}_i$ is $\lambda^\prime_* = 1 + \rho
\left(N-1\right)$.}  This implies that in the zeroth approximation $V^{(1)}_i\approx U^{(1)}_i$. When $N$ is large, in many systems this in fact is a good approximation. In our case $N = 96$, so it is large enough. Table \ref{table.1pc.diff} gives cross-sectional summaries of $|\sqrt{N}V^{(1)}_i - 1|$ (assuming $V^{(1)}_i$ are normalized such that $\sum_{i=1}^N V^{(1)}_i > 0$). It is evident that $V^{(1)}_i\approx U^{(1)}_i$ is indeed a pretty good approximation and, not surprisingly, the more total occurrence counts we have, the better this approximation works.

\subsubsection{Factoring Out ``Overall" Mode}\label{overall.stripped}

{}The ``overall" mode is clearly present and across all cancer types. Therefore it makes sense to factor it out altogether before performing any analysis on the data. Factoring out the ``overall" mode is nothing but cross-sectionally demeaning the matrix $R_{is}$, i.e., instead of $R_{is}$ we use
\begin{equation}
 R_{is}^\prime = R_{is} - {\overline R}_s = R_{is} - {1\over N} \sum_{j=1}^N R_{js}
\end{equation}
The results are given in Tables \ref{table.genome.fac.counts.overall.stripped} and \ref{table.eig.cor.overall.stripped}. The summary of the absolute values of pair-wise inner products (in the units of 1\%) between the first principal components of the sample correlation matrices $[\Psi(\alpha)]_{ij}$ for individual cancer types reads: Min = 0.122, 1st Qu. = 10.95, Median = 27.06, Mean = 30.83, 3rd Qu. = 42.86, Max = 90.74, StDev = 22.89, MAD = 23.55, and $E_1 = 7.06$. These results lead us to the following nontrivial conclusion: there appear to be common signatures for these 14 cancer types other than the ``overall" mode. The value $E_1 = 7.06$ suggests that the number of these signatures $K_1$ should be roughly 7. This is consistent with the values of $K$ in the last row of Table \ref{table.genome.fac.counts.overall.stripped}. However, Table \ref{table.eig.cor.overall.stripped} makes it clear that we no longer have a large gap between the first and higher eigenvalues, so higher principal components contribute substantially and it is difficult to expect out-of-sample stability.

\subsubsection{No Log}\label{nolog.genome}

{}Thus far we have been using the log-based definition (\ref{log.def}). Let us now check what transpires if we use the $R_{is} = G_{is}$ definition. The results are given in Tables \ref{table.genome.fac.counts.nolog} and \ref{table.eig.cor.genome.nolog}. Overall, the average correlations decrease and the values of $K$ increase. The summary of the absolute values of pair-wise inner products (in the units of 1\%) between the first principal components of the sample correlation matrices $[\Psi(\alpha)]_{ij}$ for individual cancer types reads: Min = 71.37, 1st Qu. = 83.86, Median = 96.53, Mean = 92.05, 3rd Qu. = 97.95, Max = 99.59, StDev = 8.023, MAD = 2.934, and $E_1 = 1.46$. These results suggest that the log-based definition (\ref{log.def}) does work better, as we anticipated above based on the skewed nature of distributions of counts.

\subsubsection{No Log with ``Overall" Mode Factored Out}\label{nolog.genome.overall.stripped}

{}For the sake of completeness, let us also look at what happens if we use the $R_{is} = G_{is}$ definition and factor out the ``overall" mode by cross-sectionally demeaning the so-defined $R_{is}$. The results are given in Tables \ref{table.genome.fac.counts.nolog.overall.stripped} and \ref{table.eig.cor.genome.nolog.overall.stripped}. The summary of the absolute values of pair-wise inner products (in the units of 1\%) between the first principal components of the sample correlation matrices $[\Psi(\alpha)]_{ij}$ for individual cancer types reads: Min = 0.960, 1st Qu. = 31.80, Median = 44.93, Mean = 45.67, 3rd Qu. = 59.74, Max = 86.75, StDev = 20.57, MAD = 20.84, and $E_1 = 5.61$. These results show that with the no log definition we capture fewer independent signatures. This should come as no surprise -- the skewed nature of distributions of counts obscures the underlying signatures. One way to see this is that for several cancer types sizable average correlation is still present in Table \ref{table.eig.cor.genome.nolog.overall.stripped} despite removing the ``overall" mode.

\section{Nonnegative Matrix Factorization}\label{sec.7}

\subsection{First, A Multiplicative Model}
{}Using the statistical factor model approach allows us to: i) fix the number of factors $K$; and ii) remove the ``overall" mode. The number of factors $K_1$ excluding the ``overall" mode predicted by the eRank based method agrees with that obtained via $E_1$ in Subsection \ref{overall.stripped}. However, a priori the statistical factor model approach would appear to lack biologic interpretation. If we apply it directly to the no log definition $R_{is} = G_{is}$ (irrespective of the ``overall" mode), the matrices $\Omega_{iA}$ and $F_{As}$ generally can have negative elements. If we apply it to the log-based definition (\ref{log.def}), then we can re-exponentiate (\ref{lin.reg}) via (recall that $\sigma_i^2$ is the sample variance, and the factor model is for the correlation matrix, which is why $\sigma_i$ appears in the exponent)\footnote{\, Technically speaking, we should subtract the 1 we added inside the log in the definition (\ref{log.def}). Then we have to deal with negative values. This would obscure our discussion here with no benefit.}
\begin{equation}
 {\widehat G}_{is} = \exp\left(\sigma_i\varepsilon_{is} + \sigma_i\sum_{A=1}^K \Omega_{iA}~F_{As}\right) = \gamma_{is}
 \prod_{A=1}^K \left(Z_{As}\right)^{\nu_{iA}}
\end{equation}
where $\gamma_{is} = \exp\left(\sigma_i\varepsilon_{is}\right)$, $Z_{As} = \exp(F_{As})$ and $\nu_{iA} = \sigma_i\Omega_{iA}$. So, ignoring the ``multiplicative error" term $\gamma_{is}$ for a moment, ${\widehat G}_{is}$ provides a {\em positive} decomposition of the matrix $1+G_{is}$, except that it is a {\em multiplicative} decomposition (as opposed to an additive one, as in NMF). So, instead of ``weights", here we have the powers $\nu_{is}$ for the ``exposures" $Z_{As}$. In fact, such a multiplicative model may not be too farfetched. The processes inside DNA do appear to have ``exponential" tendencies. We intend to discuss this approach in more detail in a forthcoming paper. Instead, here we will apply the improvement we get from factoring out the ``overall" mode to NMF.

\subsection{NMF: Vanilla Counts Matrix}\label{NMF.vanilla}

{}The commonly accepted method for extracting cancer signatures from $G_{is}$ \cite{Alexandrov.NMF} is via nonnegative matrix factorization (NMF) \cite{Paatero}, \cite{LeeSeung}. Under NMF the matrix $G$ is approximated via $G \approx W~H$, where $W_{iA}$ is an $N\times K$ matrix, $H_{As}$ is a $K\times d$ matrix, and both $W$ and $H$ are nonnegative. The appeal of NMF is its biologic interpretation whereby the $K$ columns of the matrix $W$ are interpreted as the weights with which the $K$ cancer signatures contribute into the $N=96$ mutation categories, and the columns of the matrix $H$ are interpreted as the exposures to the $K$ signatures in each sample.

{}Usually, NMF is applied either to individual cancer types or to a ``big matrix" obtained by combining the samples from all cancer types. Here we apply NMF in a novel fashion to the $96 \times 14$ matrix obtained by aggregating samples by cancer type. The advantage of this approach is that we get to include low count samples without destabilizing the results (in-sample), and this way we also avoid undesirable proliferation of signatures which can occur when the number of samples is large.\footnote{\, Such ubiquity of signatures generally makes them less useful.}

\begin{sidewaysfigure}[ht]
\centering
\includegraphics[scale=0.7]{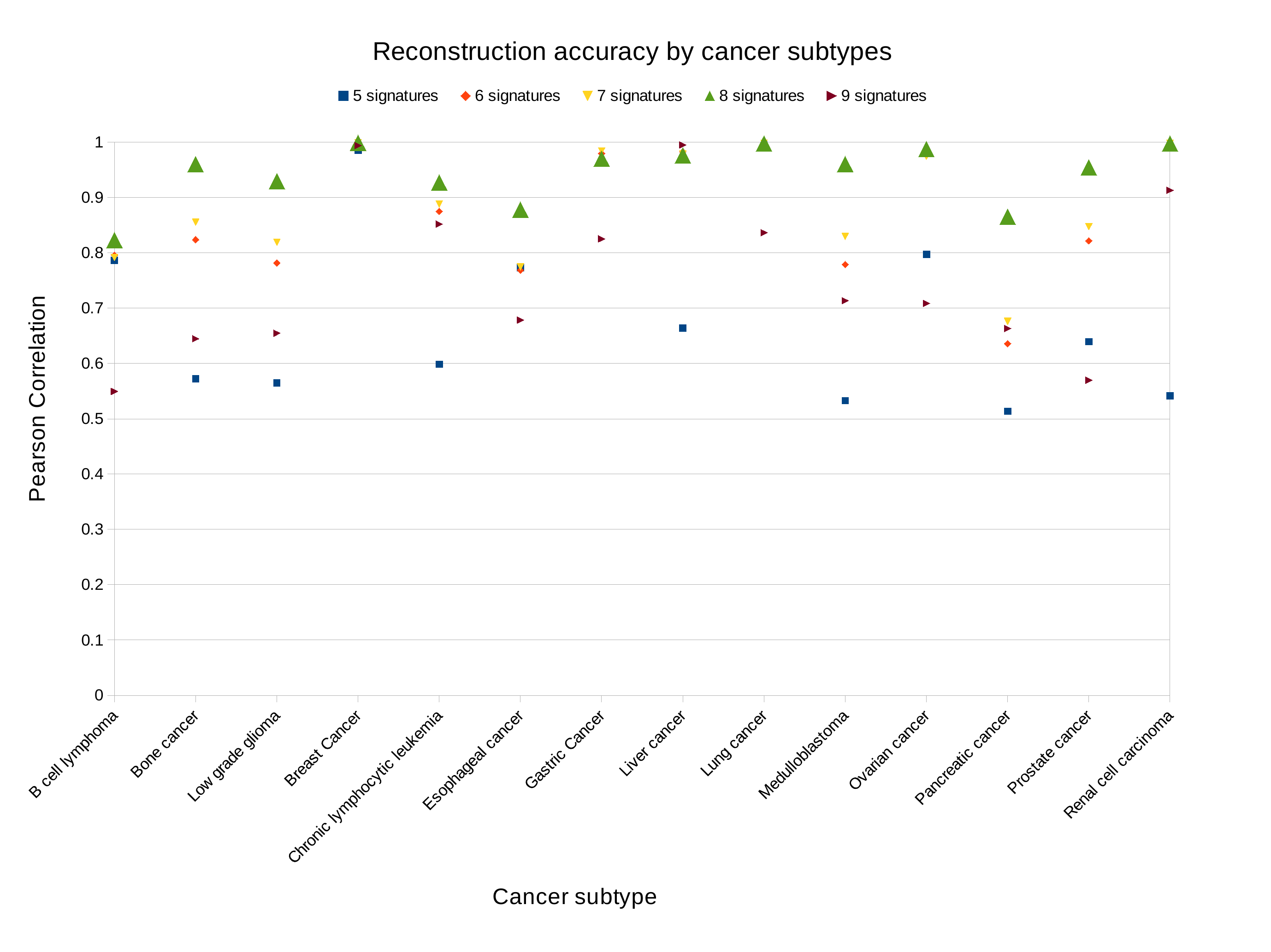}
\caption{Pearson correlations between the vanilla matrix $G$ and the reconstructed matrix $G^* =W~H$ for 5 to 9 signatures. See Subsection \ref{NMF.vanilla} for details.}
\label{recon.cor}
\end{sidewaysfigure}

\begin{sidewaysfigure}[ht]
\centering
\includegraphics[scale=0.7]{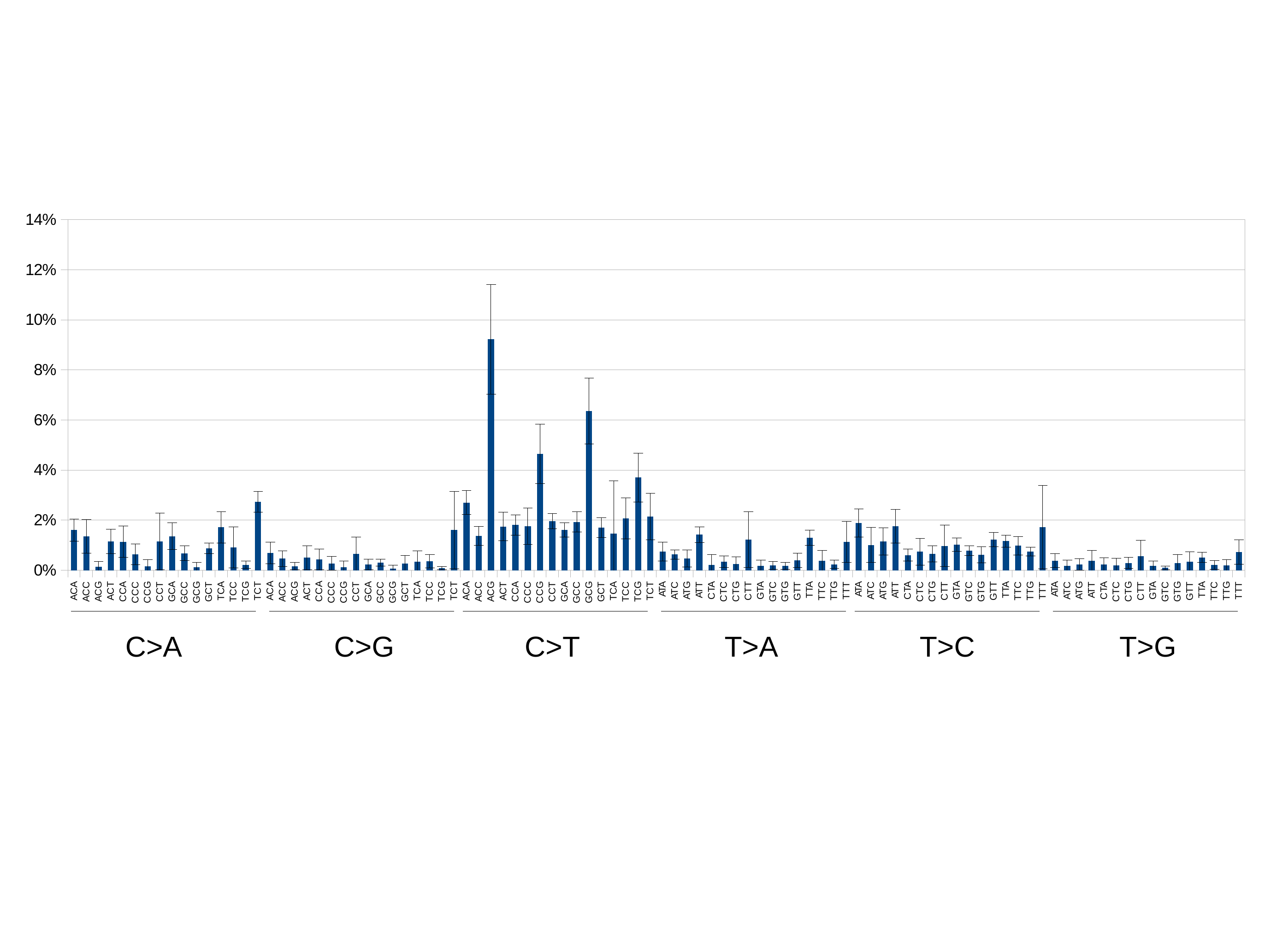}
\caption{Signature 1 extracted from the vanilla matrix $G$ using NMF. See Subsection \ref{NMF.vanilla} for details.}
\label{sig1}
\end{sidewaysfigure}

\begin{sidewaysfigure}[ht]
\centering
\includegraphics[scale=0.7]{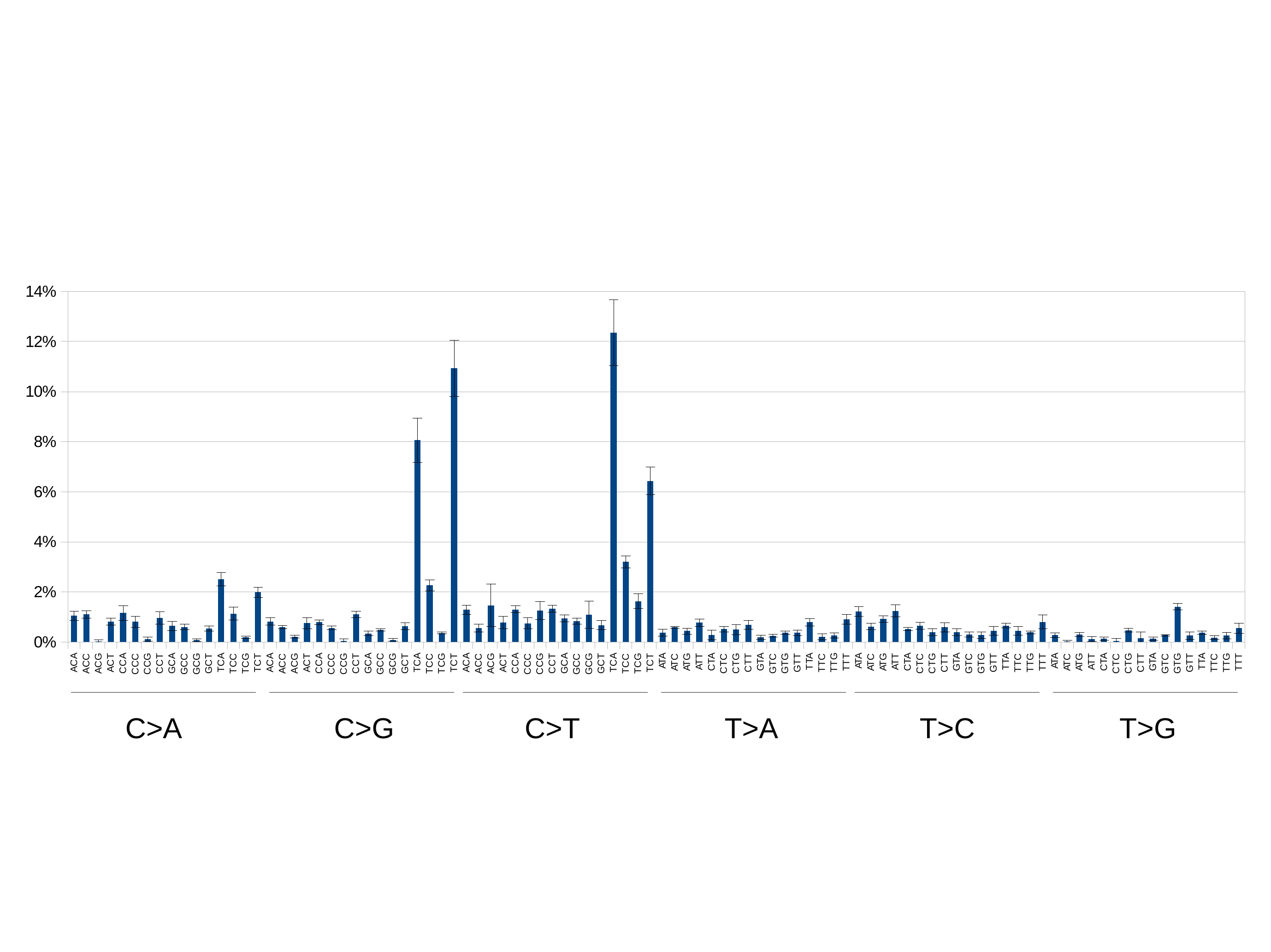}
\caption{Signature 2 extracted from the vanilla matrix $G$ using NMF. See Subsection \ref{NMF.vanilla} for details.}
\label{sig2}
\end{sidewaysfigure}

\begin{sidewaysfigure}[ht]
\centering
\includegraphics[scale=0.7]{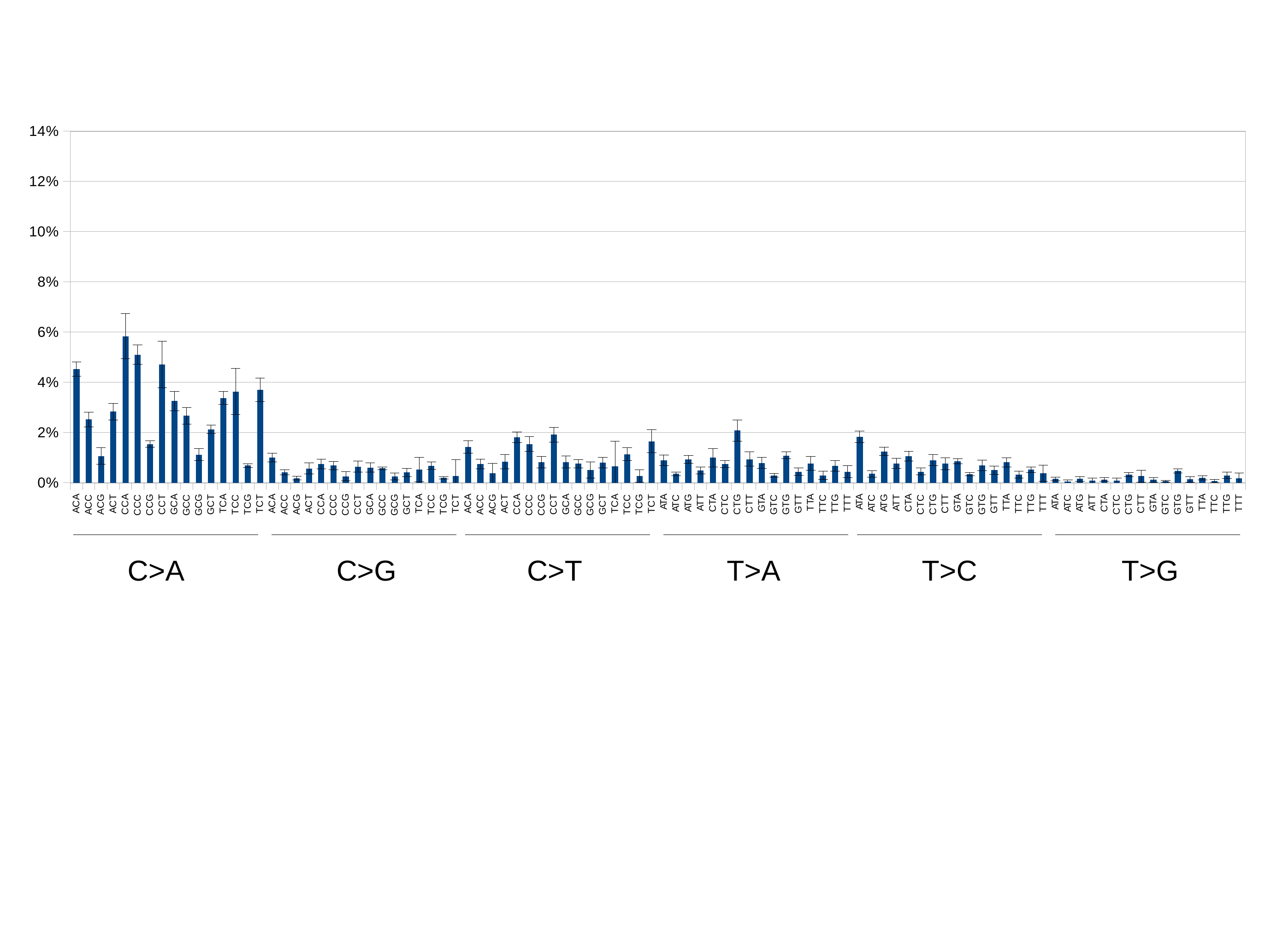}
\caption{Signature 3 extracted from the vanilla matrix $G$ using NMF. See Subsection \ref{NMF.vanilla} for details.}
\label{sig3}
\end{sidewaysfigure}

\begin{sidewaysfigure}[ht]
\centering
\includegraphics[scale=0.7]{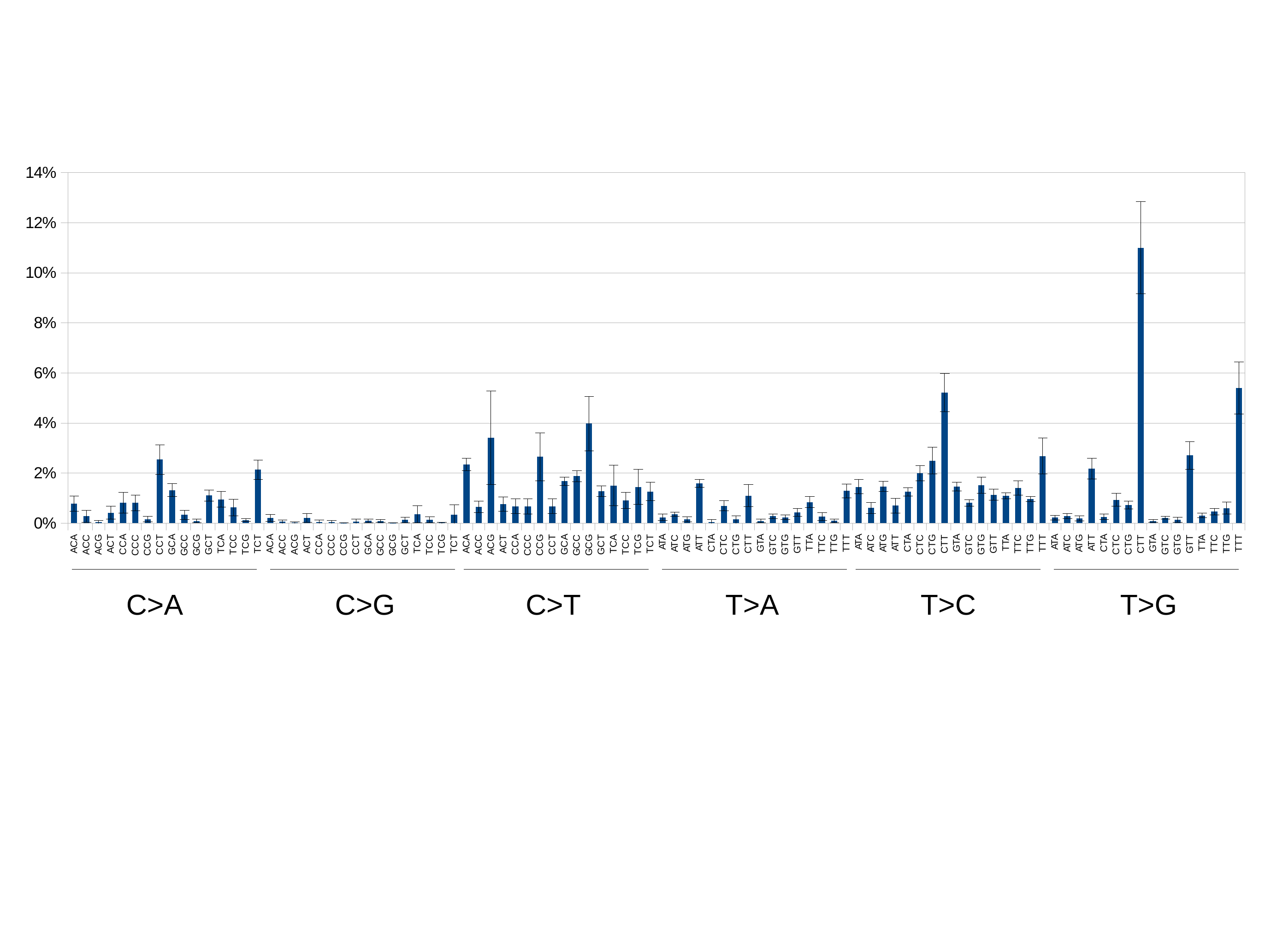}
\caption{Signature 4 extracted from the vanilla matrix $G$ using NMF. See Subsection \ref{NMF.vanilla} for details.}
\label{sig4}
\end{sidewaysfigure}

\begin{sidewaysfigure}[ht]
\centering
\includegraphics[scale=0.7]{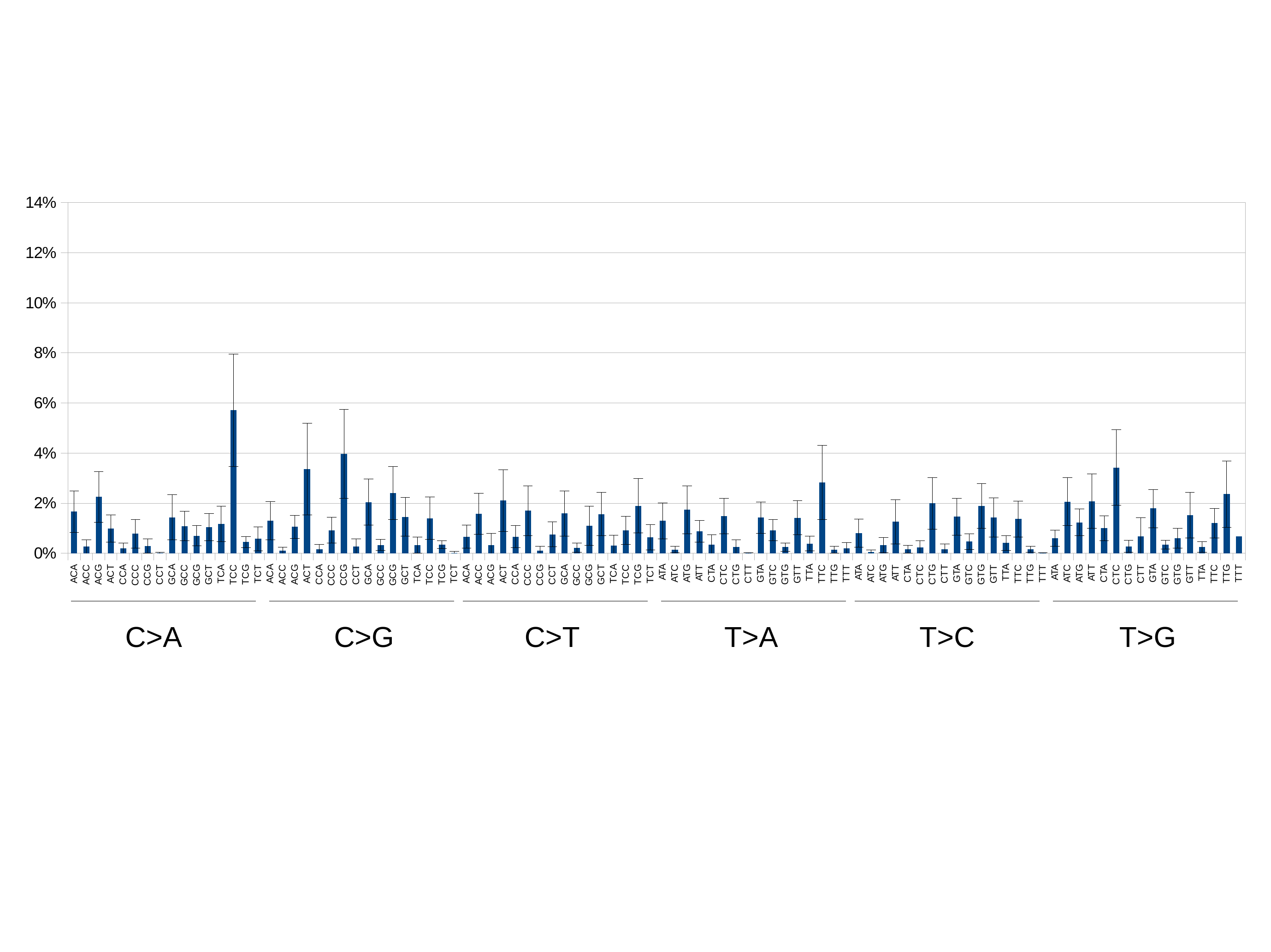}
\caption{Signature 5 extracted from the vanilla matrix $G$ using NMF. See Subsection \ref{NMF.vanilla} for details.}
\label{sig5}
\end{sidewaysfigure}

\begin{sidewaysfigure}[ht]
\centering
\includegraphics[scale=0.7]{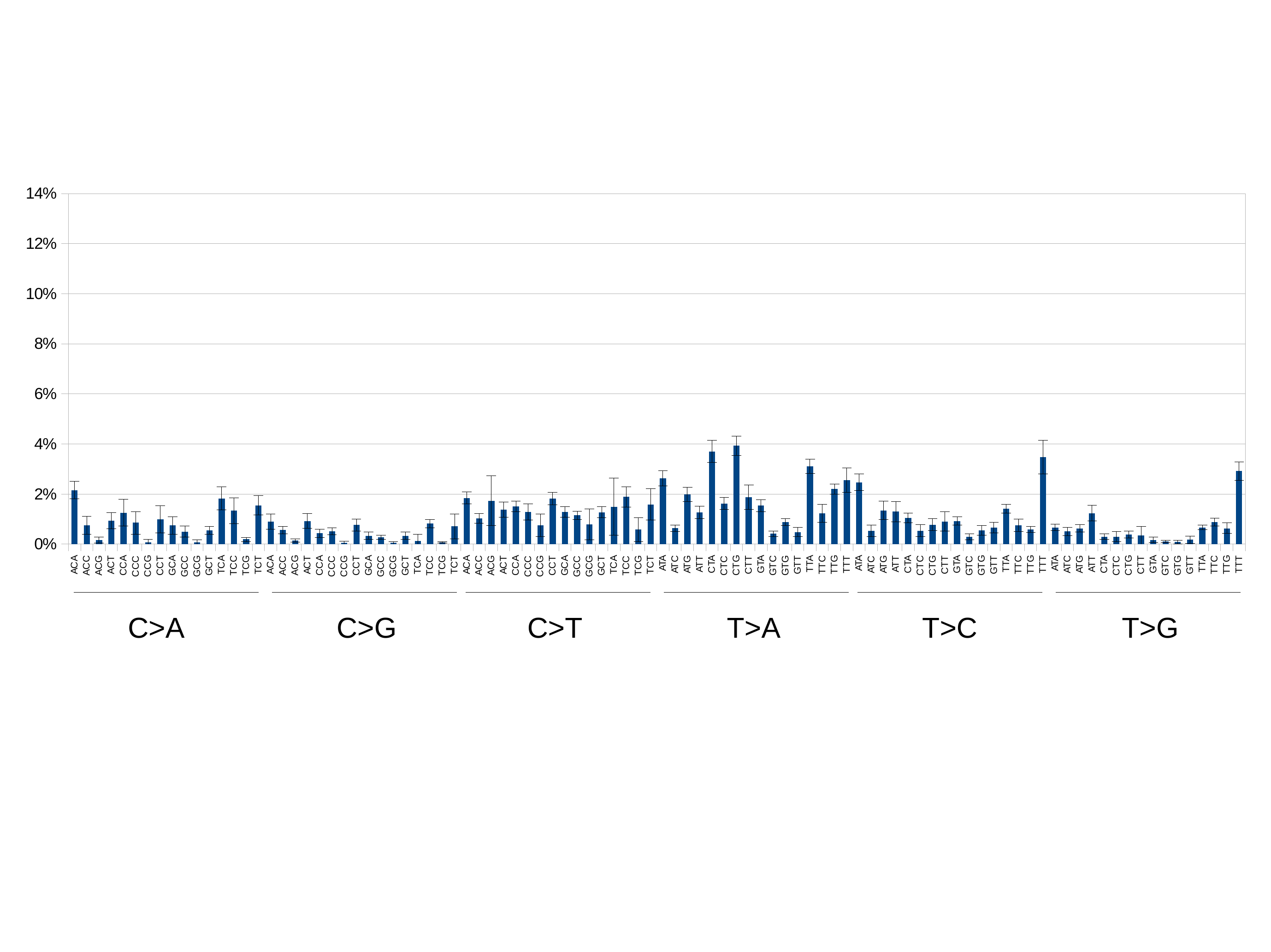}
\caption{Signature 6 extracted from the vanilla matrix $G$ using NMF. See Subsection \ref{NMF.vanilla} for details.}
\label{sig6}
\end{sidewaysfigure}

\begin{sidewaysfigure}[ht]
\centering
\includegraphics[scale=0.7]{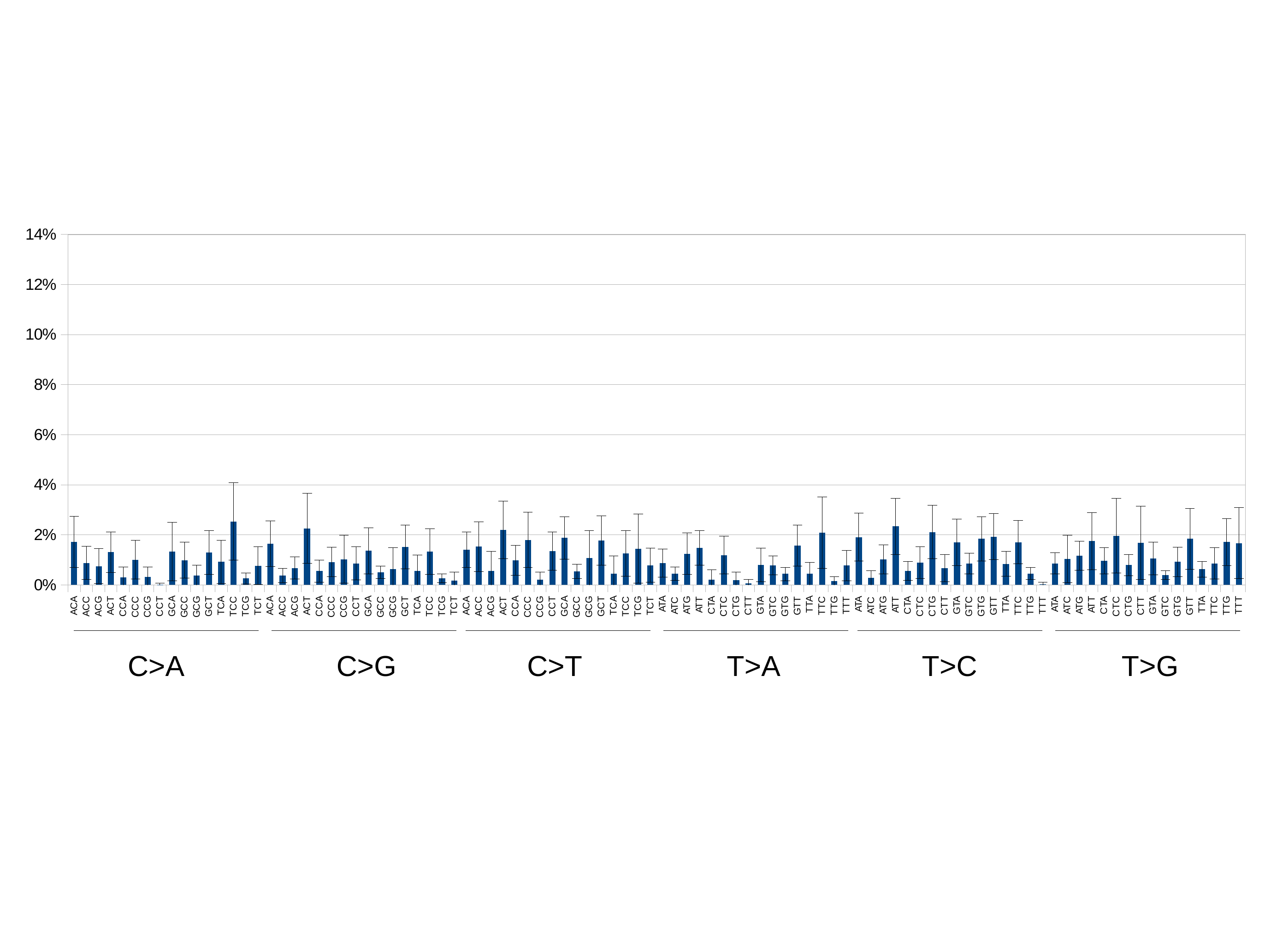}
\caption{Signature 7 extracted from the vanilla matrix $G$ using NMF. See Subsection \ref{NMF.vanilla} for details.}
\label{sig7}
\end{sidewaysfigure}

\begin{sidewaysfigure}[ht]
\centering
\includegraphics[scale=0.7]{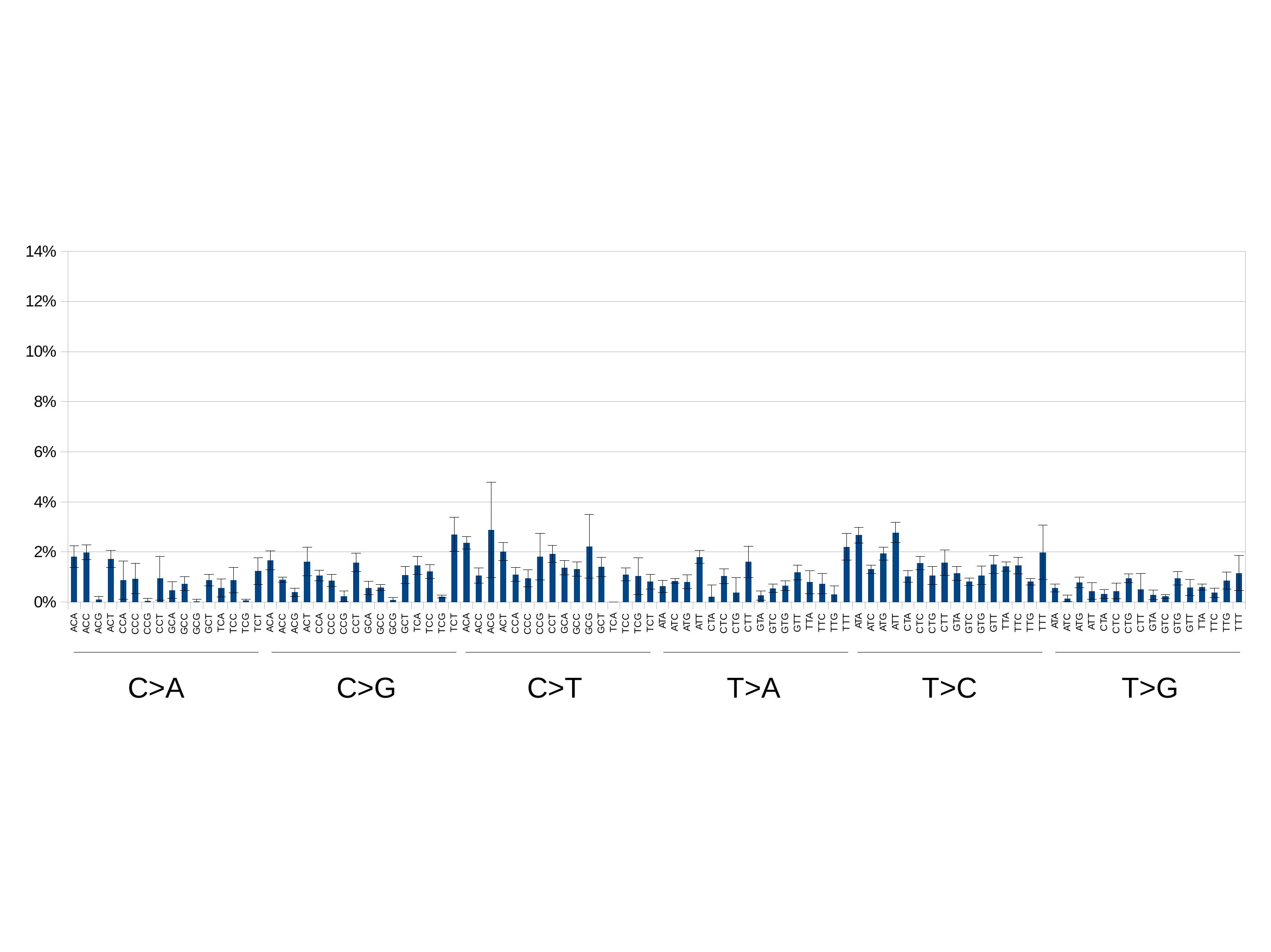}
\caption{Signature 8 extracted from the vanilla matrix $G$ using NMF. See Subsection \ref{NMF.vanilla} for details.}
\label{sig8}
\end{sidewaysfigure}

\begin{sidewaysfigure}[ht]
\centering
\includegraphics[scale=0.7]{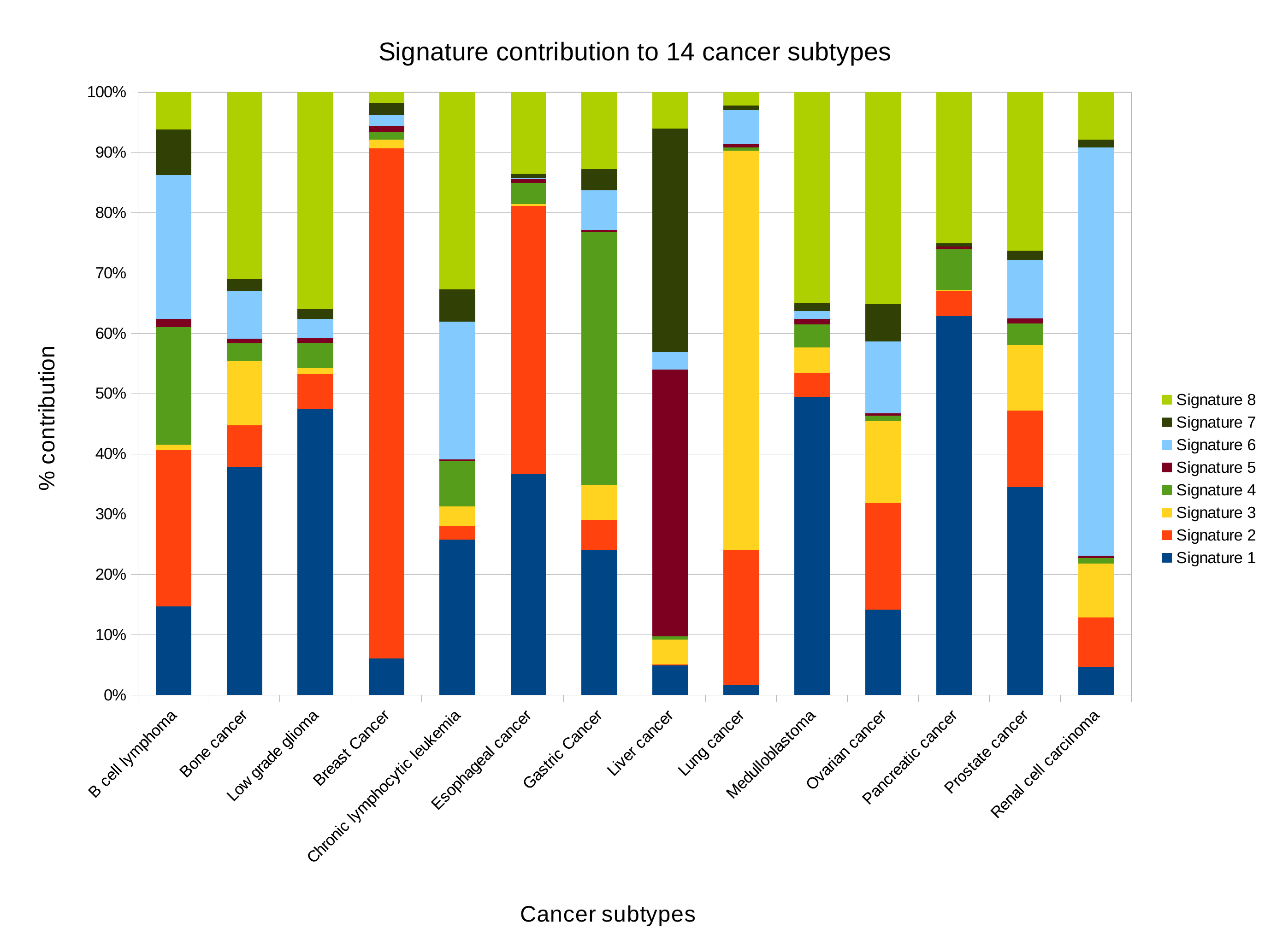}
\caption{Signature contributions (percentiles) into different cancer types (vanilla matrix $G$). See Subsection \ref{NMF.vanilla} for details.}
\label{sig.contr}
\end{sidewaysfigure}

{}We use organic R code for running NMF (and check that it produces the same results as the R package ``NMF", https://cran.r-project.org/package=NMF). We run NMF for 100 ``samplings" using random starting $W$ and $H$ for each ``sampling".\footnote{\, Each ``sampling" finds a local optimum -- NMF does not guarantee global convergence.} Figure \ref{recon.cor} gives the Pearson correlations between the vanilla matrix $G$ and the reconstructed matrix $G^* =W~H$ for 5 to 9 signatures.\footnote{\, This range is based on the values of $K$ in the last rows of Tables \ref{table.genome.fac.counts} and \ref{table.genome.fac.counts.overall.stripped}.} The highest reconstruction accuracy is achieved for $K=8$ signatures, which is what we anticipated above for the vanilla matrix ($K_1 = 7$ plus the ``overall" mode). Figures \ref{sig1}-\ref{sig8} plot the 8 signatures. For each signature, the corresponding weights in the $W$ columns (for each of the 96 mutation categories) are averages over the 100 ``samplings", and the error bars are the standard deviations.\footnote{\, We use the k-means clustering to sort the resultant signatures across the 100 ``samplings".} We discuss the interpretation of the signatures below. Here we note that the error bars for the vanilla matrix are substantial. Also, Signature 8 has substantial presence in most cancer types. This is the noise largely stemming from the ``overall" mode. Figure \ref{sig.contr} summarizes signature contributions.

\subsection{NMF: ``Overall" Mode Factored Out}\label{NMF.stripped}

{}We now repeat the NMF procedure of the last subsection using the data with the ``overall" mode factored about. For this purpose, we simply re-exponentiate the column-wise demeaned matrix $R^\prime_{is}$, i.e., we take
\begin{equation}
 {\widetilde G}_{is} = \exp(R^\prime_{is})
\end{equation}
and run NMF on ${\widetilde G}_{is}$. We can include an overall normalization by taking ${\widetilde G}_{is} = \exp(\mbox{Mean}(R_{is}) + R^\prime_{is})$, or ${\widetilde G}_{is} = \exp(\mbox{Median}(R_{is}) + R^\prime_{is})$, or ${\widetilde G}_{is} = \exp(\mbox{Median}({\overline R}_s) + R^\prime_{is})$ (recall that ${\overline R}_s$ is the vector of column means of $R_{is}$), etc., to make it look more like the original matrix $G_{is}$, however, this does not affect the signatures extracted via NMF.\footnote{\, This is because each column of $W$, being weights, is normalized to add up to 1.} Again, technically speaking, after re-exponentiating we should subtract the extra 1 we added in the definition (\ref{log.def}) (assuming we include one of the aforesaid overall normalizations). However, this does not seem to affect the results much.

\begin{sidewaysfigure}[ht]
\centering
\includegraphics[scale=0.7]{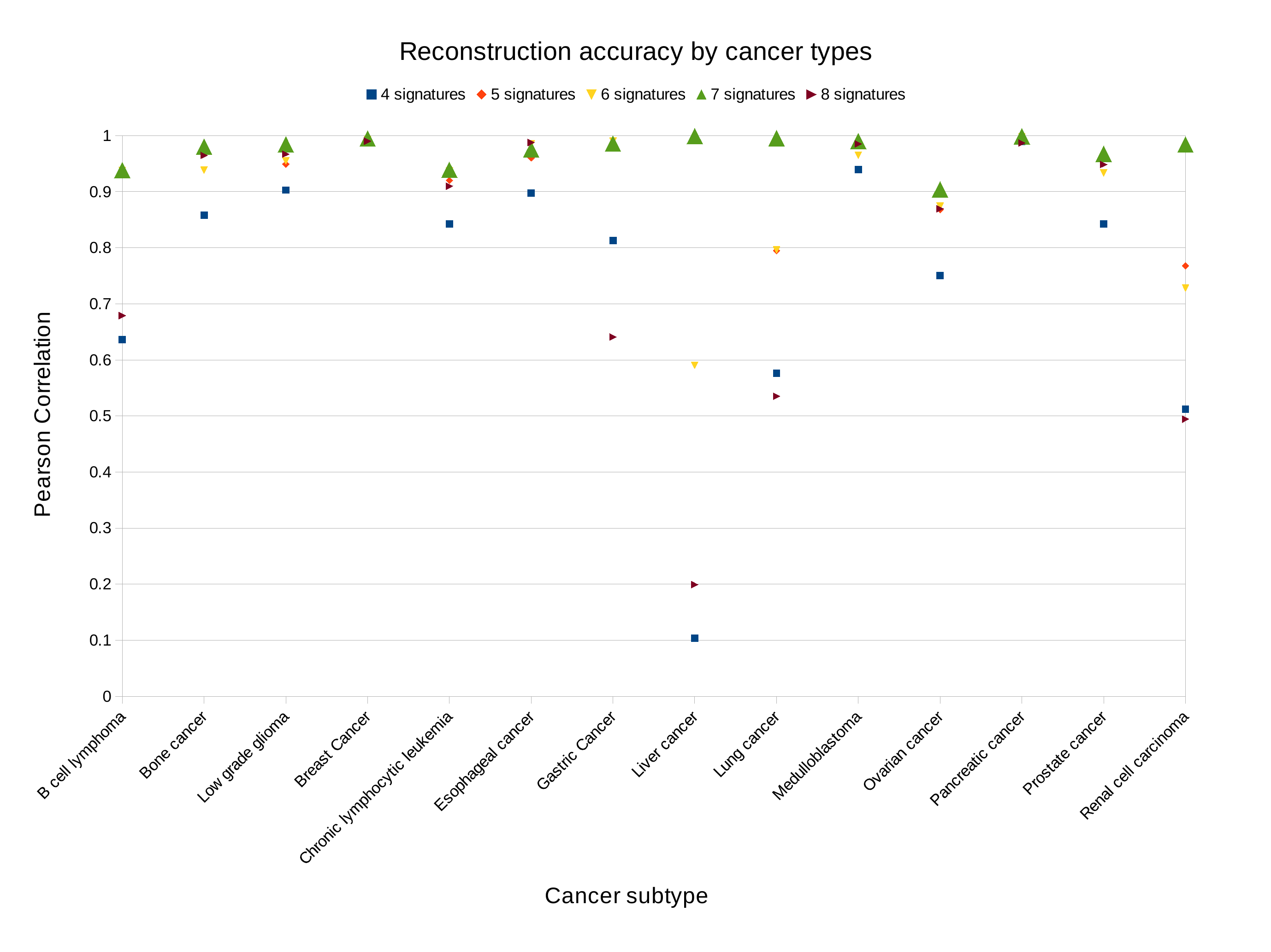}
\caption{Pearson correlations between the ``de-noised" matrix ${\widetilde G}$ and the reconstructed matrix ${\widetilde G}^* =W~H$ for 4 to 8 signatures. See Subsection \ref{NMF.stripped} for details.}
\label{recon.cor.stripped}
\end{sidewaysfigure}

\begin{sidewaysfigure}[ht]
\centering
\includegraphics[scale=0.7]{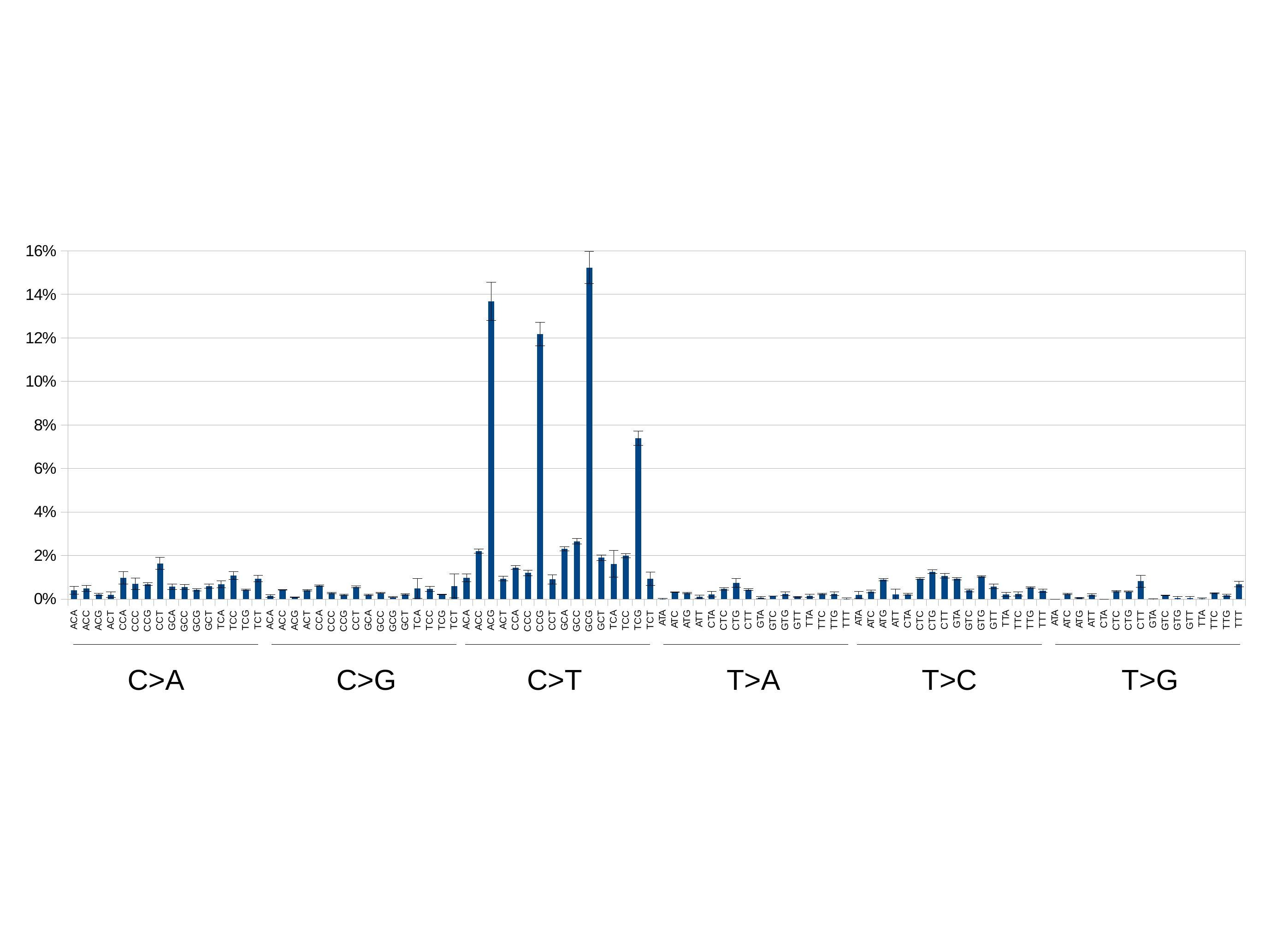}
\caption{Signature 1 extracted from the ``de-noised" matrix ${\widetilde G}$ using NMF. See Subsection \ref{NMF.stripped} for details.}
\label{sig1.stripped}
\end{sidewaysfigure}

\begin{sidewaysfigure}[ht]
\centering
\includegraphics[scale=0.7]{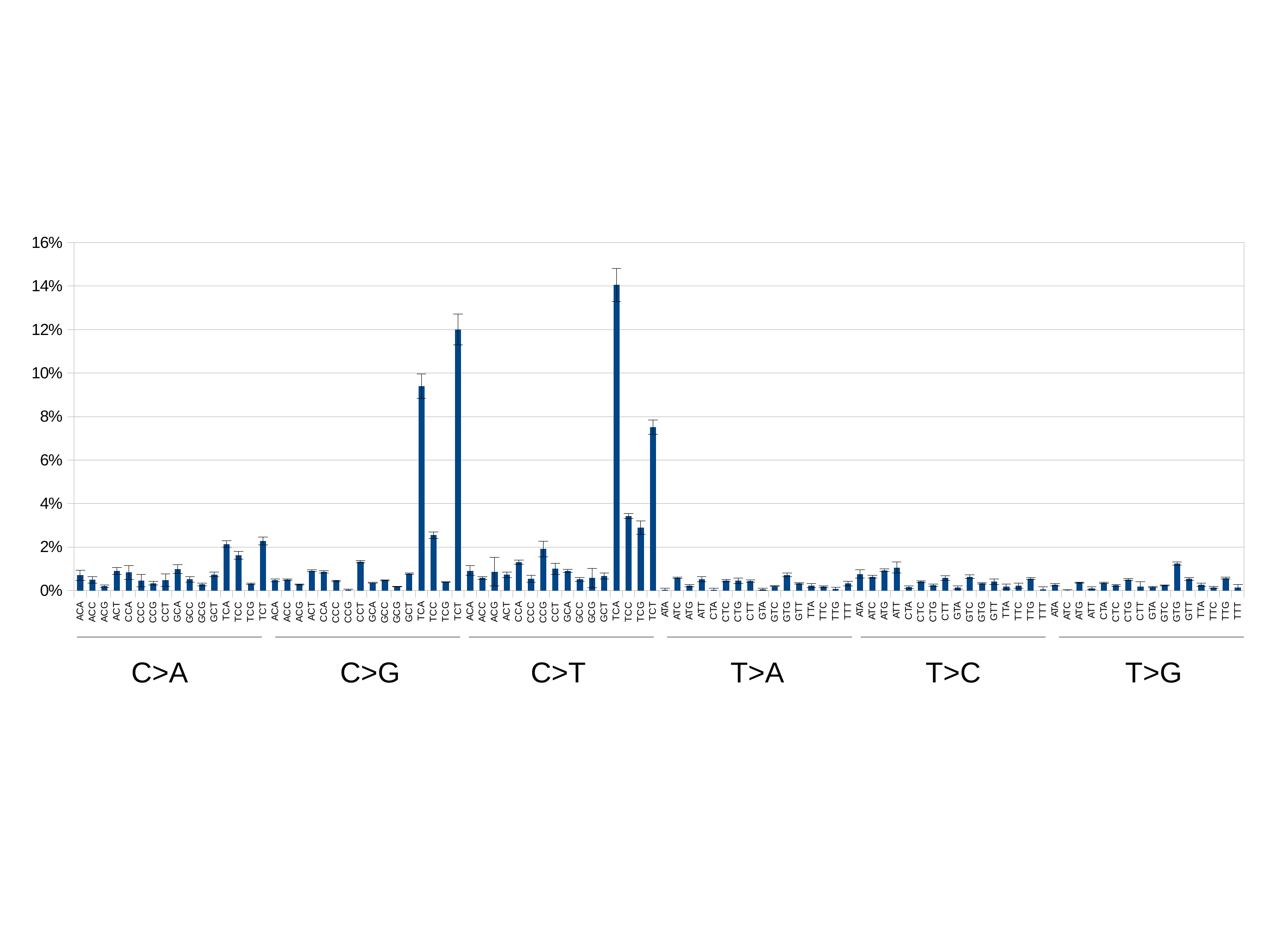}
\caption{Signature 2 extracted from the ``de-noised" matrix ${\widetilde G}$ using NMF. See Subsection \ref{NMF.stripped} for details.}
\label{sig2.stripped}
\end{sidewaysfigure}

\begin{sidewaysfigure}[ht]
\centering
\includegraphics[scale=0.7]{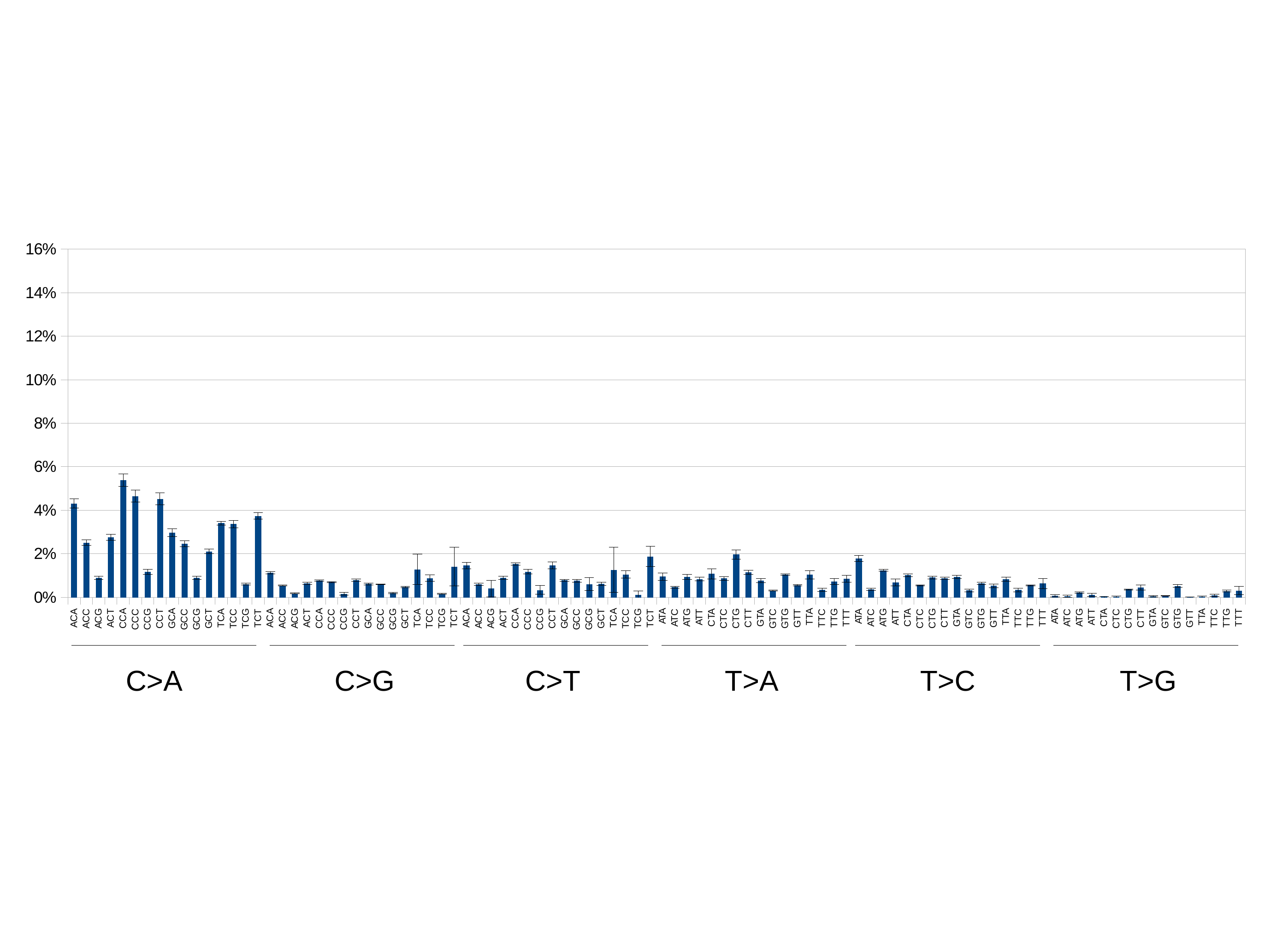}
\caption{Signature 3 extracted from the ``de-noised" matrix ${\widetilde G}$ using NMF. See Subsection \ref{NMF.stripped} for details.}
\label{sig3.stripped}
\end{sidewaysfigure}

\begin{sidewaysfigure}[ht]
\centering
\includegraphics[scale=0.7]{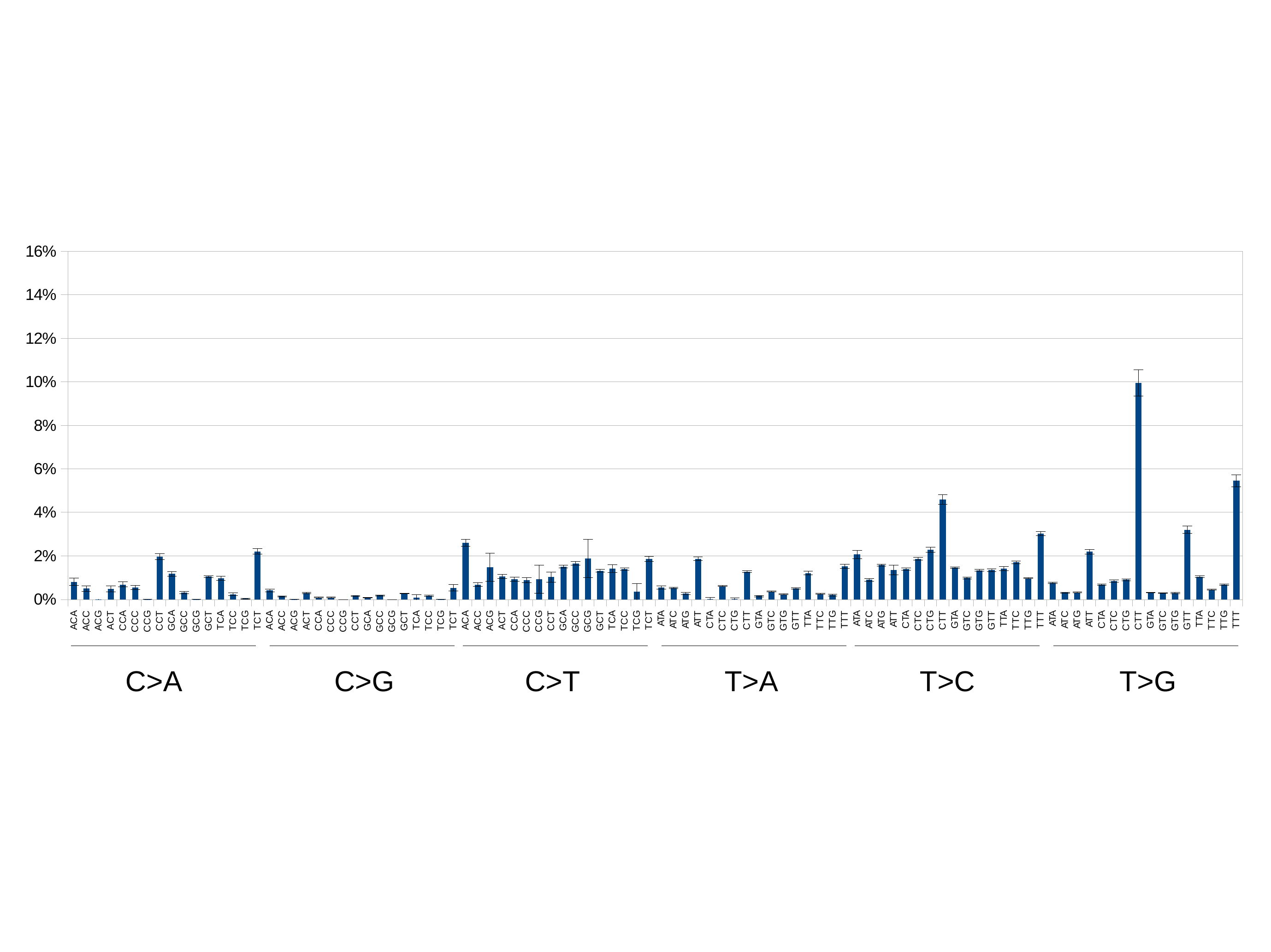}
\caption{Signature 4 extracted from the ``de-noised" matrix ${\widetilde G}$ using NMF. See Subsection \ref{NMF.stripped} for details.}
\label{sig4.stripped}
\end{sidewaysfigure}

\begin{sidewaysfigure}[ht]
\centering
\includegraphics[scale=0.7]{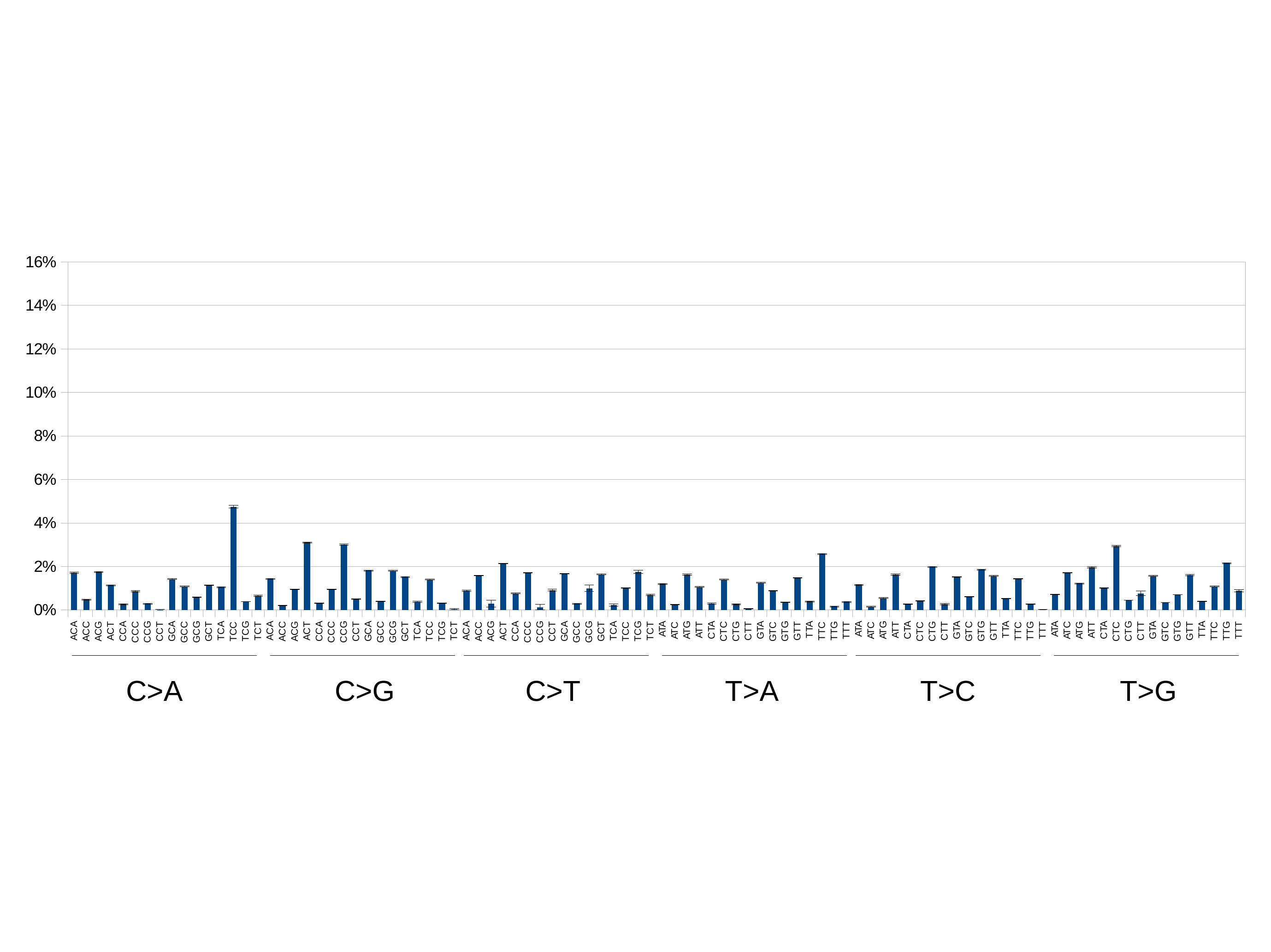}
\caption{Signature 5 extracted from the ``de-noised" matrix ${\widetilde G}$ using NMF. See Subsection \ref{NMF.stripped} for details.}
\label{sig5.stripped}
\end{sidewaysfigure}

\begin{sidewaysfigure}[ht]
\centering
\includegraphics[scale=0.7]{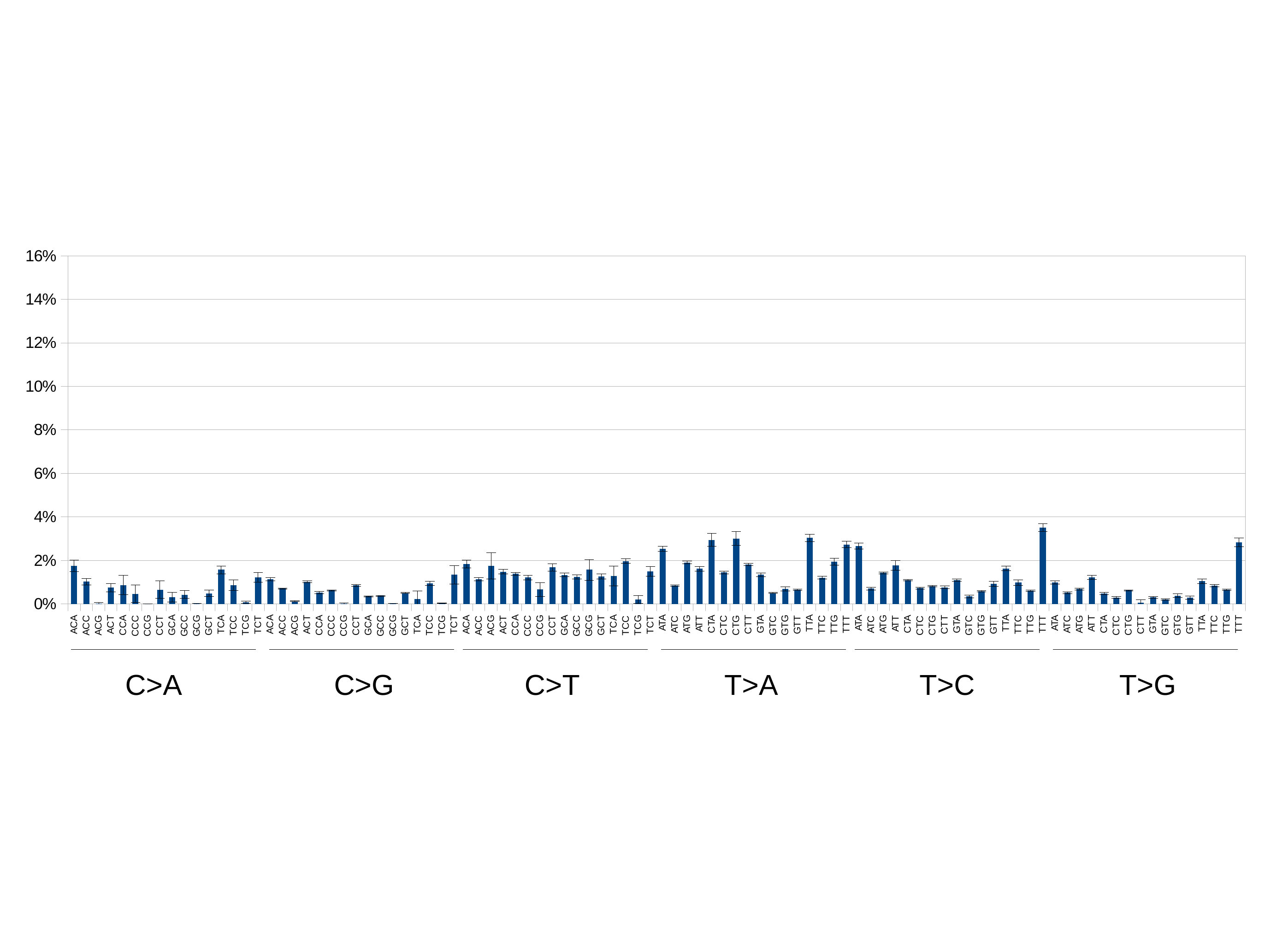}
\caption{Signature 6 extracted from the ``de-noised" matrix ${\widetilde G}$ using NMF. See Subsection \ref{NMF.stripped} for details.}
\label{sig6.stripped}
\end{sidewaysfigure}

\begin{sidewaysfigure}[ht]
\centering
\includegraphics[scale=0.7]{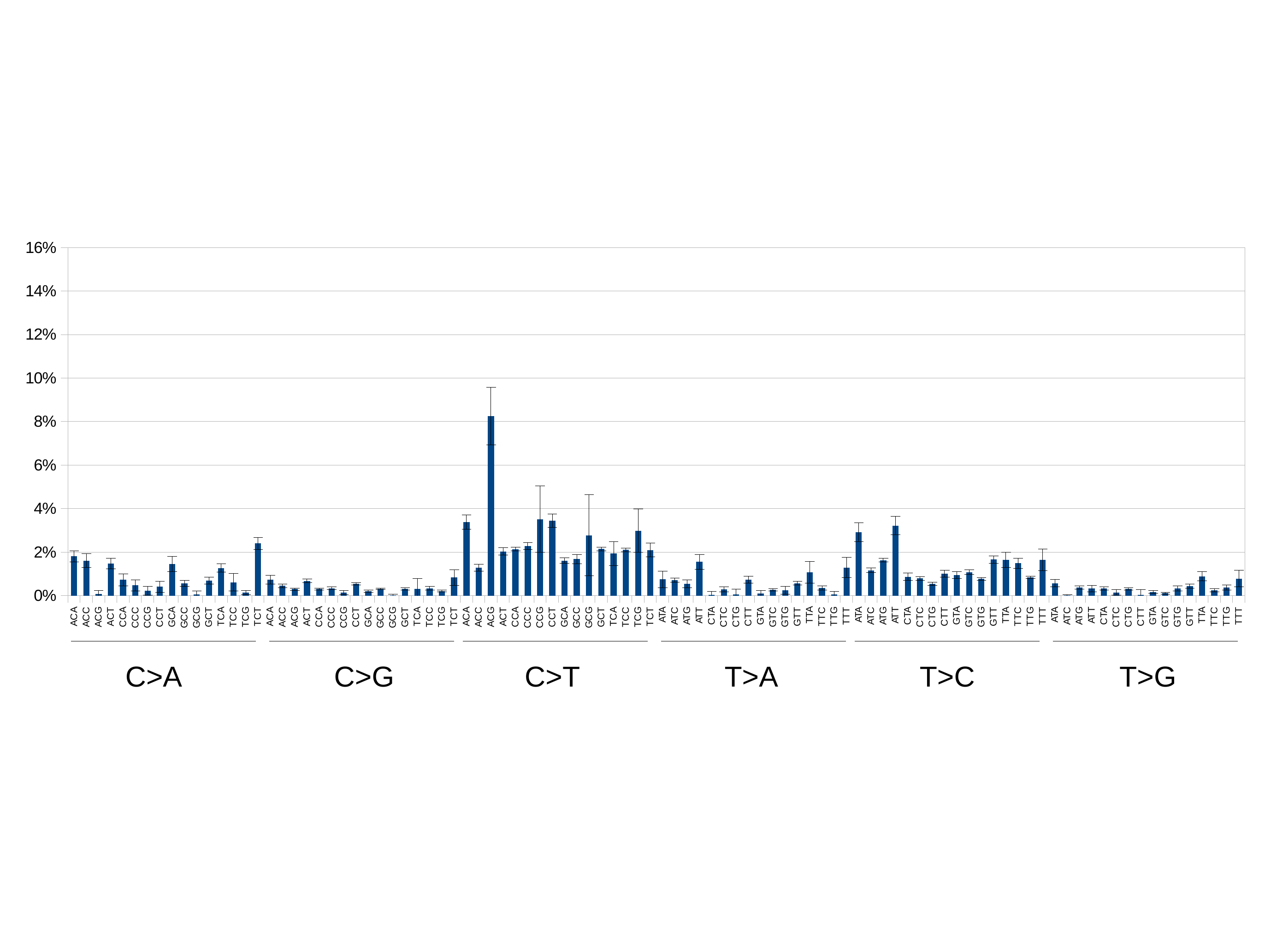}
\caption{Signature 7 extracted from the ``de-noised" matrix ${\widetilde G}$ using NMF. See Subsection \ref{NMF.stripped} for details.}
\label{sig7.stripped}
\end{sidewaysfigure}

{}Figure \ref{recon.cor.stripped} gives the Pearson correlations between the vanilla matrix ${\widetilde G}$ and the reconstructed matrix ${\widetilde G}^* =W~H$ for 4 to 8 signatures. The highest reconstruction accuracy is achieved for $K_1=7$ signatures, which is what we anticipated above. Figures \ref{sig1.stripped}-\ref{sig7.stripped} plot the 7 signatures, and the error bars are the standard deviations for each mutation category as a result of the 100 ``samplings". Here we note that the error bars for the ``de-noised" matrix ${\widetilde G}$ (Figures \ref{sig1.stripped}-\ref{sig7.stripped}) are substantially smaller than for the vanilla matrix $G$ (Figures \ref{sig1}-\ref{sig8}) due to factoring out the ``overall" mode.

\begin{sidewaysfigure}[ht]
\centering
\includegraphics[scale=0.7]{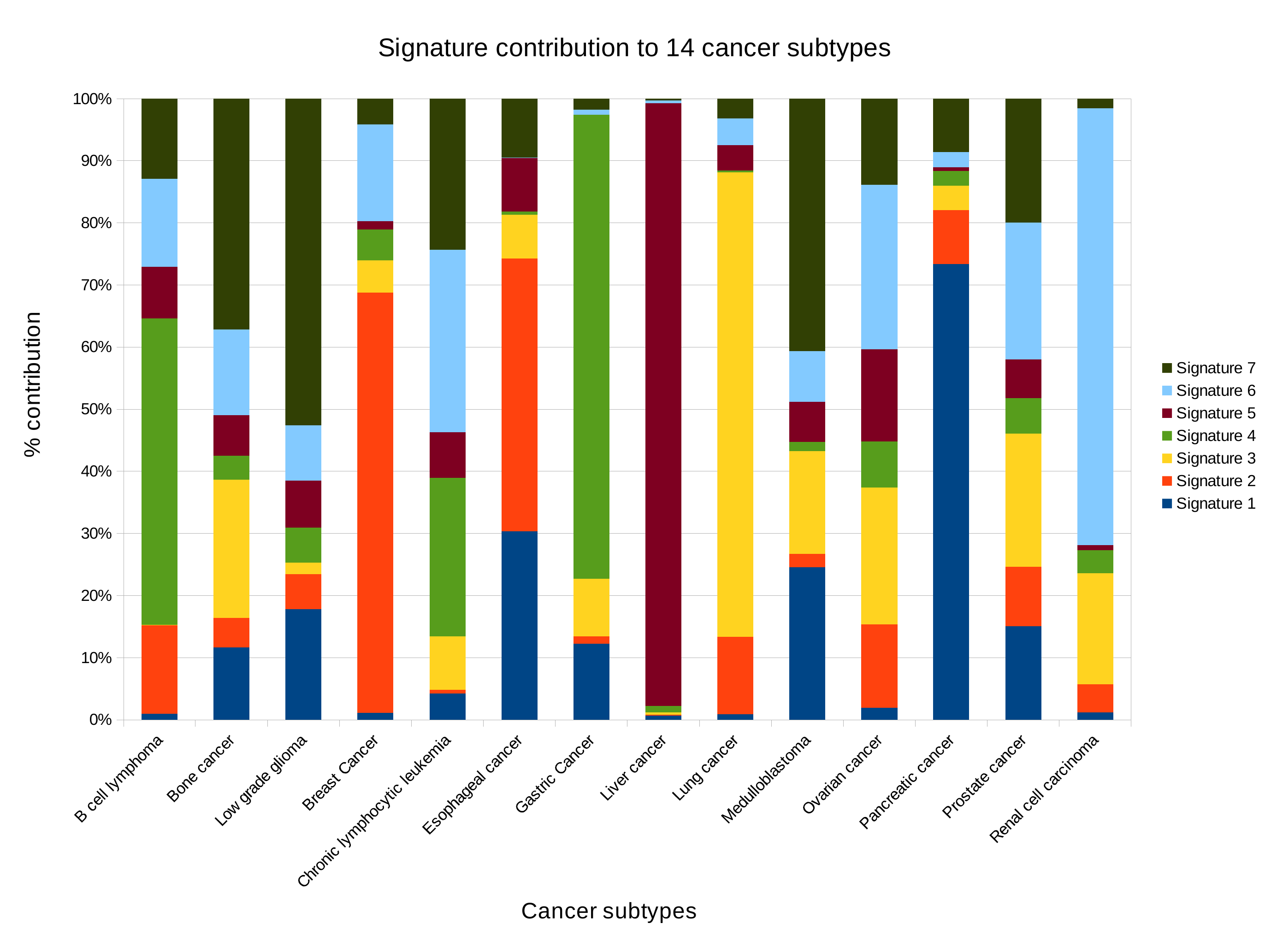}
\caption{Signature contributions (percentiles) into different cancer types (``de-noised" matrix ${\widetilde G}$). See Subsection \ref{NMF.stripped} for details.}
\label{sig.contr.stripped}
\end{sidewaysfigure}

{}Figure \ref{sig.contr.stripped} summarizes signature contributions. Our Signatures 1-4 are previously known signatures, to wit, mutational signatures 1 (spontaneous cytosine deamination), 2+13 (APOBEC mediated cytosine deamination),\footnote{\, This was reported as a single signature (almost identical to our Signature 2) in \cite{Alexandrov}, however, subsequently, it was split into 2 distinct signatures, which usually appear in the same samples. For detailed comments, see http://cancer.sanger.ac.uk/cosmic/signatures.} 4 (tobacco carcinogen related exposure) and 17 (appearing in oesophagus cancer, breast cancer, liver cancer, lung adenocarcinoma, B-cell lymphoma, stomach cancer and melanoma; mutational process unknown) of \cite{Nik-Zainal}, \cite{Alexandrov}.\footnote{\, Recovering these signatures is not surprising as we use data from these references.} Our Signatures 5-7 are new. New Signature 5 dominates liver cancer (with over 96\% contribution), with almost no peak variability. New Signature 6 to a lesser degree dominates renal cell carcinoma (with over 70\% contribution). New Signature 7 appears mostly in bone cancer, brain lower grade glioma and medulloblastoma (and also 5 other cancers to lesser degrees). The super-dominant liver cancer signature is exciting. Tables \ref{weights.errors.1} and \ref{weights.errors.2} give weights with errors for the 7 signatures.

\section{Concluding Remarks}\label{sec.8}

$\bullet$ {\bf Out-of-sample (in)stability.} This is a sticking point for any statistically based method, which includes NMF. Usually, ``stability" is addressed in the context of NMF by perturbing the matrix $G$ and checking whether the signatures are stable. However, this does not address out-of-sample stability. Out-of-sample stability is well-understood and is the bread-and-butter in the context of quantitative trading. Since there one deals with time series and forecasting, if a given model lacks out-of-sample stability, it is pretty much useless. This is because time flows only in one direction and if a model built using parameters computed based on a time period in the past does not perform well during a future time period -- that is, {\em out-of-sample} -- it has no predictive (i.e., forecasting) power. In quantitative finance money is at stake so methods for checking out-of-sample stability are rather well understood.

{}Based on those methods, a true test for out-of-sample stability in the context of cancer signatures would be to take a set of samples, split it into 2 (or more) non-overlapping subsets, independently extract signatures based on these subsets and compare them. In fact, to have any kind of statistical significance, we would need an ensemble of such non-overlapping sets. E.g., we could take some number of samples and split them randomly into two halves a number of times. For this to be meaningful, we need a sizable number of samples to begin with. The data we work with in this paper is rather limited in this sense because it includes only published genome samples. Not only is the number of cancer types limited to 14, but the number of samples within each cancer type is also limited. E.g., for prostate cancer we have mere 5 samples and any meaningful out-of-sample stability test for that cancer type is unattainable. On the other hand, for liver cancer we have a substantial number of samples (389), which appears to be a contributing factor to the extraction of a super-dominant signature for this cancer type, albeit not the leading factor -- without ``de-noising" this signature is not as dominant, nor was it found in \cite{Fujimoto}, where most of the liver cancer samples are published. We need as much data as possible to study out-of-sample stability in any meaningful fashion. The (still embargoed) ICGC data appears to hold promise in this regard.

\noindent $\bullet$ {\bf What about individual cancer types?} We ran our NMF analysis on the data aggregated by cancer type. Can we do the same for individual cancer types? The answer is yes -- after all, this is how NMF is usually applied -- but with caveats. One of the advantages of aggregating the data by cancer type is that it reduces the noise level. Individual cancer type samples generally are too noisy. Low count samples exacerbate this issue. Table \ref{table.genome.fac.counts.overall.stripped} is an apt testament to this. Once we remove the ``overall" mode (which artificially lowers the value of $K$), we get too many factors based on the statistical factor model analysis, and we therefore can expect a proliferation of signatures as well. As mentioned above, too many signatures are useless. In fact, high values of $K$ for individual cancer types indicate out-of-sample instability of any potential signatures. There are methods to reduce noise for individual cancer types, however, they are outside of the scope hereof and will be reported elsewhere. A practical motivation for considering individual cancer types is that within each cancer type there may be biologic factors one may wish to understand, e.g., mutational spectra of liver cancers can have substantial regional dependence as they are mutagenized by exposures to different chemicals.\footnote{\, We would like to thank Steven Rozen for emphasizing this point to us.} However, aggregation by regions within a cancer type may still be warranted to reduce noise.\footnote{\, Or else expectedly unstable factors in individual cancer types appear too copious (Table \ref{table.genome.fac.counts.overall.stripped}).}

\noindent $\bullet$ {\bf Exome data.} The volume of published exome data is substantially higher than that of the published genome data. In this regard, it would make sense to apply our methods to the exome data. The caveat is that the exome data is much more sparsely populated than the genome data, which has the same effect as the low count samples in the context of the genome data. Aggregation by cancer type is a natural remedy to this. The main issue is that meticulously ascertaining which samples are published is time consuming. We plan to discuss the exome data separately.

\noindent $\bullet$ {\bf ``Minimization" and eRank based algorithms.} The former typically leads to lower values of $K$ than the latter. For the data at hand, the eRank based algorithm is right on the money for fixing the number of signatures. In this regard, it appears that the eRank based algorithm should be the go-to method, however, the``minimization" based algorithm is still useful as the two algorithms set the expected range of the values of $K$ where the search should be performed. Having a (tight) range of $K$ helps reduce computational cost -- as mentioned above, NMF, being an iterative procedure, is computationally costly. Speaking of which, we observed a reduction of the number of iterations (within each ``sampling" -- see above) by about a factor of 10 between the vanilla and ``de-noised" matrices. Not only does ``de-noising" improve the quality of the resultant signatures, but it also provides substantial computational cost savings. This is not surprising in retrospect. ``De-noising" (not aggregation by cancer type) is the key factor in improving the overall stability -- the signatures based on aggregated data without ``de-noising" have much larger error bars.

\noindent $\bullet$ {\bf Framework.} This paper is not intended to be exhaustive in any way. As mentioned above, the data we work with here is limited, etc. Rather, the purpose of this paper is to set forth the framework of factor models for cancer signatures, including its application as an improvement to NMF. We hope it facilitates further research and helps identify the (hopefully not too many) underlying cancer signatures.

\section*{Acknowledgments}
{}We would like to thank Steven Rozen for valuable discussions and comments.

\appendix
\section{Genome Sample IDs}\label{app.IDs}

{}In this Appendix we give the sample IDs with the corresponding publication references for the genome data we use.\\
{\tiny
$\blacklozenge$ {\bf B Cell Lymphoma}:\\
$\bullet$ \cite{Alexandrov}:\\
4101316,
4105105,
4108101,
4112512,
4116738,
4119027,
4121361,
4125240,
4133511,
4135350,
4142267,
4158726,
4159170,
4163639,
4175837,
4177856,
4182393,
4189200,
4189998,
4190495,
4193278,
4194218,
4194891.\\
$\bullet$ \cite{Love}:\\
G1.\\
$\blacklozenge$ {\bf Bone Cancer}:\\
$\bullet$ \cite{Tirode}:\\
IC009T,
IC015T,
IC024T,
IC034T,
IC044T,
IC046T,
IC049T,
IC053T,
IC054T,
IC057T,
IC058T,
IC066T,
IC067T,
IC071T,
IC076T,
IC077T,
IC080T,
IC082T,
IC086T,
IC092T,
IC093T,
IC096T,
IC1057T,
IC105T,
IC106T,
IC111T,
IC112T,
IC114T,
IC116T,
IC121T,
IC128T,
IC130T,
IC136T,
IC147T,
IC149T,
IC151T,
IC158T,
IC165T,
IC168T,
IC174T,
IC193T,
IC196T,
IC197T,
IC198T,
IC204T,
IC213T,
IC215T,
IC224T,
IC242T,
IC248T,
IC254T,
IC262T,
IC263T,
IC264T,
IC267T,
IC268T,
IC270T,
IC271T,
IC272T,
IC273T,
IC274T,
IC275T,
IC277T,
IC278T,
IC279T,
IC280T,
IC282T,
IC283T,
IC284T,
IC286T,
IC288T,
IC294T,
IC295T,
IC296T,
IC297T,
IC299T,
IC300T,
IC301T,
IC302T,
IC303T,
IC305T,
IC306T,
IC309T,
IC310T,
IC311T,
IC315T,
IC316T,
IC318T,
IC319T,
IC323T,
IC324T,
IC325T,
IC340T,
IC343T,
IC349T,
IC831T,
IC929T,
IC973T.\\
$\blacklozenge$ {\bf Brain Lower Grade Glioma}:\\
$\bullet$ \cite{Alexandrov}:\\
PA10,
PA102,
PA103,
PA105,
PA107,
PA109,
PA11,
PA110,
PA112,
PA116,
PA117,
PA12,
PA131,
PA134,
PA136,
PA138,
PA14,
PA143,
PA145,
PA148,
PA149,
PA157,
PA166,
PA17,
PA20,
PA21,
PA22,
PA25,
PA3,
PA36,
PA4,
PA41,
PA43,
PA46,
PA48,
PA5,
PA53,
PA54,
PA55,
PA56,
PA58,
PA59,
PA62,
PA63,
PA64,
PA65,
PA69,
PA70,
PA73,
PA75,
PA79,
PA8,
PA81,
PA82,
PA83,
PA84,
PA85,
PA86,
PA87,
PA9,
PA90,
PA93,
PA96.\\
$\bullet$ \cite{Zhang}:\\
SJLGG001,
SJLGG002,
SJLGG003,
SJLGG004,
SJLGG005,
SJLGG006,
SJLGG006R,
SJLGG007,
SJLGG008,
SJLGG009,\\
SJLGG010,
SJLGG011,
SJLGG012,
SJLGG013,
SJLGG015,
SJLGG016,
SJLGG018,
SJLGG019,
SJLGG020,
SJLGG021,
SJLGG022,
SJLGG024,
SJLGG025,
SJLGG026,
SJLGG027,
SJLGG028,
SJLGG029,
SJLGG030,
SJLGG031,
SJLGG032,
SJLGG033,
SJLGG034,
SJLGG035,
SJLGG037,
SJLGG038,
SJLGG039,
SJLGG040,
SJLGG042.\\
$\blacklozenge$ {\bf Breast Cancer}:\\
$\bullet$ \cite{Nik-Zainal}:\\
PD3851a,
PD4085a,
PD4088a,
PD4103a,
PD4120a,
PD4194a,
PD4192a,
PD4198a,
PD4199a,
PD4248a,
PD4086a,
PD4109a,
PD4107a,
PD3890a,
PD3905a,
PD4005a,
PD4006a,
PD3904a,
PD3945a,
PD4115a,
PD4116a.\\
$\bullet$ \cite{Alexandrov}:\\
PD3989a,
PD4069a,
PD4072a,
PD4080a,
PD4224a,
PD4225a,
PD4255a,
PD4261a,
PD4266a,
PD4267a,
PD4315a,
PD4604a,
PD4605a,
PD4606a,
PD4607a,
PD4608a,
PD4613a,
PD4826a,
PD4833a,
PD4836a,
PD4841a,
PD4847a,
PD4951a,
PD4952a,
PD4953a,
PD4954a,
PD4955a,
PD4957a,
PD4958a,
PD4959a,
PD4962a,
PD4963a,
PD4965a,
PD4966a,
PD4967a,
PD4968a,
PD4970a,
PD4971a,
PD4972a,
PD4975a,
PD4976a,
PD4980a,
PD4981a,
PD4982a,
PD4983a,
PD4985a,
PD4986a,
PD5928a,
PD5934a,
PD5935a,
PD5936a,
PD5942a,
PD5944a,
PD5947a,
PD5951a,
PD5956a,
PD6018a,
PD6041a,
PD6042a,
PD6043a,
PD6044a,
PD6045a,
PD6046a,
PD6049a,
PD6409a,
PD6410a,
PD6411a,
PD6413a,
PD6417a,
PD6418a,
PD6422a,
PD6466b,
PD6719a,
PD6720a,
PD6721a,
PD6722a,
PD7199a,
PD7201a,
PD7207a,
PD7208a,
PD7209a,
PD7210a,
PD7212a,
PD7214a,
PD7215a,
PD7216a,
PD7217a,
PD7218a,
PD7219a,
PD7221a,
PD7321a,
PD7404a,
PD7409a,
PD7431a,
PD7433a,
PD8618a,
PD8622a,
PD8623a.\\
$\blacklozenge$ {\bf Chronic Lymphocytic Leukemia}:\\
$\bullet$ \cite{Alexandrov}:\\
001-0002-03TD,
003-0005-09TD,
012-02-1TD,
004-0012-05TD,
005-0015-01TD,
006-0018-01TD,
007-0020-01TD,
008-0022-01TD,
009-0026-02TD,
013-0035-01TD,
016-0040-02TD,
017-0042-01TD,
018-0046-01TD,
019-0047-01TD,
020-0049-01TD,
022-0053-01TD,
023-0056-01TD,
027-0063-01TD,
029-0065-01TD,
030-0066-01TD,
032-0069-01TD,
033-0070-01TD,
038-0087-01TD,
039-0076-01TD,
040-0088-01TD,
041-0090-01TD,
042-0091-01TD,
043-0094-01TD,
044-0092-01TD,
045-0082-04TD,
048-0089-01TD,
049-0086-01TD,
051-0099-05TD,
052-0103-01TD,
053-0104-02TD,
054-0114-05TD,
063-0127-01TD,
064-0128-01TD,
082-02-1TD,
083-01-2TD,
090-02-1TD,
091-01-6TD,
100-02-2TD,
110-0218-04TD,
117-01-1TD,
124-01-1TD,
136-02-3TD,
141-02-3TD,
144-01-1TD,
145-01-1TD,
146-01-5TD,
148-02-3TD,
152-01-4TD,
155-01-1TD,
156-01-1TD,
157-01-1TD,
159-01-1TD,
165-01-5TD,
166-01-4TD,
168-02-2TD,
170-01-3TD,
171-01-2TD,
172-01-1TD,
173-01-3TD,
174-01-3TD,
175-01-3TD,
178-01-2TD,
181-01-3TD,
182-01-4TD,
184-01-4TD,
185-01-6TD,
186-01-6TD,
188-01-2TD,
189-01-1TD,
191-01-3TD,
192-01-4TD,
193-01-1TD,
194-01-2TD,
195-01-5TD,
197-01-3TD,
264-01-7TD,
266-01-7TD,
267-01-6TD,
270-01-2TD,
272-01-2TD,
273-01-5TD,
274-01-3TD,
275-01-2TD,
276-01-4TD,
278-01-4TD,
279-01-4TD,
280-01-4TD,
282-01-12TD,
290-1950-01TD,
319-01-1TD,
321-01-1TD,
322-01-1TD,
323-01-1TD,
324-01-1TD,
325-01-1TD,
326-01-1TD,
328-01-1TD,
375-1099-15TD,
618-1503-04TD,
642-1991-01TD,
680-1992-01TD,
758-2041-01TD,
761-01-1TD,
785-1836-01TD.\\
$\bullet$ \cite{Puente2011}:\\
CLL4-ARTICLE.\\
$\bullet$ \cite{Puente2015}:\\
125,
128,
137,
141,
151,
178,
192,
26,
277,
282,
294,
306,
308,
318,
342,
343,
367,
393,
467,
473,
477,
519,
523,
564.\\
$\blacklozenge$ {\bf Esophageal Cancer}:\\
$\bullet$ \cite{Cheng}:\\
ESCC-ESCC-001T,
ESCC-ESCC-002T,
ESCC-ESCC-003T,
ESCC-ESCC-004T,
ESCC-ESCC-005T,
ESCC-ESCC-006T,\\
ESCC-ESCC-008T,
ESCC-ESCC-009T,
ESCC-ESCC-010T,
ESCC-ESCC-011T,
ESCC-ESCC-012T,
ESCC-ESCC-013T,\\
ESCC-ESCC-014T,
ESCC-ESCC-015T,
ESCC-ESCC-016T,
ESCC-ESCC-017T,
ESCC-ESCC-018T.\\
{\bf Gastric Cancer}:\\
$\bullet$ \cite{Wang}:\\
pfg005T,
pfg008T,
pfg022T,
pfg023T,
pfg030T,
pfg031T,
pfg032T,
pfg034T,
pfg035T,
pfg036T,
pfg038T,
pfg039T,
pfg043T,
pfg050T,
pfg052T,
pfg053T,
pfg054T,
pfg057T,
pfg058T,
pfg059T,
pfg060T,
pfg062T,
pfg064T,
pfg065T,
pfg068T,
pfg069T,
pfg072T,
pfg073T,
pfg076T,
pfg081T,
pfg082T,
pfg088T,
pfg089T,
pfg092T,
pfg094T,
pfg097T,
pfg099T,
pfg100T,
pfg102T,
pfg103T,
pfg104T,
pfg105T,
pfg106T,
pfg107T,
pfg108T,
pfg115T,
pfg116T,
pfg118T,
pfg119T,
pfg120T,
pfg121T,
pfg122T,
pfg123T,
pfg124T,
pfg125T,
pfg127T,
pfg129T,
pfg130T,
pfg132T,
pfg135T,
pfg136T,
pfg138T,
pfg142T,
pfg143T,
pfg144T,
pfg145T,
pfg146T,
pfg151T,
pfg156T,
pfg157T,
pfg160T,
pfg164T,
pfg166T,
pfg167T,
pfg173T,
pfg180T,
pfg181T,
pfg182T,
pfg205T,
pfg212T,
pfg213T,
pfg217T,
pfg220T,
pfg222T,
pfg228T,
pfg258T,
pfg272T,
pfg277T,
pfg282T,
pfg311T,
pfg316T,
pfg317T,
pfg344T,
pfg373T,
pfg375T,
pfg378T,
pfg398T,
pfg413T,
pfg416T,
pfg424T.\\
$\blacklozenge$ {\bf Liver Cancer}:\\
$\bullet$ \cite{Sung}:\\
HK101T,
HK105T,
HK106T,
HK108T,
HK113T,
HK114T,
HK115T,
HK116T,
HK117T,
HK11T,
HK122T,
HK126T,
HK131T,
HK13T,
HK145T,
HK14T,
HK154T,
HK159T,
HK169T,
HK172T,
HK174T,
HK177T,
HK179T,
HK17T,
HK180T,
HK181T,
HK182T,
HK186T,
HK193T,
HK198T,
HK19T,
HK200T,
HK203T,
HK204T,
HK205T,
HK206T,
HK207T,
HK21T,
HK22T,
HK23T,
HK260T,
HK261T,
HK262T,
HK266T,
HK267T,
HK268T,
HK26T,
HK272T,
HK273T,
HK274T,
HK276T,
HK29T,
HK30T,
HK32T,
HK34T,
HK35T,
HK36T,
HK38T,
HK39T,
HK41T,
HK43T,
HK45T,
HK46T,
HK49T,
HK53T,
HK55T,
HK58T,
HK60T,
HK62T,
HK63T,
HK64T,
HK65T,
HK67T,
HK68T,
HK70T,
HK71T,
HK73T,
HK75T,
HK76T,
HK79T,
HK81T,
HK82T,
HK84T,
HK87T,
HK90T,
HK92T,
HK95T,
HK98T.\\
$\bullet$ \cite{Fujimoto}:\\
RK001\_C01,
RK002\_C,
RK003\_C,
RK004\_C01,
RK005\_C,
RK006\_C1,
RK006\_C2,
RK007\_C01,
RK010\_C,
RK012\_C01,
RK014\_C01,
RK015\_C,
RK016\_C01,
RK018\_C01,
RK019\_C,
RK020\_C01,
RK021\_C01,
RK022\_C01,
RK023\_C,
RK024\_C,
RK025\_C,
RK026\_C01,
RK027\_C01,
RK028\_C01,
RK029\_C,
RK030\_C01,
RK031\_C01,
RK032\_C01,
RK033\_C01,
RK034\_C,
RK035\_C01,
RK036\_C01,\\
RK037\_C01,
RK038\_C01,
RK040\_C01,
RK041\_C01,
RK042\_C,
RK043\_C01,
RK044\_C01,
RK046\_C01,
RK046\_C02,
RK047\_C01,\\
RK048\_C,
RK049\_C01,
RK050\_C,
RK051\_C01,
RK052\_C01,
RK053\_C01,
RK054\_C01,
RK055\_C01,
RK056\_C01,
RK057\_C01,\\
RK058\_C01,
RK059\_C01,
RK060\_C01,
RK061\_C01,
RK062\_C01,
RK063\_C,
RK064\_C01,
RK065\_C01,
RK066\_C01,
RK067\_C01,\\
RK068\_C,
RK069\_C01,
RK070\_C01,
RK071\_C01,
RK072\_C01,
RK073\_C01,
RK074\_C01,
RK075\_C01,
RK076\_C01,
RK077\_C01,\\
RK079\_C01,
RK080\_C01,
RK081\_C01,
RK082\_C01,
RK083\_C01,
RK084\_C01,
RK085\_C01,
RK086\_C01,
RK087\_C01,
RK088\_C01,
RK089\_C01,
RK090\_C01,
RK091\_C01,
RK092\_C01,
RK093\_C01,
RK095\_C01,
RK096\_C01,
RK098\_C01,
RK099\_C01,
RK100\_C01,
RK101\_C01,
RK102\_C01,
RK103\_C01,
RK104\_C01,
RK105\_C01,
RK106\_C01,
RK107\_C01,
RK108\_C01,
RK109\_C01,
RK110\_C01,
RK111\_C01,
RK112\_C01,
RK113\_C01,
RK115\_C01,
RK116\_C01,
RK117\_C01,
RK118\_C01,
RK119\_C01,
RK120\_C01,
RK121\_C01,
RK122\_C01,
RK123\_C01,
RK124\_C01,
RK125\_C01,
RK126\_C01,
RK128\_C01,
RK130\_C01,
RK131\_C01,
RK133\_C01,
RK134\_C01,
RK135\_C01,
RK136\_C01,
RK137\_C01,
RK138\_C01,
RK139\_C01,
RK140\_C01,
RK141\_C01,
RK142\_C01,
RK143\_C01,
RK144\_C01,
RK145\_C01,
RK146\_C01,
RK147\_C01,
RK148\_C01,
RK149\_C01,
RK150\_C01,
RK151\_C01,
RK152\_C01,
RK153\_C01,
RK154\_C01,
RK155\_C01,
RK156\_C01,
RK157\_C01,
RK159\_C01,
RK162\_C01,
RK163\_C01,
RK164\_C01,
RK165\_C01,
RK166\_C01,
RK167\_C01,
RK169\_C01,
RK170\_C01,
RK171\_C01,
RK172\_C01,
RK175\_C01,
RK176\_C01,
RK177\_C01,
RK178\_C01,
RK179\_C01,
RK180\_C01,
RK180\_C02,
RK181\_C01,
RK182\_C01,
RK183\_C01,
RK184\_C01,
RK185\_C01,
RK186\_C01,
RK187\_C01,
RK188\_C01,
RK189\_C01,
RK190\_C01,
RK191\_C01,
RK193\_C01,
RK194\_C01,
RK195\_C01,
RK196\_C01,
RK197\_C01,
RK198\_C01,
RK199\_C01,
RK200\_C01,
RK201\_C01,
RK202\_C01,
RK204\_C02,
RK205\_C01,
RK206\_C01,
RK207\_C01,
RK208\_C01,
RK209\_C01,
RK210\_C01,
RK211\_C01,
RK212\_C01,
RK213\_C01,
RK214\_C01,
RK215\_C01,
RK216\_C01,
RK217\_C01,
RK219\_C01,
RK220\_C01,
RK221\_C01,
RK222\_C01,
RK223\_C01,
RK224\_C01,
RK225\_C01,
RK226\_C01,
RK227\_C01,
RK228\_C01,
RK229\_C01,
RK230\_C01,
RK232\_C01,
RK233\_C01,
RK233\_C02,
RK234\_C01,
RK235\_C01,
RK236\_C01,
RK237\_C01,
RK240\_C01,
RK241\_C01,
RK243\_C01,
RK244\_C01,
RK245\_C01,
RK254\_C01,
RK256\_C01,
RK257\_C01,
RK258\_C01,
RK259\_C01,
RK260\_C01,
RK261\_C01,
RK261\_C02,
RK262\_C01,
RK263\_C01,
RK264\_C01,
RK265\_C01,
RK266\_C01,
RK267\_C01,
RK268\_C01,
RK269\_C01,
RK270\_C01,
RK272\_C01,
RK275\_C01,
RK277\_C01,
RK278\_C01,
RK279\_C01,
RK280\_C01,
RK281\_C01,
RK282\_C01,
RK283\_C01,
RK284\_C01,
RK285\_C01,
RK287\_C01,
RK288\_C01,
RK289\_C01,
RK290\_C01,
RK297\_C01,
RK298\_C01,
RK303\_C01,
RK304\_C01,
RK305\_C01,
RK306\_C01,
RK307\_C01,
RK308\_C01,
RK309\_C01,
RK310\_C01,
RK312\_C01,
RK316\_C01,
RK317\_C01,
RK326\_C01,
RK337\_C01,
RK338\_C01,
HX10T,
HX11T,
HX12T,
HX13T,
HX14T,
HX15T,
HX16T,
HX17T,
HX18T,
HX19T,
HX20T,
HX21T,
HX22T,
HX23T,
HX24T,
HX25T,
HX26T,
HX27T,
HX28T,
HX29T,
HX30T,
HX31T,
HX32T,
HX33T,
HX34T,
HX35T,
HX36T,
HX37T,
HX4T,
HX5T,
HX9T.\\
$\blacklozenge$ {\bf Lung Cancer}:\\
$\bullet$ \cite{Imielinski}:\\
LU-A08-43,
LUAD-2GUGK,
LUAD-5V8LT,
LUAD-AEIUF,
LUAD-D02326,
LUAD-E00934,
LUAD-E01014,
LUAD-E01278,
LUAD-E01317,
LUAD-FH5PJ,
LUAD-QY22Z,
LUAD-S00488,
LUAD-S01302,
LUAD-S01331,
LUAD-S01341,
LUAD-S01345,
LUAD-S01346,
LUAD-S01356,
LUAD-S01381,
LUAD-S01404,
LUAD-S01405,
LUAD-S01467,
LUAD-S01478,
LUAD-U6SJ7.\\
$\blacklozenge$ {\bf Medulloblastoma}:\\
$\bullet$ \cite{Jones}:\\
LFS\_MB1,
LFS\_MB2,
LFS\_MB4,
MB1,
MB101,
MB102,
MB104,
MB106,
MB107,
MB108,
MB110,
MB112,
MB113,
MB114,
MB115,
MB117,
MB119,
MB12,
MB121,
MB122,
MB124,
MB125,
MB126,
MB127,
MB128,
MB129,
MB130,
MB131,
MB132,
MB134,
MB139,
MB15,
MB16,
MB17,
MB18,
MB19,
MB2,
MB20,
MB21,
MB23,
MB24,
MB26,
MB28,
MB3,
MB31,
MB32,
MB34,
MB35,
MB36,
MB37,
MB38,
MB39,
MB40,
MB45,
MB46,
MB49,
MB5,
MB50,
MB51,
MB518,
MB53,
MB56,
MB57,
MB58,
MB59,
MB6,
MB60,
MB61,
MB612,
MB63,
MB64,
MB66,
MB67,
MB69,
MB7,
MB70,
MB74,
MB75,
MB77,
MB78,
MB79,
MB8,
MB800,
MB81,
MB82,
MB83,
MB84,
MB85,
MB86,
MB88,
MB89,
MB9,
MB90,
MB91,
MB92,
MB94,
MB95,
MB96,
MB98,
MB99.\\
$\blacklozenge$ {\bf Ovarian Cancer}:\\
$\bullet$ \cite{Patch}:\\
AOCS-001-1,
AOCS-004-1,
AOCS-005-1,
AOCS-034-1,
AOCS-055-1,
AOCS-056-1,
AOCS-057-1,
AOCS-058-1,
AOCS-059-1,
AOCS-060-1,
AOCS-061-1,
AOCS-063-1,
AOCS-064-1,
AOCS-065-1,
AOCS-075-1,
AOCS-076-1,
AOCS-077-1,
AOCS-078-1,
AOCS-079-1,
AOCS-080-1,
AOCS-081-1,
AOCS-083-1,
AOCS-084-1,
AOCS-085-1,
AOCS-086-1,
AOCS-088-1,
AOCS-090-1,
AOCS-091-1,
AOCS-092-1,
AOCS-093-1,
AOCS-093-12,
AOCS-094-1,
AOCS-095-1,
AOCS-096-1,
AOCS-097-1,
AOCS-104-1,
AOCS-105-1,
AOCS-106-1,
AOCS-107-1,
AOCS-108-1,
AOCS-109-1,
AOCS-111-1,
AOCS-112-1,
AOCS-113-1,
AOCS-114-1,
AOCS-115-1,
AOCS-116-1,
AOCS-122-1,
AOCS-123-1,
AOCS-124-1,
AOCS-125-1,
AOCS-126-1,
AOCS-128-1,
AOCS-130-1,
AOCS-131-1,
AOCS-132-1,
AOCS-133-1,
AOCS-137-12,
AOCS-139-1,
AOCS-143-1,
AOCS-144-1,
AOCS-145-1,
AOCS-146-1,
AOCS-147-1,
AOCS-148-1,
AOCS-149-1,
AOCS-152-1,
AOCS-153-1,
AOCS-157-1,
AOCS-158-1,
AOCS-159-1,
AOCS-160-1,
AOCS-161-1,
AOCS-162-1,
AOCS-163-1,
AOCS-164-1,
AOCS-165-1,
AOCS-166-1,
AOCS-168-1,
AOCS-169-1,
AOCS-170-1,
AOCS-170-12,
AOCS-171-1,
AOCS-171-12.\\
$\blacklozenge$ {\bf Pancreatic Cancer}:\\
$\bullet$ \cite{Waddell}:\\
TD\_ICGC\_0002,
TD\_ICGC\_0004,
TD\_ICGC\_0005,
TD\_ICGC\_0006,
TD\_ICGC\_0007,
TD\_ICGC\_0008,
TD\_ICGC\_0009,
TD\_ICGC\_0016,
TD\_ICGC\_0025,
TD\_ICGC\_0026,
TD\_ICGC\_0031,
TD\_ICGC\_0032,
TD\_ICGC\_0033,
TD\_ICGC\_0034,
TD\_ICGC\_0035,
TD\_ICGC\_0036,
TD\_ICGC\_0037,
TD\_ICGC\_0040,
TD\_ICGC\_0042,
TD\_ICGC\_0051,
TD\_ICGC\_0052,
TD\_ICGC\_0054,
TD\_ICGC\_0055,
TD\_ICGC\_0059,
TD\_ICGC\_0061,
TD\_ICGC\_0062,
TD\_ICGC\_0063,
TD\_ICGC\_0066,
TD\_ICGC\_0069,
TD\_ICGC\_0072,
TD\_ICGC\_0075,
TD\_ICGC\_0076,
TD\_ICGC\_0077,
TD\_ICGC\_0087,
TD\_ICGC\_0088,
TD\_ICGC\_0089,
TD\_ICGC\_0098,
TD\_ICGC\_0103,
TD\_ICGC\_0105,
TD\_ICGC\_0109,
TD\_ICGC\_0114,
TD\_ICGC\_0115,
TD\_ICGC\_0116,
TD\_ICGC\_0118,
TD\_ICGC\_0119,
TD\_ICGC\_0121,
TD\_ICGC\_0131,
TD\_ICGC\_0134,
TD\_ICGC\_0135,
TD\_ICGC\_0137,
TD\_ICGC\_0138,
TD\_ICGC\_0139,
TD\_ICGC\_0140,
TD\_ICGC\_0141,
TD\_ICGC\_0143,
TD\_ICGC\_0144,
TD\_ICGC\_0146,
TD\_ICGC\_0149,
TD\_ICGC\_0153,
TD\_ICGC\_0154,
TD\_ICGC\_0169,
TD\_ICGC\_0182,
TD\_ICGC\_0185,
TD\_ICGC\_0188,
TD\_ICGC\_0199,
TD\_ICGC\_0201,
TD\_ICGC\_0205,
TD\_ICGC\_0206,
TD\_ICGC\_0207,
TD\_ICGC\_0212,
TD\_ICGC\_0214,
TD\_ICGC\_0215,
TD\_ICGC\_0223,
TD\_ICGC\_0235,
TD\_ICGC\_0239,
TD\_ICGC\_0242,
TD\_ICGC\_0255,
TD\_ICGC\_0257,
TD\_ICGC\_0283,
TD\_ICGC\_0285,
TD\_ICGC\_0289,
TD\_ICGC\_0290,
TD\_ICGC\_0295,
TD\_ICGC\_0296,
TD\_ICGC\_0300,
TD\_ICGC\_0303,
TD\_ICGC\_0304,
TD\_ICGC\_0309,
TD\_ICGC\_0312,
TD\_ICGC\_0315,
TD\_ICGC\_0321,
TD\_ICGC\_0326,
TD\_ICGC\_0327,
TD\_ICGC\_0328,
TD\_ICGC\_0329,
TD\_ICGC\_0338,
TD\_ICGC\_0420,
TD\_ICGC\_0532,
TD\_ICGC\_0548,
TD\_ICGC\_0549.\\
$\blacklozenge$ {\bf Prostate Cancer}:\\
$\bullet$ \cite{Gundem}:\\
A10E-0015\_CRUK\_PC\_0015\_T1,\\
A22C-0016\_CRUK\_PC\_0016\_T1,\\
A29C-0017\_CRUK\_PC\_0017\_T1,\\
A31C-0018\_CRUK\_PC\_0018\_T1,\\
A32C-0019\_CRUK\_PC\_0019\_T1.\\
$\blacklozenge$ {\bf Renal Cell Carcinoma}:\\
$\bullet$ \cite{Scelo}:\\
C0001T,
C0002T,
C0004T,
C0005T,
C0006T,
C0007T,
C0008T,
C0009T,
C0010T,
C0011T,
C0012T,
C0013T,
C0014T,
C0015T,
C0016T,
C0017T,
C0018T,
C0019T,
C0020T,
C0021T,
C0022T,
C0023T,
C0024T,
C0025T,
C0026T,
C0027T,
C0028T,
C0029T,
C0030T,
C0031T,
C0032T,
C0033T,
C0034T,
C0035T,
C0036T,
C0037T,
C0038T,
C0039T,
C0040T,
C0041T,
C0042T,
C0043T,
C0044T,
C0045T,
C0046T,
C0047T,
C0048T,
C0049T,
C0050T,
C0051T,
C0052T,
C0053T,
C0054T,
C0055T,
C0056T,
C0057T,
C0058T,
C0059T,
C0060T,
C0061T,
C0062T,
C0063T,
C0064T,
C0065T,
C0066T,
C0067T,
C0068T,
C0069T,
C0070T,
C0071T,
C0072T,
C0073T,
C0074T,
C0075T,
C0076T,
C0077T,
C0078T,
C0079T,
C0080T,
C0081T,
C0082T,
C0083T,
C0084T,
C0085T,
C0086T,
C0089T,
C0091T,
C0092T,
C0094T,
C0096T,
C0097T,
C0098T,
C0099T,
C0100T.
}

\section{R Source Code}\label{app.code}

{}In this appendix we give the R (R Package for Statistical Computing, http://www.r-project.org) source code for building purely statistical factor models (using principal components) based on the algorithm we discuss in Sections \ref{sub.statfm} and \ref{sub.fix.K}, including the ``minimization" and eRank based algorithms for fixing the number of factors $K$ in Section \ref{sub.fix.K}. The two functions below are self-explanatory and straightforward as they follow the formulas therein. This source code is an adaptation of that in Appendix A of \cite{KYb} reflecting peculiarities in the mutation count data.\footnote{\, E.g., we can have cases with vanishing serial variance, and the code deals with this accordingly. In financial applications this would correspond to a stock that does not trade at all.}

{}The function {\tt{\small bio.cov.pc(ret, use.cor = T, excl.first = F)}} corresponds to the ``minimization" based method for fixing $K$. The input is: i) {\tt{\small ret}}, an $N\times d$ matrix $R_{is}$ (see Section \ref{sec.5}), where $N$ is the number of mutation types, and $d$ is the number of samples (or, in the case where we aggregate samples by cancer type, the number of cancer types); ii) {\tt{\small use.cor}}, where for {\tt{\small TRUE}} (default) the factors are computed based on the principal components of the sample correlation matrix $\Psi_{ij}$, whereas for {\tt{\small FALSE}} they are computed based on the sample covariance matrix $C_{ij}$; {\tt{\small excl.first}}, where for {\tt{\small TRUE}} the $K^\prime$ based method of Subsection \ref{sub.k.prime} is used. The output is a list: {\tt{\small result\$pc}} are the first $K$ principal components of a) $C_{ij}$ for {\tt{\small use.cor = F}}, and b) $\Psi_{ij}$ for {\tt{\small use.cor = T}}. For details of the other named list members, which are not needed for our purposes here, see Appendix A of \cite{KYb}. However, we keep this output as it will be useful in future research projects.

{}The second function is {\tt{\small bio.erank.pc(ret, use.cor = T, do.trunc = F, k = 0, excl.first = F)}} and corresponds to the eRank based method for fixing $K$ for the default parameter {\tt{\small k = 0}}. The input is the same as in the {\tt{\small bio.cov.pc()}} function except for the additional parameters {\tt{\small do.trunc = F}} and {\tt{\small k = 0}}. For a positive integer {\tt{\small k}} the code simply takes its value as the number of factors $K$. For {\tt{\small k = 0}} (default) the code uses the eRank method: if {\tt{\small do.trunc = F}} (default), then $K = \mbox{Round}(\mbox{eRank}(\cdot))$, while if {\tt{\small do.trunc = T}}, then $K = \mbox{floor}(\mbox{eRank}(\cdot))$. (The argument of $\mbox{eRank}(\cdot)$ is the matrix $C_{ij}$ if {\tt{\small use.cor = F}}, and the matrix $\Psi_{ij}$ if {\tt{\small use.cor = T}}). The output is the same as in the {\tt{\small bio.cov.pc()}} function. We use defaults unless stated otherwise.

{}The source code we give in this appendix is not written to be ``fancy" or optimized for speed or in any other way -- e.g, using the functions {\tt{\small qrm.calc.eigen()}} and {\tt{\small qrm.calc.eigen.eff()}}, as applicable, provided in Appendix B and Appendix C of \cite{KYb}, respectively, instead of the built-in R function {\tt{\small eigen()}} may speed up the code. The sole purpose of the code herein is to illustrate the algorithms described in the main text in a simple-to-understand fashion. See Appendix \ref{app.disc} for some important legalese.\\
\\
{\tt{\small
bio.cov.pc <- function (ret, use.cor = T, excl.first = F)\\
\{\\
\indent print("Running bio.cov.pc()...")\\
\\
\indent tr <- apply(ret, 1, sd)\\
\indent take <- tr == 0\\
\indent tr[take] <- 1\\
\indent if(use.cor)\\
\indent \indent ret <- ret / tr\\
\\
\indent d <- ncol(ret)\\
\indent x <- ret\\
\indent x <- x - rowMeans(x)\\
\indent x <- x \%*\% t(x) / (ncol(x) - 1)\\
\\
\indent tv <- diag(x)\\
\indent x <- eigen(x)\\
\\
\indent if(excl.first)\\
\indent \{\\
\indent \indent k1 <- 2\\
\indent \indent y1 <- sqrt(x\$values[1]) * matrix(x\$vectors[, 1], nrow(ret), 1)\\
\indent \indent x1 <- y1 \%*\% t(y1)\\
\indent \indent tv <- tv - diag(x1)\\
\indent \}\\
\indent else\\
\indent \{\\
\indent \indent k1 <- 1\\
\indent \indent x1 <- 0\\
\indent \}\\
\\
\indent g.prev <- 999\\
\indent for(k in k1:(d-1))\\
\indent \{\\
\indent \indent u <- x\$values[k1:k]\\
\indent \indent v <- x\$vectors[, k1:k]\\
\indent \indent v <- t(sqrt(u) * t(v))\\
\indent \indent x.f <- v \%*\% t(v)\\
\indent \indent x.s <- tv - diag(x.f)\\
\indent \indent take <- x.s < 1e-10\\
\indent \indent x.s[take] <- 0\\
\indent \indent z <- x.s / tv\\
\indent \indent z <- z[is.finite(z)]\\
\indent \indent g <- abs(sqrt(min(z)) + sqrt(max(z)) - 1)\\
\\
\indent \indent if(is.na(g))\\
\indent \indent \indent break\\
\\
\indent \indent if(g > g.prev)\\
\indent \indent \indent break\\
\\
\indent \indent g.prev <- g\\
\indent \indent spec.risk <- sqrt(x.s)\\
\indent \indent if(excl.first)\\
\indent \indent \indent fac.load <- cbind(y1, v)\\
\indent \indent else\\
\indent \indent \indent fac.load <- v\\
\indent \indent fac.cov <- diag(1, k)\\
\indent \indent cov.mat <- diag(x.s) + x.f + x1\\
\indent \}\\
\\
\indent y.s <- 1 / spec.risk\^{}2\\
\indent v <- fac.load\\
\indent v1 <- y.s * v\\
\indent inv.cov <- diag(y.s) - v1 \%*\% \\
\indent\indent solve(diag(1, ncol(v)) + t(v)\%*\% v1) \%*\% t(v1)\\
\\
\indent if(use.cor)\\
\indent \{\\
\indent \indent spec.risk <- tr * spec.risk\\
\indent \indent fac.load <- tr * fac.load\\
\indent \indent cov.mat <- tr * t(tr * cov.mat)\\
\indent \indent inv.cov <- t(inv.cov / tr) / tr\\
\indent \}\\
\\
\indent result <- new.env()\\
\indent result\$spec.risk <- spec.risk\\
\indent result\$fac.load <- fac.load\\
\indent result\$fac.cov <- fac.cov\\
\indent result\$cov.mat <- cov.mat\\
\indent result\$inv.cov <- inv.cov\\
\indent result\$pc <- x\$vectors[, 1:ncol(fac.load)]\\
\indent result <- as.list(result)\\
\indent return(result)\\
\}\\
\\
bio.erank.pc <- function (ret, use.cor = T, do.trunc = F, k = 0,\\
\indent excl.first = F)\\
\{\\
\indent print("Running bio.erank.pc()...")\\
\\
\indent calc.erank <- function(x, excl.first)\\
\indent \{\\
\indent \indent take <- x > 0\\
\indent \indent x <- x[take]\\
\indent \indent if(excl.first)\\
\indent \indent \indent x <- x[-1]\\
\indent \indent p <- x / sum(x)\\
\indent \indent h <- - sum(p * log(p))\\
\indent \indent er <- exp(h)\\
\indent \indent if(excl.first)\\
\indent \indent \indent er <- er + 1\\
\indent \indent return(er)\\
\indent \}\\
\\
\indent if(use.cor)\\
\indent \{\\
\indent \indent tr <- apply(ret, 1, sd)\\
\indent \indent take <- tr == 0\\
\indent \indent tr[take] <- 1\\
\indent \indent ret <- ret / tr\\
\indent \}\\
\\
\indent x <- ret\\
\indent x <- x - rowMeans(x)\\
\indent x <- x \%*\% t(x) / (ncol(x) - 1)\\
\\
\indent tv <- diag(x)\\
\indent y <- eigen(x)\\
\\
\indent if(k == 0)\\
\indent \{\\
\indent \indent er <- calc.erank(y\$values, excl.first)\\
\\
\indent \indent if(do.trunc)\\
\indent \indent \indent k <- trunc(er)\\
\indent \indent else\\
\indent \indent \indent k <- round(er)\\
\indent \}\\
\\
\indent k <- min(k, ncol(ret) - 2)\\
\indent fac.load <- t(t(y\$vectors[, 1:k]) * sqrt(y\$values[1:k]))\\
\indent fac.cov <- diag(1, k)\\
\indent x.f <- fac.load \%*\% t(fac.load)\\
\indent x.s <- tv - diag(x.f)\\
\indent take <- x.s < 1e-10\\
\indent x.s[take] <- 0\\
\indent spec.risk <- sqrt(x.s)\\
\indent cov.mat <- diag(x.s) + x.f\\
\\
\indent y.s <- 1 / spec.risk\^{}2\\
\indent v <- fac.load\\
\indent v1 <- y.s * v\\
\indent inv.cov <- diag(y.s) - v1 \%*\% solve(diag(1, k) + t(v) \%*\% v1) \%*\% t(v1)\\
\\
\indent if(use.cor)\\
\indent \{\\
\indent \indent spec.risk <- tr * spec.risk\\
\indent \indent fac.load <- tr * fac.load\\
\indent \indent cov.mat <- tr * t(tr * cov.mat)\\
\indent \indent inv.cov <- t(inv.cov / tr) / tr\\
\indent \}\\
\\
\indent result <- new.env()\\
\indent result\$spec.risk <- spec.risk\\
\indent result\$fac.load <- fac.load\\
\indent result\$fac.cov <- fac.cov\\
\indent result\$cov.mat <- cov.mat\\
\indent result\$inv.cov <- inv.cov\\
\indent result\$pc <- y\$vectors[, 1:k]\\
\indent result <- as.list(result)\\
\indent return(result)\\
\}
}}

\section{DISCLAIMERS}\label{app.disc}

{}Wherever the context so requires, the masculine gender includes the feminine and/or neuter, and the singular form includes the plural and {\em vice versa}. The author of this paper (``Author") and his affiliates including without limitation Quantigic$^\circledR$ Solutions LLC (``Author's Affiliates" or ``his Affiliates") make no implied or express warranties or any other representations whatsoever, including without limitation implied warranties of merchantability and fitness for a particular purpose, in connection with or with regard to the content of this paper including without limitation any code or algorithms contained herein (``Content").

{}The reader may use the Content solely at his/her/its own risk and the reader shall have no claims whatsoever against the Author or his Affiliates and the Author and his Affiliates shall have no liability whatsoever to the reader or any third party whatsoever for any loss, expense, opportunity cost, damages or any other adverse effects whatsoever relating to or arising from the use of the Content by the reader including without any limitation whatsoever: any direct, indirect, incidental, special, consequential or any other damages incurred by the reader, however caused and under any theory of liability; any loss of profit (whether incurred directly or indirectly), any loss of goodwill or reputation, any loss of data suffered, cost of procurement of substitute goods or services, or any other tangible or intangible loss; any reliance placed by the reader on the completeness, accuracy or existence of the Content or any other effect of using the Content; and any and all other adversities or negative effects the reader might encounter in using the Content irrespective of whether the Author or his Affiliates is or are or should have been aware of such adversities or negative effects.

{}The R code included in Appendix \ref{app.code} hereof is part of the copyrighted R code of Quantigic$^\circledR$ Solutions LLC and is provided herein with the express permission of Quantigic$^\circledR$ Solutions LLC. The copyright owner retains all rights, title and interest in and to its copyrighted source code included in Appendix \ref{app.code} hereof and any and all copyrights therefor.

\newpage

\begin{table}[ht]
\noindent
\caption{Summaries of skewness for the mutation occurrence counts aggregated by cancer type for the 14 cancers in Table \ref{table.genome.summary}. $G$ is a $96 \times 14$ matrix of the so-aggregated counts. M = Mean/Median ratio, S = skewness (i.e., $\mu_3/\mu_2^{3/2}$, where $\mu_k$ is the $k$-th central moment). T = summary across cancer types, X = summary across mutation categories. 1st Qu. = 1st Quartile, 3rd Qu. = 3rd Quartile.}
\begin{tabular}{l l l l l l l} 
\\
\hline\hline 
Quantity & Min & 1st Qu. & Median & Mean & 3rd Qu. & Max\\[0.5ex] 
\hline 
$G$, M, T & 3.35 & 9.13 & 11.6 & 15.4 & 18.4 & 86.0\\
$\ln(1+G)$, M, T & 0.98 & 1.05 & 1.07 & 1.08 & 1.11 & 1.21\\
$G$, S, T & 0.69 & 1.82 & 2.40 & 2.33 & 3.00 & 3.32\\
$\ln(1 + G)$, S, T & -0.57 & -0.24 & -0.09 & -0.08 & 0.06 & 0.53\\
$G$, M, X & 1.06 & 1.14 & 1.33 & 1.39 & 1.47 & 2.21\\
$\ln(1+G)$, M, X & 0.97 & 0.98 & 0.99 & 0.99 & 1.00 & 1.04\\
$G$, S, X & 0.69 & 1.44 & 2.48 & 2.37 & 3.07 & 4.45\\
$\ln(1 + G)$, S, X & -0.76 & -0.67 & -0.22 & -0.15 & 0.17 & 0.85\\
 [1ex] 
\hline 
\end{tabular}
\label{table.skewness} 
\end{table}

\begin{table}[ht]
\noindent
\caption{Genome data summary. All Brain Lower Glioma samples are Pilocytic Astrocytoma samples. See Subsection \ref{data.summary} for the data source definitions.}
\begin{tabular}{l l l l} 
\\
\hline\hline 
Cancer Type & Total Counts & \# of Samples & Source \\[0.5ex] 
\hline 
B Cell Lymphoma & 43626 & 24 & A1-2\\
Bone Cancer & 36374 & 98 & B1\\
Brain Lower Grade Glioma & 3572 & 101 & A1, C1\\
Breast Cancer & 254381 & 119 & A1, D1\\
Chronic Lymphocytic Leukemia & 19489 & 134 & A1, E1-2\\
Esophageal Cancer & 1064 & 17 & F1\\
Gastric Cancer & 1996615 & 100 & G1\\
Liver Cancer & 3017487 & 389 & H1-2\\
Lung Cancer & 449527 & 24 & I1\\
Medulloblastoma & 44689 & 100 & J1\\
Ovarian Cancer & 668918 & 84 & K1\\
Pancreatic Cancer & 5087 & 100 & L1\\
Prostate Cancer & 29142 & 5 & M1\\
Renal Cell Carcinoma & 483329 & 94 & N1\\
All Cancer Types & 7053300 & 1389 & Above\\ [1ex] 
\hline 
\end{tabular}
\label{table.genome.summary} 
\end{table}

\begin{table}[ht]
\noindent
\caption{Statistical factor counts for the genome data. $d$ = number of samples. M1 = eRank based method; M2 = eRank + $K^\prime$ based method; M3 = ``minimization" based method; M4 ``minimization" + $K^\prime$ based method. See Subsections (\ref{sub.min}), (\ref{sub.erank}) and (\ref{sub.k.prime}) for details. In the eRank based methods we take Round($\cdot$) in (\ref{eq.eRank}). The ``All Cancer Types" entry is based on the samples for all 14 cancer types in Table \ref{table.genome.summary}. The ``Aggregated by Cancer Type" entry corresponds to aggregating the occurrence counts for all samples within each of the 14 cancer types in this table and running the aforementioned algorithms for fixing $K$ on the resulting $96 \times 14$ matrix.}
\begin{tabular}{l l l l l l} 
\\
\hline\hline 
Cancer Type & $d$ & $K$ (M1) & $K$ (M2) & $K$ (M3) & $K$ (M4) \\[0.5ex] 
\hline 
B Cell Lymphoma & 24 & 5 & 18 & 5 & 11\\
Bone Cancer & 98 & 14 & 54 & 16 & 33\\
Brain Lower Grade Glioma & 101 & 37 & 51 & 23 & 30\\
Breast Cancer & 119 & 6 & 37 & 6 & 22\\
Chronic Lymphocytic Leukemia & 134 & 3 & 43 & 4 & 23\\
Esophageal Cancer & 17 & 12 & 13 & 8 & 8\\
Gastric Cancer & 100 & 3 & 12 & 1 & 9\\
Liver Cancer & 389 & 2 & 38 & 1 & 20\\
Lung Cancer & 24 & 2 & 9 & 1 & 7\\
Medulloblastoma & 100 & 11 & 52 & 10 & 33\\
Ovarian Cancer & 84 & 3 & 21 & 2 & 13\\
Pancreatic Cancer & 100 & 36 & 52 & 23 & 33\\
Prostate Cancer & 5 & 2 & 3 & 2 & 3\\
Renal Cell Carcinoma & 94 & 3 & 23 & 2 & 17\\
All Cancer Types & 1389 & 2 & 15 & 1 & 18\\
Aggregated by Cancer Type & 14 & 1 & 7 & 1 & 5\\
 [1ex] 
\hline 
\end{tabular}
\label{table.genome.fac.counts} 
\end{table}

\begin{table}[ht]
\noindent
\caption{Average pair-wise correlation (Avg.Cor, in the units of 1\%) and the first 5 eigenvalues (Eig.1-5) of the sample correlation matrix based on the genome data. (For Prostate Cancer the fifth eigenvalue is null as $d=5$.)}
\begin{tabular}{l l l l l l l} 
\\
\hline\hline 
Cancer Type & Avg.Cor & Eig.1 & Eig.2 & Eig.3 & Eig.4 & Eig.5\\[0.5ex] 
\hline 
B Cell Lymphoma & 66.6 & 65.0 & 5.67 & 3.08 & 2.54 & 2.23\\
Bone Cancer & 48.1 & 48.1 & 3.31 & 2.03 & 1.88 & 1.81\\
Brain Lower Grade Glioma & 17.7 & 20.3 & 4.52 & 3.91 & 3.47 & 3.23\\
Breast Cancer & 65.2 & 64.1 & 6.18 & 3.01 & 2.07 & 1.11\\
Chronic Lymphocytic Leukemia & 79.6 & 77.6 & 1.73 & 1.23 & 1.17 & 1.07\\
Esophageal Cancer & 16.7 & 23.3 & 9.58 & 7.59 & 7.34 & 6.98\\
Gastric Cancer & 80.9 & 78.2 & 6.41 & 3.95 & 1.62 & 0.68\\
Liver Cancer & 87.9 & 84.7 & 1.77 & 1.09 & 0.90 & 0.76\\
Lung Cancer & 80.0 & 78.3 & 6.03 & 4.47 & 1.93 & 1.17\\
Medulloblastoma & 54.2 & 53.6 & 3.33 & 2.01 & 1.76 & 1.63\\
Ovarian Cancer & 75.6 & 73.8 & 6.04 & 2.64 & 1.99 & 1.31\\
Pancreatic Cancer & 17.0 & 21.3 & 5.38 & 3.71 & 3.27 & 2.84\\
Prostate Cancer & 68.1 & 68.5 & 11.5 & 8.60 & 7.43 & 0\\
Renal Cell Carcinoma & 78.2 & 75.9 & 5.89 & 1.86 & 1.63 & 0.97\\
All Cancer Types & 88.1 & 84.9 & 5.47 & 0.77 & 0.58 & 0.37\\
Aggregated by Cancer Type & 96.1 & 92.4 & 1.26 & 0.89 & 0.51 & 0.34\\
[1ex] 
\hline 
\end{tabular}
\label{table.eig.cor} 
\end{table}

\begin{table}[ht]
\noindent
\caption{Summaries of the absolute values of pair-wise inner products (in the units of 1\%) between the $a$-th principal components (PCs), $a=1,2,3$, and also for a combination of the 2nd and 3rd PCs, of the sample correlation matrices $[\Psi(\alpha)]_{ij}$ for the 14 individual cancer types in Table \ref{table.genome.fac.counts}. 1st Qu. = 1st Quartile, 3rd Qu. = 3rd Quartile, StDev = standard deviation, MAD = mean absolute deviation.}
\begin{tabular}{l l l l l l l l l} 
\\
\hline\hline 
prin.comp & Min & 1st Qu. & Median & Mean & 3rd Qu. & Max & StDev & MAD \\[0.5ex] 
\hline 
1st & 77.82 & 90.17 & 97.93 & 94.45 & 99.12 & 99.72 & 6.205 & 2.304\\
2nd & 0.171 & 6.035 & 14.72 & 20.08 & 29.94 & 69.35 & 17.63 & 14.14\\
3rd & 0.160 & 4.855 & 10.60 & 15.76 & 21.98 & 71.63 & 15.90 & 11.01\\
2nd+3rd & 0.160 & 5.190 & 12.40 & 17.96 & 24.28 & 82.30 & 17.31 & 12.01\\ [1ex] 
\hline 
\end{tabular}
\label{table.summary.corr} 
\end{table}

\begin{table}[ht]
\noindent
\caption{Cross-sectional summaries of $|\sqrt{N} V^{(1)}_i - 1|$ (in the units of 1\%), where $V^{(1)}_i$ is the first principal component of the sample covariance matrix $\Psi_{ij}$ for: each of the 14 cancer types; all cancer types; and data aggregated by cancer type (see Table \ref{table.genome.fac.counts}).}
\begin{tabular}{l l l l l l l} 
\\
\hline\hline 
Cancer Type & Min & 1st Qu. & Median & Mean & 3rd Qu. & Max \\[0.5ex] 
\hline 
B Cell Lymphoma & 0.169 & 3.102 & 6.030 & 7.777 & 9.975 & 44.02\\
Bone Cancer & 0.402 & 7.227 & 12.61 & 14.57 & 19.18 & 48.52\\
Brain Lower Grade Glioma & 0.208 & 10.26 & 24.91 & 28.82 & 41.19 & 105.6\\
Breast Cancer & 0.093 & 3.653 & 6.898 & 9.521 & 10.91 & 47.82\\
Chronic Lymph. Leukemia & 0.026 & 3.196 & 5.242 & 7.232 & 7.128 & 54.85\\
Esophageal Cancer & 0.092 & 19.17 & 42.68 & 47.30 & 65.43 & 138.7\\
Gastric Cancer & 0.061 & 2.084 & 4.586 & 5.078 & 6.487 & 15.95\\
Liver Cancer & 0.033 & 1.419 & 2.297 & 2.954 & 3.255 & 19.76\\
Lung Cancer & 0.214 & 4.399 & 5.646 & 8.916 & 7.841 & 49.05\\
Medulloblastoma & 0.393 & 5.862 & 10.42 & 12.09 & 15.90 & 46.33\\
Ovarian Cancer & 0.040 & 2.395 & 4.648 & 6.792 & 6.728 & 53.78\\
Pancreatic Cancer & 0.013 & 20.64 & 39.54 & 40.09 & 56.66 & 115.4\\
Prostate Cancer & 0.151 & 3.871 & 10.01 & 13.42 & 15.49 & 104.9\\
Renal Cell Carcinoma & 0.168 & 2.938 & 5.127 & 6.250 & 7.046 & 42.73\\
All Cancer Types & 0.043 & 1.442 & 2.552 & 3.461 & 4.024 & 17.58\\
Aggr. by Cancer Type & 0.048 & 0.565 & 1.059 & 1.232 & 1.383 & 6.576\\
 [1ex] 
\hline 
\end{tabular}
\label{table.1pc.diff} 
\end{table}

\begin{table}[ht]
\noindent
\caption{Statistical factor counts for the genome data with the ``overall" mode factored out -- see Subsection \ref{overall.stripped}. The notations are the same as in Table \ref{table.genome.fac.counts}.}
\begin{tabular}{l l l l l l} 
\\
\hline\hline 
Cancer Type & $d$ & $K$ (M1) & $K$ (M2) & $K$ (M3) & $K$ (M4) \\[0.5ex] 
\hline 
B Cell Lymphoma & 24 & 17 & 19 & 11 & 13\\
Bone Cancer & 98 & 49 & 56 & 28 & 32\\
Brain Lower Grade Glioma & 101 & 43 & 51 & 24 & 34\\
Breast Cancer & 119 & 35 & 42 & 18 & 30\\
Chronic Lymphocytic Leukemia & 134 & 26 & 50 & 15 & 37\\
Esophageal Cancer & 17 & 12 & 13 & 7 & 8\\
Gastric Cancer & 100 & 12 & 16 & 7 & 14\\
Liver Cancer & 389 & 36 & 44 & 24 & 40\\
Lung Cancer & 24 & 8 & 13 & 4 & 9\\
Medulloblastoma & 100 & 46 & 55 & 28 & 32\\
Ovarian Cancer & 84 & 20 & 27 & 12 & 21\\
Pancreatic Cancer & 100 & 41 & 53 & 23 & 31\\
Prostate Cancer & 5 & 3 & 3 & 3 & 3\\
Renal Cell Carcinoma & 94 & 23 & 33 & 13 & 30\\
All Cancer Types & 1389 & 21 & 41 & 14 & 32\\
Aggregated by Cancer Type & 14 & 7 & 8 & 5 & 6\\
 [1ex] 
\hline 
\end{tabular}
\label{table.genome.fac.counts.overall.stripped} 
\end{table}

\begin{table}[ht]
\noindent
\caption{Average pair-wise correlation and the first 5 eigenvalues of the sample correlation matrix for the genome data with the ``overall" mode factored out -- see Subsection \ref{overall.stripped}. The notations are the same as in Table \ref{table.eig.cor}.}
\begin{tabular}{l l l l l l l} 
\\
\hline\hline 
Cancer Type & Avg.Cor & Eig.1 & Eig.2 & Eig.3 & Eig.4 & Eig.5\\[0.5ex] 
\hline 
B Cell Lymphoma & -0.98 & 19.9 & 9.51 & 6.79 & 6.15 & 6.02\\
Bone Cancer & -1.04 & 11.4 & 4.32 & 3.92 & 3.08 & 3.02\\
Brain Lower Grade Glioma & -0.01 & 14.0 & 4.70 & 4.20 & 3.62 & 3.49\\
Breast Cancer & -0.31 & 15.5 & 14.3 & 5.27 & 3.51 & 2.84\\
Chronic Lymphocytic Leukemia & -0.92 & 31.8 & 6.13 & 3.08 & 2.96 & 2.58\\
Esophageal Cancer & 0.16 & 19.3 & 10.9 & 9.37 & 8.65 & 8.08\\
Gastric Cancer & -0.42 & 30.8 & 25.2 & 7.25 & 4.02 & 3.12\\
Liver Cancer & -0.87 & 15.1 & 11.3 & 8.17 & 5.59 & 3.81\\
Lung Cancer & -0.44 & 44.1 & 13.5 & 6.74 & 5.49 & 4.90\\
Medulloblastoma & -1.04 & 13.6 & 4.84 & 4.00 & 3.54 & 2.95\\
Ovarian Cancer & -0.81 & 22.8 & 17.3 & 8.23 & 4.17 & 4.05\\
Pancreatic Cancer & -0.01 & 17.2 & 4.65 & 3.98 & 3.29 & 3.15\\
Prostate Cancer & -0.67 & 30.6 & 24.8 & 22.6 & 18.0 & 0\\
Renal Cell Carcinoma & 0.15 & 25.6 & 11.9 & 7.78 & 5.05 & 3.85\\
All Cancer Types & -0.89 & 32.9 & 11.7 & 5.68 & 4.41 & 3.23\\
Aggregated by Cancer Type & -0.91 & 28.9 & 21.0 & 15.6 & 7.94 & 5.77\\
[1ex] 
\hline 
\end{tabular}
\label{table.eig.cor.overall.stripped} 
\end{table}

\begin{table}[ht]
\noindent
\caption{Statistical factor counts for the genome data with $R_{is} = G_{is}$ (no log -- see Subsection \ref{nolog.genome} for details). The notations are the same as in Table \ref{table.genome.fac.counts}.}
\begin{tabular}{l l l l l l} 
\\
\hline\hline 
Cancer Type & $d$ & $K$ (M1) & $K$ (M2) & $K$ (M3) & $K$ (M4) \\[0.5ex] 
\hline 
B Cell Lymphoma & 24 & 3 & 10 & 2 & 6\\
Bone Cancer & 98 & 11 & 37 & 8 & 20\\
Brain Lower Grade Glioma & 101 & 34 & 46 & 18 & 27\\
Breast Cancer & 119 & 5 & 17 & 2 & 11\\
Chronic Lymphocytic Leukemia & 134 & 4 & 24 & 4 & 14\\
Esophageal Cancer & 17 & 11 & 13 & 8 & 9\\
Gastric Cancer & 100 & 4 & 7 & 3 & 4\\
Liver Cancer & 389 & 3 & 11 & 2 & 7\\
Lung Cancer & 24 & 2 & 6 & 2 & 5\\
Medulloblastoma & 100 & 7 & 38 & 5 & 22\\
Ovarian Cancer & 84 & 4 & 10 & 2 & 5\\
Pancreatic Cancer & 100 & 25 & 43 & 15 & 30\\
Prostate Cancer & 5 & 3 & 3 & 2 & 3\\
Renal Cell Carcinoma & 94 & 4 & 9 & 2 & 5\\
All Cancer Types & 1389 & 6 & 10 & 4 & 6\\
Aggregated by Cancer Type & 14 & 2 & 3 & 1 & 3\\
 [1ex] 
\hline 
\end{tabular}
\label{table.genome.fac.counts.nolog} 
\end{table}

\begin{table}[ht]
\noindent
\caption{Average pair-wise correlation and the first 5 eigenvalues of the sample correlation matrix for the genome data with $R_{is} = G_{is}$ (no log -- see Subsection \ref{nolog.genome} for details). The notations are the same as in Table \ref{table.eig.cor}.}
\begin{tabular}{l l l l l l l} 
\\
\hline\hline 
Cancer Type & Avg.Cor & Eig.1 & Eig.2 & Eig.3 & Eig.4 & Eig.5\\[0.5ex] 
\hline 
 Cell Lymphoma & 76.6 & 75.1 & 6.67 & 4.59 & 2.89 & 1.75\\
Bone Cancer & 51.1 & 50.9 & 6.19 & 3.33 & 2.81 & 2.59\\
Brain Lower Grade Glioma & 17.7 & 20.5 & 5.15 & 4.74 & 4.31 & 3.67\\
Breast Cancer & 60.8 & 62.0 & 13.4 & 3.50 & 2.69 & 1.09\\
Chronic Lymphocytic Leukemia & 74.3 & 72.5 & 4.23 & 2.20 & 1.79 & 1.34\\
Esophageal Cancer & 16.8 & 25.0 & 10.9 & 7.72 & 7.42 & 6.65\\
Gastric Cancer & 57.5 & 57.8 & 14.3 & 13.3 & 2.72 & 1.87\\
Liver Cancer & 72.9 & 71.5 & 10.4 & 4.08 & 1.85 & 1.60\\
Lung Cancer & 79.2 & 78.5 & 8.28 & 4.82 & 1.09 & 0.92\\
Medulloblastoma & 63.5 & 62.3 & 4.27 & 2.86 & 2.29 & 1.84\\
Ovarian Cancer & 63.9 & 62.8 & 15.8 & 4.63 & 2.42 & 1.94\\
Pancreatic Cancer & 20.5 & 28.5 & 7.30 & 4.10 & 3.75 & 3.09\\
Prostate Cancer & 65.9 & 66.7 & 13.5 & 9.39 & 6.44 & 0\\
Renal Cell Carcinoma & 60.8 & 60.0 & 15.4 & 9.97 & 2.21 & 0.97\\
All Cancer Types & 49.2 & 49.4 & 15.3 & 8.92 & 5.97 & 3.88\\
Aggregated by Cancer Type & 74.7 & 73.7 & 16.9 & 2.91 & 1.43 & 0.75\\
[1ex] 
\hline 
\end{tabular}
\label{table.eig.cor.genome.nolog} 
\end{table}

\begin{table}[ht]
\noindent
\caption{Statistical factor counts for the genome data with $R_{is} = G_{is}$ (no log) with the ``overall" mode factored out -- see Subsection \ref{nolog.genome.overall.stripped} for details. The notations are the same as in Table \ref{table.genome.fac.counts}.}
\begin{tabular}{l l l l l l} 
\\
\hline\hline 
Cancer Type & $d$ & $K$ (M1) & $K$ (M2) & $K$ (M3) & $K$ (M4) \\[0.5ex] 
\hline 
B Cell Lymphoma & 24 & 3 & 7 & 2 & 6\\
Bone Cancer & 98 & 19 & 33 & 10 & 21\\
Brain Lower Grade Glioma & 101 & 29 & 45 & 17 & 33\\
Breast Cancer & 119 & 2 & 14 & 1 & 9\\
Chronic Lymphocytic Leukemia & 134 & 9 & 22 & 5 & 14\\
Esophageal Cancer & 17 & 9 & 12 & 6 & 10\\
Gastric Cancer & 100 & 4 & 6 & 2 & 5\\
Liver Cancer & 389 & 3 & 10 & 3 & 6\\
Lung Cancer & 24 & 3 & 6 & 2 & 6\\
Medulloblastoma & 100 & 10 & 34 & 6 & 26\\
Ovarian Cancer & 84 & 5 & 11 & 3 & 7\\
Pancreatic Cancer & 100 & 6 & 38 & 5 & 23\\
Prostate Cancer & 5 & 3 & 3 & 2 & 3\\
Renal Cell Carcinoma & 94 & 4 & 6 & 2 & 3\\
All Cancer Types & 1389 & 7 & 9 & 5 & 6\\
Aggregated by Cancer Type & 14 & 3 & 3 & 2 & 3\\
 [1ex] 
\hline 
\end{tabular}
\label{table.genome.fac.counts.nolog.overall.stripped} 
\end{table}

\begin{table}[ht]
\noindent
\caption{Average pair-wise correlation and the first 5 eigenvalues of the sample correlation matrix for the genome data with $R_{is} = G_{is}$ (no log) with the ``overall" mode factored out -- see Subsection \ref{nolog.genome.overall.stripped} for details. The notations are the same as in Table \ref{table.eig.cor}.}
\begin{tabular}{l l l l l l l} 
\\
\hline\hline 
Cancer Type & Avg.Cor & Eig.1 & Eig.2 & Eig.3 & Eig.4 & Eig.5\\[0.5ex] 
\hline 
B Cell Lymphoma & 17.6 & 63.4 & 16.6 & 4.87 & 3.32 & 1.96\\
Bone Cancer & 2.01 & 32.1 & 10.9 & 4.87 & 4.56 & 3.76\\
Brain Lower Grade Glioma & 4.56 & 25.2 & 5.27 & 4.32 & 3.69 & 3.60\\
Breast Cancer & 58.1 & 87.2 & 3.07 & 1.32 & 0.98 & 0.57\\
Chronic Lymphocytic Leukemia & 1.00 & 47.3 & 9.59 & 4.27 & 3.70 & 3.31\\
Esophageal Cancer & 7.75 & 33.6 & 12.1 & 7.90 & 7.03 & 6.19\\
Gastric Cancer & 14.1 & 59.2 & 20.6 & 5.32 & 3.07 & 2.02\\
Liver Cancer & 8.01 & 69.5 & 11.4 & 3.74 & 2.86 & 1.53\\
Lung Cancer & 18.4 & 70.9 & 12.2 & 5.79 & 1.98 & 1.41\\
Medulloblastoma & 8.37 & 50.6 & 7.63 & 3.29 & 3.14 & 2.56\\
Ovarian Cancer & 7.54 & 56.4 & 15.9 & 7.69 & 3.77 & 2.12\\
Pancreatic Cancer & 25.6 & 64.6 & 3.24 & 2.59 & 2.01 & 1.78\\
Prostate Cancer & 1.60 & 44.7 & 22.7 & 18.0 & 10.7 & 0\\
Renal Cell Carcinoma & 19.0 & 54.6 & 23.6 & 10.2 & 2.00 & 0.76\\
All Cancer Types & 3.79 & 34.8 & 21.9 & 11.2 & 8.54 & 4.42\\
Aggregated by Cancer Type & 0.75 & 53.0 & 33.0 & 5.77 & 2.23 & 1.35\\
[1ex] 
\hline 
\end{tabular}
\label{table.eig.cor.genome.nolog.overall.stripped} 
\end{table}

\begin{table}[ht]
\noindent
\caption{Weights ``sig$\star$" and errors ``err$\star$" (in the units of 1\%, rounded to 2 digits) for the first 48 mutation categories for the 7 signatures we extract based on the ``de-noised" matrix ${\widetilde G}$. See Subsection \ref{NMF.stripped} for details.}
{\tiny
\begin{tabular}{l l l l l l l l l l l l l l l} 
\\
\hline\hline 
Mutation & sig1 & sig2 & sig3 & sig4 & sig5 & sig6 & sig7 & err1 & err2 & err3 & err4 & err5 & err6 & err7\\[0.5ex] 
\hline 
\\
C $>$ A: ACA & 0.39 & 0.7 & 4.31 & 0.8 & 1.72 & 1.75 & 1.81 & 0.18 & 0.22 & 0.21 & 0.17 & 0.03 & 0.27 & 0.25\\
C $>$ A: ACC & 0.49 & 0.49 & 2.51 & 0.5 & 0.48 & 1.03 & 1.61 & 0.13 & 0.15 & 0.12 & 0.13 & 0.02 & 0.14 & 0.31\\
C $>$ A: ACG & 0.21 & 0.2 & 0.9 & 0 & 1.75 & 0.02 & 0.09 & 0.06 & 0.06 & 0.08 & 0 & 0.02 & 0.04 & 0.14\\
C $>$ A: ACT & 0.2 & 0.9 & 2.77 & 0.48 & 1.14 & 0.74 & 1.47 & 0.12 & 0.16 & 0.14 & 0.13 & 0.02 & 0.19 & 0.24\\
C $>$ A: CCA & 0.98 & 0.83 & 5.39 & 0.68 & 0.27 & 0.87 & 0.73 & 0.29 & 0.32 & 0.29 & 0.12 & 0.03 & 0.44 & 0.28\\
C $>$ A: CCC & 0.7 & 0.46 & 4.65 & 0.55 & 0.86 & 0.47 & 0.47 & 0.26 & 0.29 & 0.28 & 0.1 & 0.03 & 0.4 & 0.26\\
C $>$ A: CCG & 0.69 & 0.34 & 1.17 & 0 & 0.29 & 0 & 0.23 & 0.06 & 0.09 & 0.11 & 0.01 & 0.02 & 0.01 & 0.2\\
C $>$ A: CCT & 1.64 & 0.48 & 4.53 & 1.97 & 0.01 & 0.66 & 0.41 & 0.27 & 0.29 & 0.28 & 0.14 & 0.02 & 0.4 & 0.25\\
C $>$ A: GCA & 0.57 & 0.99 & 2.97 & 1.17 & 1.42 & 0.31 & 1.46 & 0.13 & 0.21 & 0.18 & 0.09 & 0.03 & 0.23 & 0.35\\
C $>$ A: GCC & 0.54 & 0.51 & 2.46 & 0.31 & 1.08 & 0.43 & 0.56 & 0.12 & 0.13 & 0.14 & 0.05 & 0.02 & 0.18 & 0.13\\
C $>$ A: GCG & 0.43 & 0.28 & 0.9 & 0 & 0.6 & 0 & 0.07 & 0.06 & 0.07 & 0.08 & 0 & 0.01 & 0.02 & 0.14\\
C $>$ A: GCT & 0.6 & 0.74 & 2.12 & 1.05 & 1.14 & 0.49 & 0.69 & 0.1 & 0.11 & 0.11 & 0.04 & 0.01 & 0.14 & 0.16\\
C $>$ A: TCA & 0.69 & 2.13 & 3.41 & 0.97 & 1.05 & 1.57 & 1.27 & 0.16 & 0.15 & 0.08 & 0.1 & 0.02 & 0.18 & 0.19\\
C $>$ A: TCC & 1.09 & 1.63 & 3.37 & 0.24 & 4.75 & 0.86 & 0.62 & 0.18 & 0.18 & 0.18 & 0.06 & 0.05 & 0.24 & 0.4\\
C $>$ A: TCG & 0.43 & 0.31 & 0.6 & 0.03 & 0.39 & 0.08 & 0.13 & 0.03 & 0.04 & 0.04 & 0.02 & 0.01 & 0.06 & 0.09\\
C $>$ A: TCT & 0.94 & 2.27 & 3.74 & 2.21 & 0.66 & 1.22 & 2.4 & 0.14 & 0.17 & 0.15 & 0.12 & 0.03 & 0.22 & 0.28\\
C $>$ G: ACA & 0.14 & 0.47 & 1.12 & 0.42 & 1.43 & 1.14 & 0.74 & 0.06 & 0.06 & 0.05 & 0.06 & 0.02 & 0.07 & 0.2\\
C $>$ G: ACC & 0.42 & 0.51 & 0.54 & 0.14 & 0.21 & 0.7 & 0.45 & 0.02 & 0.03 & 0.02 & 0.02 & 0.01 & 0.02 & 0.08\\
C $>$ G: ACG & 0.08 & 0.28 & 0.18 & 0 & 0.96 & 0.13 & 0.3 & 0.02 & 0.02 & 0.02 & 0.01 & 0.01 & 0.02 & 0.05\\
C $>$ G: ACT & 0.4 & 0.91 & 0.64 & 0.3 & 3.1 & 1.02 & 0.67 & 0.03 & 0.04 & 0.05 & 0.03 & 0.03 & 0.05 & 0.08\\
C $>$ G: CCA & 0.61 & 0.87 & 0.78 & 0.1 & 0.32 & 0.52 & 0.31 & 0.04 & 0.05 & 0.03 & 0.02 & 0.01 & 0.05 & 0.05\\
C $>$ G: CCC & 0.27 & 0.44 & 0.7 & 0.09 & 0.95 & 0.62 & 0.34 & 0.03 & 0.04 & 0.02 & 0.02 & 0.01 & 0.02 & 0.06\\
C $>$ G: CCG & 0.19 & 0.03 & 0.17 & 0 & 3 & 0.02 & 0.15 & 0.03 & 0.04 & 0.07 & 0 & 0.03 & 0.03 & 0.1\\
C $>$ G: CCT & 0.56 & 1.32 & 0.78 & 0.16 & 0.51 & 0.85 & 0.54 & 0.05 & 0.05 & 0.06 & 0.02 & 0.01 & 0.04 & 0.05\\
C $>$ G: GCA & 0.2 & 0.35 & 0.61 & 0.08 & 1.82 & 0.35 & 0.2 & 0.03 & 0.02 & 0.04 & 0.01 & 0.02 & 0.02 & 0.05\\
C $>$ G: GCC & 0.28 & 0.47 & 0.6 & 0.18 & 0.4 & 0.37 & 0.31 & 0.02 & 0.02 & 0.01 & 0.01 & 0 & 0.02 & 0.03\\
C $>$ G: GCG & 0.09 & 0.18 & 0.2 & 0 & 1.8 & 0.01 & 0.02 & 0.02 & 0.02 & 0.04 & 0 & 0.02 & 0.01 & 0.05\\
C $>$ G: GCT & 0.21 & 0.78 & 0.47 & 0.27 & 1.52 & 0.51 & 0.31 & 0.03 & 0.03 & 0.03 & 0.01 & 0.02 & 0.03 & 0.05\\
C $>$ G: TCA & 0.49 & 9.4 & 1.28 & 0.09 & 0.38 & 0.23 & 0.3 & 0.46 & 0.57 & 0.7 & 0.12 & 0.03 & 0.36 & 0.48\\
C $>$ G: TCC & 0.47 & 2.55 & 0.88 & 0.16 & 1.4 & 0.94 & 0.34 & 0.12 & 0.15 & 0.15 & 0.03 & 0.02 & 0.09 & 0.09\\
C $>$ G: TCG & 0.21 & 0.39 & 0.17 & 0 & 0.32 & 0.04 & 0.21 & 0.01 & 0.02 & 0.02 & 0 & 0 & 0.01 & 0.04\\
C $>$ G: TCT & 0.61 & 12 & 1.41 & 0.53 & 0.04 & 1.35 & 0.84 & 0.56 & 0.71 & 0.89 & 0.16 & 0.04 & 0.42 & 0.36\\
C $>$ T: ACA & 0.98 & 0.92 & 1.46 & 2.6 & 0.89 & 1.83 & 3.38 & 0.19 & 0.23 & 0.14 & 0.16 & 0.04 & 0.18 & 0.33\\
C $>$ T: ACC & 2.2 & 0.57 & 0.6 & 0.68 & 1.58 & 1.14 & 1.29 & 0.09 & 0.06 & 0.05 & 0.08 & 0.01 & 0.07 & 0.15\\
C $>$ T: ACG & 13.7 & 0.86 & 0.4 & 1.49 & 0.3 & 1.75 & 8.25 & 0.88 & 0.66 & 0.38 & 0.65 & 0.16 & 0.6 & 1.33\\
C $>$ T: ACT & 0.94 & 0.73 & 0.9 & 1.05 & 2.13 & 1.48 & 2.03 & 0.11 & 0.13 & 0.08 & 0.09 & 0.01 & 0.11 & 0.17\\
C $>$ T: CCA & 1.45 & 1.3 & 1.53 & 0.93 & 0.76 & 1.38 & 2.14 & 0.08 & 0.09 & 0.05 & 0.09 & 0.02 & 0.06 & 0.08\\
C $>$ T: CCC & 1.2 & 0.53 & 1.18 & 0.88 & 1.72 & 1.21 & 2.28 & 0.13 & 0.16 & 0.1 & 0.12 & 0.02 & 0.11 & 0.16\\
C $>$ T: CCG & 12.2 & 1.91 & 0.34 & 0.92 & 0.13 & 0.67 & 3.51 & 0.54 & 0.36 & 0.2 & 0.65 & 0.12 & 0.32 & 1.52\\
C $>$ T: CCT & 0.9 & 1 & 1.47 & 1.03 & 0.92 & 1.68 & 3.44 & 0.22 & 0.26 & 0.15 & 0.23 & 0.04 & 0.17 & 0.3\\
C $>$ T: GCA & 2.31 & 0.9 & 0.78 & 1.51 & 1.66 & 1.34 & 1.61 & 0.09 & 0.06 & 0.04 & 0.06 & 0.01 & 0.08 & 0.13\\
C $>$ T: GCC & 2.65 & 0.52 & 0.76 & 1.65 & 0.29 & 1.24 & 1.68 & 0.12 & 0.08 & 0.06 & 0.09 & 0.03 & 0.09 & 0.22\\
C $>$ T: GCG & 15.2 & 0.58 & 0.61 & 1.88 & 1 & 1.57 & 2.77 & 0.75 & 0.44 & 0.3 & 0.88 & 0.15 & 0.47 & 1.86\\
C $>$ T: GCT & 1.9 & 0.66 & 0.61 & 1.32 & 1.64 & 1.26 & 2.14 & 0.13 & 0.14 & 0.08 & 0.07 & 0.01 & 0.12 & 0.09\\
C $>$ T: TCA & 1.62 & 14.1 & 1.26 & 1.42 & 0.24 & 1.28 & 1.93 & 0.61 & 0.77 & 1.04 & 0.17 & 0.05 & 0.46 & 0.54\\
C $>$ T: TCC & 1.99 & 3.42 & 1.05 & 1.4 & 1.02 & 1.96 & 2.11 & 0.1 & 0.12 & 0.17 & 0.05 & 0.02 & 0.1 & 0.09\\
C $>$ T: TCG & 7.39 & 2.9 & 0.12 & 0.36 & 1.76 & 0.2 & 2.99 & 0.32 & 0.3 & 0.17 & 0.37 & 0.07 & 0.18 & 0.99\\
C $>$ T: TCT & 0.94 & 7.51 & 1.88 & 1.86 & 0.69 & 1.49 & 2.09 & 0.3 & 0.33 & 0.47 & 0.11 & 0.03 & 0.22 & 0.32\\
[1ex] 
\hline 
\end{tabular}
}
\label{weights.errors.1} 
\end{table}

\begin{table}[ht]
\noindent
\caption{Table \ref{weights.errors.1} continued: weights and errors for the next 48 mutation categories.}
{\tiny
\begin{tabular}{l l l l l l l l l l l l l l l} 
\\
\hline\hline 
Mutation & sig1 & sig2 & sig3 & sig4 & sig5 & sig6 & sig7 & err1 & err2 & err3 & err4 & err5 & err6 & err7\\[0.5ex] 
\hline 
\\
T $>$ A: ATA & 0.01 & 0.03 & 0.95 & 0.54 & 1.2 & 2.52 & 0.74 & 0.02 & 0.07 & 0.17 & 0.08 & 0.03 & 0.12 & 0.37\\
T $>$ A: ATC & 0.32 & 0.58 & 0.46 & 0.53 & 0.26 & 0.84 & 0.7 & 0.02 & 0.03 & 0.03 & 0.03 & 0.01 & 0.03 & 0.09\\
T $>$ A: ATG & 0.28 & 0.23 & 0.93 & 0.27 & 1.63 & 1.9 & 0.54 & 0.04 & 0.06 & 0.11 & 0.04 & 0.02 & 0.07 & 0.18\\
T $>$ A: ATT & 0.1 & 0.52 & 0.84 & 1.87 & 1.07 & 1.61 & 1.55 & 0.08 & 0.11 & 0.08 & 0.09 & 0.02 & 0.11 & 0.34\\
T $>$ A: CTA & 0.22 & 0.03 & 1.08 & 0.02 & 0.29 & 2.95 & 0.03 & 0.13 & 0.07 & 0.23 & 0.06 & 0.03 & 0.29 & 0.17\\
T $>$ A: CTC & 0.46 & 0.46 & 0.87 & 0.62 & 1.41 & 1.45 & 0.29 & 0.06 & 0.05 & 0.08 & 0.03 & 0.02 & 0.06 & 0.12\\
T $>$ A: CTG & 0.75 & 0.45 & 1.97 & 0.02 & 0.27 & 3.01 & 0.07 & 0.2 & 0.13 & 0.21 & 0.06 & 0.03 & 0.31 & 0.25\\
T $>$ A: CTT & 0.43 & 0.44 & 1.15 & 1.28 & 0.05 & 1.81 & 0.73 & 0.04 & 0.05 & 0.09 & 0.05 & 0.01 & 0.05 & 0.16\\
T $>$ A: GTA & 0.06 & 0.06 & 0.76 & 0.16 & 1.25 & 1.34 & 0.11 & 0.06 & 0.05 & 0.09 & 0.03 & 0.02 & 0.08 & 0.12\\
T $>$ A: GTC & 0.14 & 0.21 & 0.31 & 0.37 & 0.88 & 0.5 & 0.28 & 0.01 & 0.02 & 0.02 & 0.01 & 0.01 & 0.02 & 0.06\\
T $>$ A: GTG & 0.25 & 0.71 & 1.04 & 0.24 & 0.35 & 0.69 & 0.24 & 0.08 & 0.08 & 0.04 & 0.02 & 0.01 & 0.1 & 0.18\\
T $>$ A: GTT & 0.1 & 0.34 & 0.55 & 0.51 & 1.48 & 0.65 & 0.57 & 0.02 & 0.03 & 0.03 & 0.03 & 0.01 & 0.03 & 0.09\\
T $>$ A: TTA & 0.15 & 0.22 & 1.04 & 1.2 & 0.4 & 3.03 & 1.07 & 0.07 & 0.1 & 0.19 & 0.08 & 0.02 & 0.17 & 0.5\\
T $>$ A: TTC & 0.23 & 0.18 & 0.35 & 0.26 & 2.58 & 1.21 & 0.35 & 0.03 & 0.04 & 0.08 & 0.02 & 0.03 & 0.07 & 0.1\\
T $>$ A: TTG & 0.25 & 0.07 & 0.73 & 0.2 & 0.16 & 1.94 & 0.06 & 0.09 & 0.07 & 0.14 & 0.04 & 0.02 & 0.16 & 0.13\\
T $>$ A: TTT & 0.01 & 0.33 & 0.85 & 1.52 & 0.37 & 2.73 & 1.29 & 0.04 & 0.1 & 0.17 & 0.09 & 0.02 & 0.16 & 0.47\\
T $>$ C: ATA & 0.19 & 0.76 & 1.79 & 2.07 & 1.15 & 2.66 & 2.92 & 0.16 & 0.19 & 0.14 & 0.19 & 0.03 & 0.13 & 0.44\\
T $>$ C: ATC & 0.34 & 0.63 & 0.37 & 0.9 & 0.15 & 0.71 & 1.16 & 0.06 & 0.07 & 0.05 & 0.05 & 0.01 & 0.06 & 0.1\\
T $>$ C: ATG & 0.89 & 0.93 & 1.24 & 1.57 & 0.56 & 1.42 & 1.63 & 0.05 & 0.07 & 0.04 & 0.05 & 0.01 & 0.04 & 0.08\\
T $>$ C: ATT & 0.22 & 1.06 & 0.68 & 1.35 & 1.63 & 1.77 & 3.22 & 0.23 & 0.25 & 0.15 & 0.23 & 0.03 & 0.22 & 0.42\\
T $>$ C: CTA & 0.22 & 0.16 & 1.02 & 1.39 & 0.27 & 1.1 & 0.87 & 0.04 & 0.06 & 0.06 & 0.05 & 0.01 & 0.03 & 0.16\\
T $>$ C: CTC & 0.94 & 0.41 & 0.55 & 1.86 & 0.42 & 0.72 & 0.8 & 0.04 & 0.04 & 0.02 & 0.06 & 0.01 & 0.04 & 0.08\\
T $>$ C: CTG & 1.26 & 0.25 & 0.91 & 2.29 & 1.98 & 0.82 & 0.54 & 0.08 & 0.05 & 0.07 & 0.12 & 0.02 & 0.04 & 0.07\\
T $>$ C: CTT & 1.07 & 0.57 & 0.88 & 4.59 & 0.27 & 0.76 & 1.01 & 0.11 & 0.1 & 0.06 & 0.22 & 0.04 & 0.06 & 0.15\\
T $>$ C: GTA & 0.93 & 0.13 & 0.93 & 1.46 & 1.53 & 1.09 & 0.94 & 0.06 & 0.07 & 0.07 & 0.03 & 0.01 & 0.05 & 0.15\\
T $>$ C: GTC & 0.4 & 0.63 & 0.32 & 0.99 & 0.61 & 0.34 & 1.08 & 0.06 & 0.09 & 0.05 & 0.04 & 0.01 & 0.06 & 0.11\\
T $>$ C: GTG & 1.04 & 0.33 & 0.64 & 1.34 & 1.85 & 0.59 & 0.77 & 0.04 & 0.03 & 0.05 & 0.05 & 0.01 & 0.03 & 0.07\\
T $>$ C: GTT & 0.58 & 0.41 & 0.53 & 1.35 & 1.58 & 0.92 & 1.66 & 0.1 & 0.12 & 0.08 & 0.07 & 0.01 & 0.11 & 0.17\\
T $>$ C: TTA & 0.21 & 0.18 & 0.84 & 1.41 & 0.53 & 1.63 & 1.64 & 0.09 & 0.12 & 0.1 & 0.09 & 0.02 & 0.1 & 0.35\\
T $>$ C: TTC & 0.23 & 0.23 & 0.34 & 1.71 & 1.43 & 0.98 & 1.49 & 0.09 & 0.12 & 0.08 & 0.07 & 0.01 & 0.13 & 0.24\\
T $>$ C: TTG & 0.54 & 0.54 & 0.55 & 0.98 & 0.27 & 0.61 & 0.84 & 0.03 & 0.04 & 0.02 & 0.02 & 0.01 & 0.02 & 0.05\\
T $>$ C: TTT & 0.38 & 0.05 & 0.64 & 3.03 & 0.01 & 3.51 & 1.64 & 0.07 & 0.12 & 0.24 & 0.1 & 0.03 & 0.17 & 0.5\\
T $>$ G: ATA & 0 & 0.27 & 0.07 & 0.76 & 0.72 & 0.99 & 0.57 & 0 & 0.05 & 0.05 & 0.04 & 0.01 & 0.08 & 0.17\\
T $>$ G: ATC & 0.23 & 0.01 & 0.05 & 0.31 & 1.72 & 0.52 & 0.01 & 0.03 & 0.03 & 0.05 & 0.03 & 0.02 & 0.05 & 0.04\\
T $>$ G: ATG & 0.06 & 0.37 & 0.21 & 0.32 & 1.22 & 0.68 & 0.38 & 0.02 & 0.02 & 0.03 & 0.02 & 0.01 & 0.04 & 0.07\\
T $>$ G: ATT & 0.19 & 0.1 & 0.12 & 2.19 & 1.97 & 1.22 & 0.32 & 0.05 & 0.06 & 0.08 & 0.11 & 0.03 & 0.09 & 0.14\\
T $>$ G: CTA & 0 & 0.34 & 0.02 & 0.67 & 1.02 & 0.47 & 0.33 & 0 & 0.03 & 0.02 & 0.03 & 0.01 & 0.05 & 0.07\\
T $>$ G: CTC & 0.36 & 0.24 & 0.02 & 0.84 & 2.94 & 0.3 & 0.15 & 0.04 & 0.04 & 0.04 & 0.06 & 0.03 & 0.05 & 0.13\\
T $>$ G: CTG & 0.34 & 0.5 & 0.36 & 0.9 & 0.46 & 0.62 & 0.31 & 0.02 & 0.04 & 0.02 & 0.04 & 0.01 & 0.03 & 0.06\\
T $>$ G: CTT & 0.82 & 0.18 & 0.44 & 9.96 & 0.77 & 0.06 & 0.05 & 0.27 & 0.21 & 0.12 & 0.61 & 0.1 & 0.14 & 0.23\\
T $>$ G: GTA & 0 & 0.17 & 0.05 & 0.32 & 1.57 & 0.31 & 0.16 & 0 & 0.03 & 0.03 & 0.02 & 0.02 & 0.04 & 0.06\\
T $>$ G: GTC & 0.17 & 0.23 & 0.07 & 0.28 & 0.35 & 0.21 & 0.12 & 0.01 & 0.02 & 0.01 & 0.01 & 0 & 0.03 & 0.04\\
T $>$ G: GTG & 0.06 & 1.24 & 0.53 & 0.3 & 0.71 & 0.38 & 0.34 & 0.06 & 0.06 & 0.07 & 0.03 & 0.01 & 0.08 & 0.12\\
T $>$ G: GTT & 0.06 & 0.53 & 0 & 3.2 & 1.61 & 0.29 & 0.44 & 0.06 & 0.06 & 0.01 & 0.17 & 0.03 & 0.08 & 0.09\\
T $>$ G: TTA & 0.02 & 0.27 & 0.03 & 1.04 & 0.39 & 1.04 & 0.89 & 0.04 & 0.08 & 0.04 & 0.05 & 0.01 & 0.11 & 0.22\\
T $>$ G: TTC & 0.27 & 0.15 & 0.1 & 0.45 & 1.09 & 0.84 & 0.24 & 0.02 & 0.03 & 0.05 & 0.02 & 0.01 & 0.05 & 0.08\\
T $>$ G: TTG & 0.17 & 0.57 & 0.29 & 0.68 & 2.15 & 0.65 & 0.38 & 0.05 & 0.04 & 0.03 & 0.03 & 0.02 & 0.04 & 0.12\\
T $>$ G: TTT & 0.69 & 0.14 & 0.31 & 5.45 & 0.89 & 2.83 & 0.78 & 0.13 & 0.14 & 0.19 & 0.27 & 0.05 & 0.19 & 0.38\\
[1ex] 
\hline 
\end{tabular}
}
\label{weights.errors.2} 
\end{table}

\newpage\clearpage

\newpage\clearpage

\newpage\clearpage

\newpage\clearpage

\newpage\clearpage

\newpage\clearpage

\newpage\clearpage

\newpage\clearpage

\newpage\clearpage

\newpage\clearpage

\newpage\clearpage

\newpage\clearpage

\newpage\clearpage

\newpage\clearpage

\newpage\clearpage

\newpage\clearpage

\newpage\clearpage

\newpage\clearpage

\newpage\clearpage

\newpage\clearpage

\newpage\clearpage

\end{document}